\documentclass[a4paper,fleqn,usenatbib]{mnras}

\usepackage{newtxtext,newtxmath}

\usepackage[T1]{fontenc}
\usepackage{ae,aecompl}

\usepackage{graphicx}	
\usepackage{amsmath}	
\usepackage{amssymb}	
\usepackage{threeparttable}

\title[Chemical evolution of a dwarf irregular galaxy]{Star formation and gas flow history of a dwarf irregular galaxy traced by gas-phase and stellar metallicities}

\author[]{
Nao Fukagawa$^{1}$$^{2}$\thanks{E-mail: nao.fukagawa@nao.ac.jp}
\\
$^{1}$The Graduate University for Advanced Studies, SOKENDAI, 2-21-1 Osawa, Mitaka, Tokyo, 181-8588, Japan\\
$^{2}$Tohoku University, 6-3 Aramaki, Aoba-ku, Sendai, Miyagi, 980-8578, Japan
}

\date{Accepted XXX. Received YYY; in original form ZZZ}

\pubyear{2019}

\begin{document}
\label{firstpage}
\pagerange{\pageref{firstpage}--\pageref{lastpage}}
\maketitle

\begin{abstract}
Studying the evolution of dwarf galaxies can provide insights 
into the characteristics of systems that can act as building blocks 
of massive galaxies.
This paper discusses the history of star formation and gas flows 
(inflow and outflow) of 
a dwarf irregular galaxy in the Local Group, NGC 6822,
from the viewpoint of gas-phase and stellar chemical abundance.
Gas-phase oxygen abundance, stellar metallicity distribution and 
gas fraction data are compared to chemical evolution models in which 
continuous star formation and gas flows are assumed.
If the galaxy is assumed to be a closed or 
an accretion-dominated system where steeper stellar initial mass 
functions are allowed, 
the observed gas-phase oxygen abundance and gas fraction can be 
explained simultaneously; 
however metallicity distributions predicted by the models seem to be 
inconsistent with the observed distribution, 
which suggests that the star formation, gas flows and/or chemical
enrichment are more complex than assumed by the models.
When NGC 6822 is assumed to be a system dominated by outflow,
the observed values of gas-phase oxygen abundance and gas fraction 
can be explained, and
the metallicity distributions predicted by some of the models are also
roughly consistent with the observed distribution in the metallicity range of $-2.0\lesssim$[Fe/H]$\lesssim-0.5$.
It should be noted that this result does not necessarily mean that
the accretion of gas is completely ruled out.
More observables, such as chemical abundance ratios, and detailed 
modelling may provide deeper insight into the evolution of the system.
\end{abstract}

\begin{keywords}
galaxies: dwarf 
\end{keywords}



\section{Introduction}
\label{sec:1}

It is important to study how dwarf galaxies evolve,
because small systems are the building blocks of larger systems in the 
framework of the hierarchical merging scenario.
Each dwarf galaxy has its own star 
formation history \citep[][]{M98,T09}. 
When supernovae inject sufficient energy to 
blow the interstellar gas away, 
a dwarf galaxy may lose material to form stars.
The accretion of gas or gas-rich systems
may also have an influence on star formation.
Investigating the evolution of individual star-forming dwarf galaxies
can provide insight into understanding the variation in 
star formation history of dwarf galaxies.

Metallicity is a fundamental property that provides information
about the evolution of galaxies.
Interstellar gas turns into stars,
which produce heavy elements and then return their gas and the synthesized 
elements into the interstellar medium.
Stars in subsequent generations are formed from gas
enriched by previous generations.
Galaxies may also acquire and/or lose gas.
The accretion of low-metallicity gas on to a galaxy 
may temporally decrease the gas-phase metallicity.
The outflow triggered by supernovae carries part of the gas, including
heavy elements, out of the galaxy.
Therefore, chemical abundances trace the histories of 
star formation and gas flows of galaxies.

There have been many studies on the chemical evolution of nearby star-forming
dwarf galaxies, including dwarf irregular galaxies and blue compact 
dwarf galaxies, based on gas-phase abundances
\citep[][references therein]{P97,T98,MF12}.
These studies explain the abundances of star-forming dwarf galaxies 
mainly by the stellar initial mass function (IMF) for the solar neighbourhood 
or steeper IMFs with which less-massive stars are formed more frequently,
galactic wind and/or discontinuous star formation. 
For some dwarf irregular galaxies in the Local Group, 
individual stars are resolved, and stellar abundances are measured
from the spectra
\citep[e.g.][]{V01,T07}.
Metallicity distributions, from which the histories of 
star formation and gas flows are obtained, and the evolution of
individual galaxies are discussed by comparing the observed distribution with
analytical chemical evolution models \citep[e.g.][]{K13,L13}.

The histories of the star formation and gas flows of dwarf irregular galaxies
are discussed from both sides of gas-phase and stellar
metallicities, but the observational data can be explained by
models with different assumptions \citep[][]{T98}, 
which brings about different suggestions.
What can be inferred about the evolution of dwarf irregular galaxies 
from both of gas-phase and stellar metallicities with unique models?
In addition to sophisticated models and simulations \citep[e.g.][]{RJ18}, 
models with a small number
of assumptions also provide insight into the dominant physical processes.

Within this context, this paper investigates
the evolution of a dwarf irregular galaxy 
by comparing the gas-phase oxygen abundance, gas fraction and 
metallicity distribution to chemical evolution models.
Among the dwarf irregular galaxies in the Local Group,
this work focuses on NGC 6822.
This galaxy is one of the nearest dwarf irregular galaxies to the
Milky Way \citep[$\sim$460~kpc;][references therein]{M12}.
The stellar mass \citep[$1\times10^8~{\rm M_{\odot}}$;][references therein]{M12} is 
about one-fifth as large as that of the Small Magellanic Cloud,
but NGC 6822 is located in a more isolated environment \citep{M98}.
The gas mass is almost comparable to the stellar mass
\citep[$\rm{M}_{\rm {HI}}=1.34\times10^8~{\rm M_{\odot}}$;][]{dW06}.
The integrated star formation rate, as measured using H~$\alpha$ images, is 
0.01~${\rm M_{\odot}\,yr^{-1}}$ \citep{HE04}, 
and a recent increase in the star formation rate has been reported
\citep[e.g.][]{C12}.
The gas-phase oxygen abundance has been measured for \textsc{H\,ii} regions
\citep[e.g.][]{S89,L06,P05}
and planetary nebulae \citep[e.g.][]{RM07,H09}.
The measured value is around 0.2--0.4 ${\rm Z_{O,\odot}}$\footnote{In this study, chemical abundances are represented by ${\rm Z_{A}=12+\log(A/H)}$ and ${\rm [A/B]=\log(A/B)-\log(A/B)_{\odot}}$, where A and B are abundances of elements A and B. If not otherwise specified, [Fe/H] is considered to be the stellar metallicity.}, 
and may vary depending on the measurement method.
There is not clear evidence for the abundance gradient
\citep[e.g.][]{L06} and the gas of the galaxy is homogeneous 
\citep[e.g.][]{H09} within 2~kpc radius from the centre.
Stellar abundances have also been investigated by observing
supergiants \citep[e.g.][]{M99,V01,P15}, 
globular clusters \citep[e.g.][]{C10,LSS18} and 
red giants \citep[e.g.][]{T01,K13,S16}.

The literature discusses the chemical history of NGC 6822 based on gas-phase 
abundances with chemical evolution models.
For example, \citet{C06} and \citet{H11} discussed 
the chemical evolution of NGC 6822
by comparing data on gas-phase abundances and abundance ratios to the models
in which the suppression of gas accretion due to reionization 
is assumed.
In these papers, the authors suggested the need to consider outflow.
The metallicity distribution has also been studied using models.
\citet{K13} showed that the shape of the observed metallicity distribution 
can be explained by an analytical model in which outflow is dominant, and 
a small amount of the gas accretes into the galaxy.
They also suggested that most of the iron produced by 
stars has been lost from NGC 6822, based on the yields of supernovae and
gas-phase and stellar abundances.
\citet{H15} explained the shape of the distribution using a model
taking gas accretion and multiple small starbursts into account, 
and interpreted the result
as star formation may be repeatedly ignited by merging with gas-rich systems or
at intersections of streams of gas.

The present study considers both sides of the gas-phase and stellar 
metallicities using chemical evolution models
in which gas accretion or outflow is assumed to dominate the system.
The structure of this paper is as follows:
observational data of NGC 6822 compared to models are 
summarized in Section \ref{sec:2};
details of the models are described in Section \ref{sec:3};
observational data are compared to the models in Section \ref{sec:4};
and a summary and discussion are presented in Section \ref{sec:5}.

\section{Observational data}
\label{sec:2}

In this study, gas-phase oxygen abundance, 
metallicity distribution and gas mass fraction 
reported previously are compared to 
model predictions.

The gas-phase oxygen abundance ($Z_{\rm O}$) is based on the value measured by
\citet[][]{L06}.
They measured chemical abundances of nebulae at radii $\lesssim$ 2~kpc
based on the intensities of emission lines.
The temperature-sensitive [\textsc{O\,iii}]$\lambda$4363 line was 
detected for five \textsc{H\,ii} regions
and the abundances were derived by the direct method.
They did not find clear evidence for the abundance gradient.
Assuming zero slope of the gradient, they took the average
oxygen abundances of the five \textsc{H\,ii} regions.
In the present study, the value $Z_{\rm O}=8.11\pm0.11$ is considered to be
the average gas-phase oxygen abundance of NGC 6822.
The value scaled to the solar abundance \citep[][]{A04} is
$Z_{\rm O}/Z_{{\rm O,\odot}}=0.28\pm0.06$.

The stellar metallicity distribution is taken from a spectroscopic study
conducted by \citet[][]{K13}.
They measured the metallicities ([Fe/H]) of individual red giant stars 
in NGC 6822 using a method based on spectral synthesis.
The metallicity distribution consists of 276 stars.
The typical error of the metallicity of the stars is 
${\rm \langle\delta[Fe/H]\rangle=0.13}$.
The mean metallicity is $\langle$[Fe/H]$\rangle=-1.05\pm0.01$.
The distribution shows that the measured metallicity ranges from 
[Fe/H]$\sim-3.1$ to $0$,
which suggests that some of the member stars have Fe abundances as high as 
the sun, while the average gas-phase oxygen abundance is about 
30 per cent of the solar abundance.
It has been reported that the mean oxygen abundance or global metallicity 
of supergiant stars in NGC 6822 agree with that of 
\textsc{Hii} regions 
(for the oxygen abundance, $Z_{\rm O}=8.36\pm0.19$ \citep[][]{V01}, and
for the global metallicity, ${\rm \log Z/Z_{\odot}}=-0.52\pm0.21$
\citep[][]{P15}).
Because red giants are older than supergiants, low metallicities of 
red giants can be interpreted as they have been formed from low-metallicity
gas where the chemical enrichment has not proceeded yet.
However, the existence of the metal rich red giant stars cannot be explained 
by the scenario alone.
Although evidence is lacking, one possible explanation of the origin of 
the Fe-rich stars is that they formed from gas enriched by supernova(e)
before the gas had been well mixed.
It is also possible that stars in other systems have accreted on to NGC 6822.
This point should be examined in future studies.

With regard to the chemical abundances, 
stellar abundance ratios give insight into 
the chemical enrichment of galaxies. 
The $\alpha$-elements, in which oxygen included, 
are mainly produced by massive stars, 
and the Fe-peak elements can also be produced by type-Ia supernovae. 
[$\alpha$/Fe] of stars in NGC 6822 has been measured for 
several globular clusters \citep[e.g.][]{LSS18} and 
A-type supergiants \citep[][]{V01}. 
Currently, the observational data about stellar abundance ratios 
may not be sufficient to compare them with models 
and constrain the star formation history. 
The observed stellar abundance ratios of dwarf irregular galaxies will 
help to understand their evolution.

The gas mass fraction ($\mu$) is defined as the fraction of gas 
(\textsc{H\,i}$+{\rm H}_2$) mass to the sum of gas and stellar masses.
For the \textsc{H\,i} gas, \citet[][]{dW06} measured \textsc{H\,i} flux
and derived \textsc{H\,i} gas mass.
Their observations covered an area of $4^{\circ}\times4^{\circ}$.
The derived \textsc{H\,i} gas mass is 
$\rm{M}_{\rm HI}=1.34\times10^8~{\rm M}_{\odot}$.
The $\rm{H}_2$ gas mass is taken from a study by \citet[][]{I97}, who suggested 
$\rm{M}_{H_2}=1.5 (+3, -1)\times10^7~{\rm M}_{\odot}$
based on ${}^{12}$CO(${\rm J}=1-0$) brightness.
For the stellar mass, assuming a mass-to-luminosity ratio of unity,
\citet[][references therein]{M12} derived the stellar mass of NGC 6822.
The value is $\rm{M}_*=1\times10^8~{\rm M}_{\odot}$.
The gas mass fraction based on these gas and stellar masses is 0.6.
The region investigated may not always be consistent among these papers.
In the present study, the value of the gas mass fraction is considered to be
representative of NGC 6822.

\section{Chemical evolution models}
\label{sec:3}

Observational data described in the previous section
are compared to three one-zone numerical models.
These models are constructed according to \citet[][]{P08}.
The assumptions are summarized in the following sections.

In this work, it is assumed that the gas is instantaneously mixed and
homogeneous (one-zone model), 
and that stars are formed continuously.
Of course, this assumption may be too simple to completely reproduce
the observed quantities of NGC 6822.
With regard to assumptions about star formation,
time variation in the star formation rate can be derived
using photometric data \citep{C06}.
However, galaxies evolve through the interplay of star formation, 
gas accretion and outflow.
The fluctuation in star formation rate may result in temporal changes
in the rates of gas accretion and outflow, which could influence
subsequent star formation activity.
Although including assumptions about this fluctuation is important, 
it requires additional parameters in the models.
In general, the star formation activity appears to be continuing
\citep{C12,W14}.
In addition, galaxies of larger mass may have more continuous 
star formation histories compared to galaxies of lower masses 
\citep[e.g.][]{W12}.
Assuming continuous star formation may not be completely against
the point that NGC 6822 is a relatively massive dwarf galaxy.

Chemical abundances and physical properties are calculated in the time range of
$t=0.00-13.80$~Gyr with intervals of 0.01~Gyr.
Previous studies \citep[e.g.][]{F14} yielded evidence that 
there are stars older than 10 Gyr in NGC 6822.
Based on the presence of the old stellar populations, 
the observed values of gas-phase oxygen abundance and 
gas mass fraction are compared to predicted values
in the universe at present ($t=13.80$~Gyr).

The interstellar gas in NGC 6822 involves a small amount of dust
\citep[${\rm M}_{{\rm dust}}=1.4\times10^4~{\rm M_{\odot}}$;][]{I96}.
Thus, a fraction of the elements in the interstellar gas is 
probably captured in the dust,
but the degree of dust depletion in low-metallicity gas is not clear.
In this study, no assumptions about dust depletion are included 
in the models.

All calculations are performed in reduced mass scaled to the total mass
of the system.
The parameters of the models are summarised in Table~\ref{tab:model}.

\begin{table*}
\centering
\caption{Summary of chemical evolution models. $f_{\rm g}, f_{\rm *}$ and 
$f_{\rm g,h}$ representing mass fractions: $f_{\rm g}=M_{\rm g}/M_{\rm total}$, 
$f_{\rm *}=M_*/M_{\rm total}$ and $f_{\rm g,h}=M_{\rm g,h}/M_{\rm total}$.} 
\label{tab:model}
\begin{threeparttable}
\begin{tabular}{lllll} 
\hline
model & paremeter(s) & infall rate (${\rm Gyr^{-1}}$) & outflow rate (${\rm Gyr^{-1}}$) & initial condition\\
\hline
A & $k_{{\rm SF}}$ & -- & -- & $f_{\rm g}=1, f_{\rm *}=0, Z_{\rm O}=0, Z_{{\rm Fe}}=0$\\
B & $k_{\rm SF}, k_{{\rm in}}$ & $F(t)=k_{{\rm in}} M_{{\rm g,h}}$ & -- & $f_{{\rm g,h}}=1, f_{\rm g}=0, f_{\rm *}=0, Z_{\rm O}=0, Z_{{\rm Fe}}=0$ \\
C & $k_{\rm SF}, \eta$ & -- & $O(t)=\eta\,\Psi(t)$ & $f_{{\rm g,h}}=0, f_{\rm g}=1, f_{\rm *}=0, Z_{\rm O}=0, Z_{{\rm Fe}}=0$ \\
\hline
\end{tabular}
\end{threeparttable}
\end{table*}

\subsection{Closed-box model (model A)}
\label{subsec:modelA}

A galaxy is assumed to evolve as a closed system of gas mass 
$M_{\rm g}$(${\rm M}_{\odot}$) and stellar mass 
$M_{\rm *}$(${\rm M}_{\odot}$).
There is no exchange of gas with the outer region.
The total mass of the system $M_{\rm total}$(${\rm M}_{\odot}$) is 
given by
$M_{\rm total} = M_{\rm g} + M_{\rm *}$.
The evolution of gas mass and stellar mass can be described as follows:
\begin{eqnarray}
\frac{{\rm d} M_{\rm g}(t)}{{\rm d} t} = - \Psi(t) + E_{\rm g}(t), \\
\frac{{\rm d} M_{\rm *}(t)}{{\rm d} t} = \Psi(t) - E_{\rm g}(t),
\end{eqnarray}
where $\Psi(t)$ and $E_{\rm g}(t)$ are the star formation rate and 
the rate of gas ejected by dying stars at time $t$~(Gyr), respectively.
The star formation rate is assumed to be 
\begin{eqnarray}
\Psi(t) = k_{\rm SF} \, M_{\rm g}(t),
\end{eqnarray}
where $k_{{\rm SF}}$ is a constant referred to as 
the star formation efficiency (SFE).
The initial mass function $\Phi(M)$ is:
\begin{eqnarray}
\Phi(M) \propto M^{-(1+x)}, 
\end{eqnarray}
where $x$ is the index. In the case of the Salpeter IMF \citep[][]{S55},
$x=1.35$.
The IMF is normalized as follows:
\begin{eqnarray} 
\Phi(M)=A \int^{M_{\rm U}}_{M_{\rm L}} \Phi(M)\, M\, {\rm d}M=1,
\end{eqnarray}
where $A$ is a normalizarion constant.
In this study, the lower and upper limits of masses of stars $M_{\rm L}$ 
and $M_{\rm U}$ are assumed to be $M_{\rm L}=0.08~{\rm M}_{\odot}$ 
and $M_{\rm U}=50~{\rm M}_{\odot}$, respectively.
Stars more massive than 50~${\rm M_{\odot}}$ may be formed in dwarf galaxies.
The upper mass of 50~${\rm M_{\odot}}$ is set in view of the mass range
provided in tables of supernova yield \citep{N13}.

The rate of mass ejection of dying stars at time $t$~(Gyr) is
\begin{eqnarray}
E_{\rm g}(t) = \int^{M_{\rm U}}_{M_t} (M-C_M)\, \Psi(t-\tau_M)\, \Phi(M)\, {\rm d}M,
\end{eqnarray}
where $M_t~({\rm M}_{\odot})$ 
represents the mass of stars whose lifetime corresponds 
to the age of the system.  
For stellar lifetime $\tau_M$~(Gyr),
the theoretical values provided by \citet[][]{S92} are adopted.
With regard to the mass of stellar remnants $C_M~({\rm M}_{\odot})$, 
it is assumed that stars of mass $M/{\rm M}_{\odot} < 9$, 
$9 < M/{\rm M}_{\odot} < 25$ and 
$25 < M/{\rm M}_{\odot}$ leave white dwarfs, neutron stars and black holes,
respectively:
\begin{eqnarray}
C_M = \left\{\begin{array}{l}
\displaystyle 0.08 M + 0.47 \,\,\, (M/{\rm M}_{\odot} < 9) \\
1.35 \,\,\, (9 < M/{\rm M}_{\odot} < 25) \\
0.24 M - 4 \,\,\, (25 < M/{\rm M}_{\odot}). \\
\end{array} \right.
\end{eqnarray}

With regard to chemical enrichment, the evolution of the mass of element $i$
is:
\begin{eqnarray}
\frac{{\rm d} (M_{\rm g} X_i)}{{\rm d} t} = - \Psi(t)\, X_i(t) + E_i(t),
\end{eqnarray}
where $X_i(t)$ is the mass fraction of element $i$ and
$E_i(t)$ is the rate of ejection of element $i$.
Massive stars produce large 
amounts of $\alpha$-elements through type-II supernovae (SNe\,II),
and type-Ia supernovae (SNe\,Ia) contribute to 
the enrichment of Fe-peak elements.
Low- and intermediate-mass stars also contribute to the enrichment of 
some elements (e.g. carbon and nitrogen).
In this study, gas-phase oxygen abundance and metallicity 
distribution are compared, 
and it is assumed that chemical abundances are mainly changed 
by SNe\,II and SNe\,Ia: 
\begin{eqnarray}
E_i(t) = E_{i,{\rm II}}(t) + E_{i,{\rm I_a}}(t),
\end{eqnarray}
where $E_{i,{\rm II}}(t)$ and $E_{i,{\rm I_a}}(t)$ are the rates of ejection of 
mass of element $i$ by SNe\,II and SNe\,Ia, respectively.
For SNe\,II,
\begin{eqnarray}
E_{i,{\rm II}}(t) = \int^{M_{\rm U}}_{M_t} \{y_i + X_i(t-\tau_M)\, (M-C_M-y_i)\}\, \Psi(t-\tau_M)\, \Phi(M)\, {\rm d}M.
\end{eqnarray}
The first bracket shows the amount of element $i$ ejected by a star 
and $y_i$(${\rm M_{\odot}}$) represents the amount of element $i$ newly 
produced by the star.
In this study, the metallicity-dependent yields of SNe\,II 
($Z=0, 0.001, 0.004, 0.008, 0.02$ and $0.05$) provided 
by \citet[][]{N13} are adopted. 
They take the mass loss of stars into account in the calculation.
A contribution to chemical enrichment by hypernovae 
is not assumed in the present study.
For SNe\,Ia, it is assumed that a proportion of stars of 
3--8~${\rm M_{\odot}}$ are binary and explode $\tau_{{\rm I_a}}$~Gyr 
after formation.
The rate of ejection of element $i$ by SNe\,Ia is described 
as follows:
\begin{eqnarray}
E_{i,{\rm I_a}}=f_{{\rm I_a}} \int^{8 {\rm M_{\odot}}}_{3 {\rm M_{\odot}}} y_{i,{\rm I_a}} \, \Phi(M) \, \Psi(t-\tau_{{\rm I_a}}) \, {\rm d}M,
\end{eqnarray}
where $y_{i,{\rm I_a}}~({\rm M_{\odot}})$ is the yield \citep[][]{N97}.
The fraction $f_{{\rm I_a}}$ and lifetime $\tau_{{\rm I_a}}$~(Gyr) are determined 
to be $f_{{\rm I_a}}=0.05$ and $\tau_{{\rm I_a}}=2.0$~(Gyr), respectively, 
based on a comparison 
between the infall model (model B) and abundances of stars in the Milky Way.
The details are explained in the Appendix~\ref{app:A}.
Predicted metallicity distributions in this work are convolved 
with an error function on the assumption that observational data include 
uncertainty (set to 0.1~dex as a typical value).
When SNe\,Ia start to contribute to chemical enrichment
(2~Gyr after star formation starts in the galaxy),
the abundance of Fe begins to increase sharply.
This effect makes the metallicity distributions from the models discontinuous,
resulting in the overproduction of stars on the metal-poor side of the
distributions.
The overproduction is particularly clear for models A and C, in which
the accretion of gas is not taken into account.
The simple assumption regarding the enrichment by SNe\,Ia partly causes
the overproduction,
and in this work the shape of the main peak of the distributions is 
compared to the observational data.

\subsection{Infall model (model B)}
A galaxy is assumed to be a system in which the accretion of 
primordial gas dominates.
Part of the metal-free gas in the gas reservoir accretes into the galaxy.
In contrast to model A in which a galaxy is assumed to be formed from a
gas cloud, in model B, the gas is provided from the gas reservoir 
to the main body.
Thus, the gas mass and star formation rate increase
with time in the early stage of the evolution.
The total mass of the system is $M_{{\rm total}}=M_{\rm g} + M_{\rm *} + M_{{\rm g,h}}$,
where $M_{{\rm g,h}}$(${\rm M_{\odot}}$) represents the gas mass of the reservoir.
The evolution of the mass of gas, stars and element $i$ in the galaxy is 
described as follows:
\begin{eqnarray}
\frac{{\rm d} M_{\rm g}}{{\rm d} t} = - \Psi(t) + E_{\rm g}(t) + F(t), \\
\frac{{\rm d} M_{\rm *}}{{\rm d} t} = \Psi(t) - E_{\rm g}(t), \\
\frac{{\rm d} (M_{\rm g} X_i)}{{\rm d} t} = - \Psi(t)\, X_i(t) + E_i(t).
\end{eqnarray}
The infall rate $F(t)$ is simply assumed to be proportional 
to the gas mass in the reservoir: 
\begin{eqnarray}
F(t) = k_{{\rm in}}\,M_{{\rm g,h}}(t).
\end{eqnarray}
The coefficient $k_{{\rm in}}$ is a constant 
referred to as the accretion efficiency (ACE).
$k_{{\rm in}}$ in this model and $\eta$ in model C may depend on 
other physical quantities, such as time \citep[e.g.][]{C06,L18}, 
but this dependence is still a matter of some debate.
In this study, these quantities are assumed to be constants.
The investigated ranges of the SFE and ACE are
$k_{\rm SF}=0.010-0.090~({\rm Gyr}^{-1})$ and
$k_{\rm in}=0.010-0.90~({\rm Gyr}^{-1})$, respectively.
Other assumptions are the same as those of model A.

\subsection{Outflow model (model C)}
A galaxy is considered an outflow-dominated system.
Part of the interstellar gas, including heavy elements, is expelled 
due to galactic wind accompanying supernovae.
Thus, the outflow rate $O(t)$ is assumed to be proportional to
the star formation rate \citep[][]{H76}: 
$O(t)=\eta\,\Psi(t)$, where $\eta$ is a constant referred to as 
the mass-loading factor.
It is assumed that the ejecta of supernovae are
mixed with the interstellar medium before the outflow
and that the chemical abundances of the outflowing gas are equal to 
those of the interstellar gas of the galaxy at time $t$.
The total mass of the system is $M_{{\rm total}}=M_{\rm g} + M_{\rm *} + M_{{\rm out}}$,
where $M_{\rm out}~({\rm M}_{\odot})$ is the mass of gas ejected from the galaxy.
The evolution is described as follows:
\begin{eqnarray}
\frac{{\rm d} M_{\rm g}}{{\rm d} t} = - \Psi(t) + E_{\rm g}(t) - O(t), \\
\frac{{\rm d} M_{\rm *}}{{\rm d} t} = \Psi(t) - E_{\rm g}(t), \\
\frac{{\rm d} (M_{\rm g} X_i)}{{\rm d} t} = - \Psi(t)\, X_i(t) + E_i(t) - O(t)\, X_i(t).
\end{eqnarray}
Other assumptions are the same as those of model A.

\section{Comparing the observational data with models}
\label{sec:4}
\subsection{Closed system}
It is assumed that NGC 6822 is a closed system.
The properties are compared to model A.

\begin{figure}
\centering
\begin{tabular}{c}
\begin{minipage}{0.50\hsize}
\centering
\includegraphics[scale=0.46]{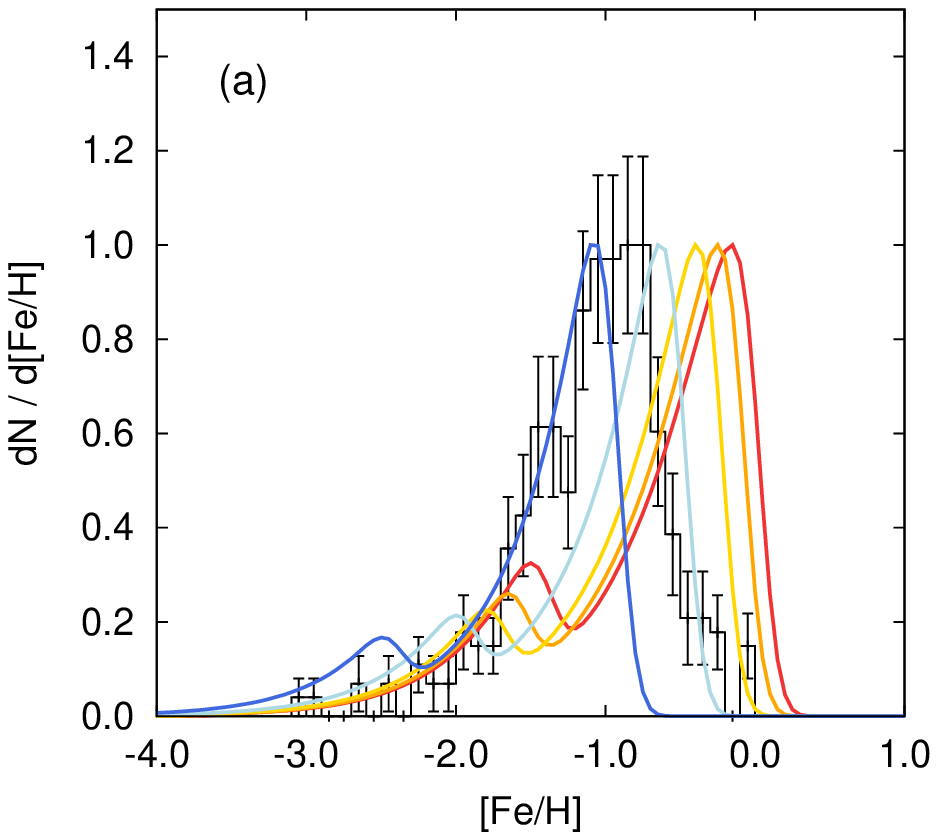}
\end{minipage}
\begin{minipage}{0.50\hsize}
\centering
\includegraphics[scale=0.46]{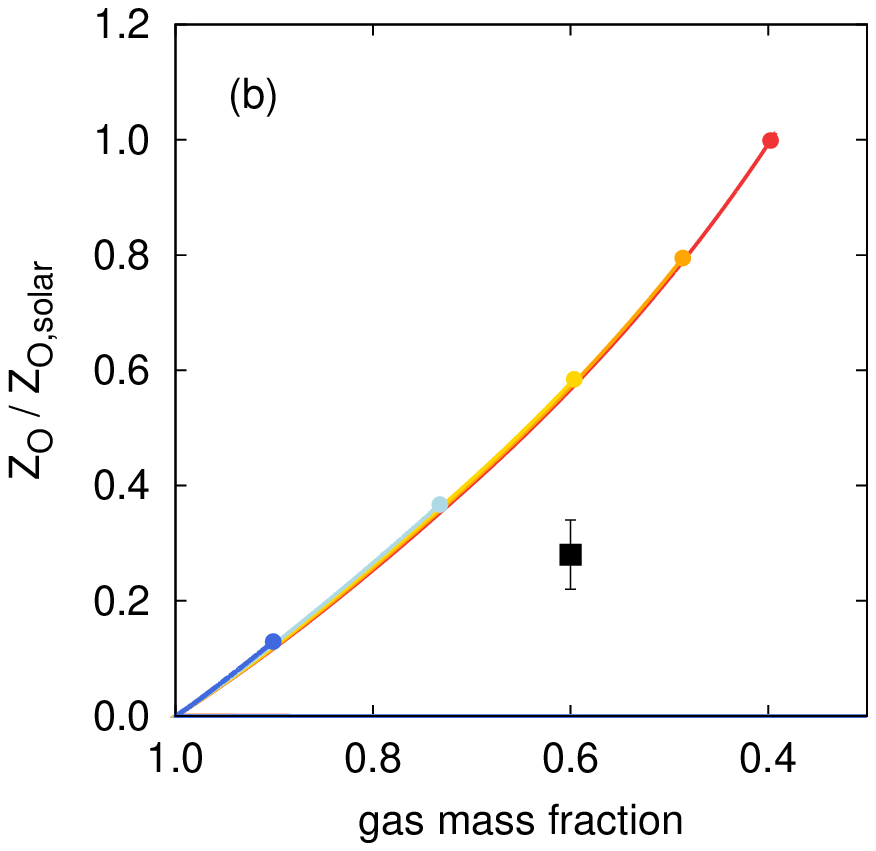} 
\end{minipage}
\end{tabular}
\caption{Comparisons between observational data of NGC 6822 and model A.
The observed metallicity distribution (panel a) is taken from \citet{K13}.
For panel b, the gas mass fraction is derived based on the stellar mass from \citet{M12}
and the gas mass \citep{dW06,I97}.
The gas-phase oxygen abundance is from \citet{L06} and is scaled to the solar value.
For the observational data, see Sec.~\ref{sec:2}.
The curves in the two panels are predictions from model A.
The colours correspond to the SFE: blue, sky blue, yellow, orange and red curves
show cases of $k_{\rm SF}=0.010, 0.030, 0.050, 0.070$ and $0.090~({\rm Gyr}^{-1})$,
respectively.
Points in panel b are the predicted gas-phase oxygen abundance and
gas mass fraction at present in the universe.
}
\label{fig:A}
\end{figure}

Firstly, star formation is assumed to obey
the IMF for the solar neighbourhood, and
the Salpeter IMF \citep[][$x=1.35$]{S55} is adopted.
Fig.~\ref{fig:A} shows the observational data and predictions of 
the models for different SFEs.
Fig.~\ref{fig:A}a shows the metallicity distributions.
According to the analytical solution of the closed-box model, metallicity
depends on the gas fraction and yield \citep[e.g.][]{P97}.
The differences in metallicity at the peak of metallicity distributions 
(${\rm [Fe/H]_{peak}}$) in Fig. \ref{fig:A}a 
are due to low SFE.
Fig.~\ref{fig:A}b shows the $\mu$--$Z_{\rm O}$ diagram.
The curves are evolutionary tracks of galaxies predicted by the models.
The gas turns into stars and the stars produce heavy elements,
so galaxies evolve along the curve from left to right on this plane.
Points on the curves show the gas-phase oxygen abundance and
the gas mass fraction in the universe at present.
Galaxies of higher SFE consume gas more rapidly by 
star formation, so they have a smaller gas mass fraction at any given time.
They also have higher gas-phase oxygen abundance at any given time,
because a large amount of oxygen is produced by SNe\,II.
Galaxies evolve along almost self-similar 
tracks on the $\mu$--$Z_{\rm O}$ diagram, regardless of the SFE.

The gaseous properties and metallicity distribution seem not to be
explained simultaneously with model A.
Regarding the gas-phase oxygen abundance and gas mass fraction,
the gas-phase oxygen abundance predicted by model A is high compared to
the observed value at the gas mass fraction of NGC 6822 ($\mu=0.6$).
For example, a gas mass fraction in a model of $k_{\rm SF}=0.050~({\rm Gyr^{-1}})$
(yellow curve in Fig.~\ref{fig:A}b) is almost consistent with 
the observed value of gas mass fraction,
but this model predicts gas-phase oxygen abundance of 
$Z_{\rm O}/Z_{\rm O,\odot}\sim0.6$, while the observed value is
$Z_{\rm O}/Z_{\rm O,\odot}\sim0.3$.
Thus, the observed low gas-phase oxygen abundance cannot
be explained only by assuming different SFEs.
These results do not conflict with those of previous studies \citep[e.g.][]{MC83}.
In light of stellar and gas-phase abundances,
Fig.~\ref{fig:A} suggests that [Fe/H$]_{\rm peak}$ and gas-phase
oxygen abundance of NGC 6822 may be explained by models of 
$k_{\rm SF}\sim0.015~({\rm Gyr^{-1}})$.
However, the gas mass fractions predicted by the models are
larger than the observed value.
Therefore, if NGC 6822 is a closed system, there may be physical processes
in which heavy elements are not produced efficiently.
One possibility is that the 
IMF is steeper than Salpeter's law \citep[e.g.][]{L79,MC83}.

\subsubsection{Cases of steeper IMF}
\label{subsec:A}

If there is a system where star formation follows a steep IMF, 
massive stars are formed less frequently.
Thus, a galaxy of a steeper IMF can have a lower average metallicity 
at a given gas fraction.
Although it is not clear whether the IMF for dwarf galaxies differs from 
that for the solar neighbourhood,
in this section the observational data of NGC 6822 are compared to
model A under the assumption that star formation can obey
a steeper IMF compared to that for the solar neighbourhood.

\begin{figure}
\centering
\begin{tabular}{c}
\begin{minipage}{0.50\hsize}
\centering
\includegraphics[scale=0.46]{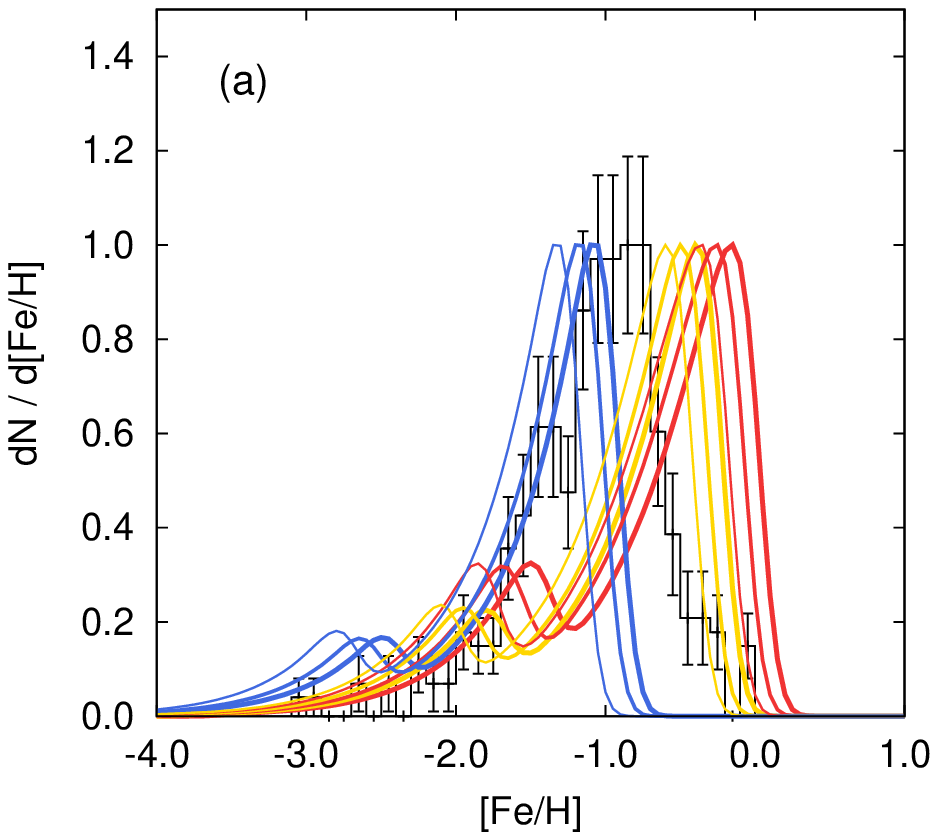}
\end{minipage}
\begin{minipage}{0.50\hsize}
\centering
\includegraphics[scale=0.46]{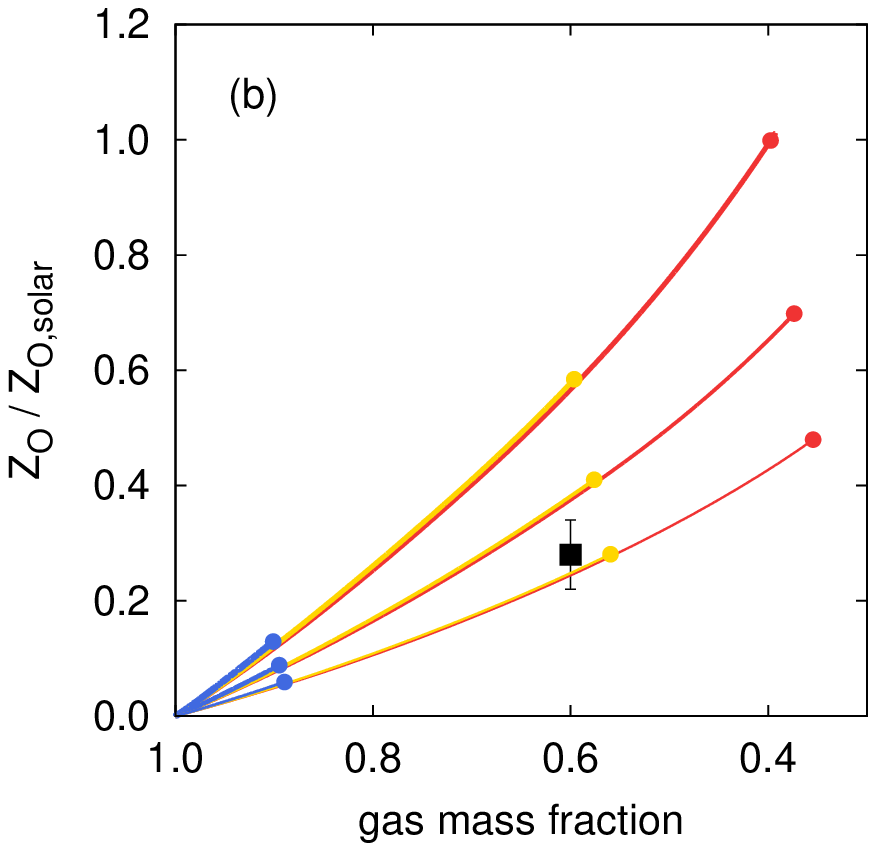} 
\end{minipage}
\end{tabular}
\caption{Metallicity distributions (panel a), gas-phase oxygen abundances and 
gas mass fractions (panel b) of galaxies of different IMFs and SFEs predicted 
by model A.
Thinner curves indicate galaxies of steeper IMF.
The index of IMF is varied by 0.10:
$x=1.35$ (thickest curves), $1.45$ and $1.55$ (thinnest curves).
The colours of the curves correspond to the SFE:
blue, yellow and red curves show galaxies of $k_{\rm SF}=0.010, 0.050$ 
and $0.090~({\rm Gyr}^{-1})$, respectively.
Points on the curves in panel b are present-day values.
The black histogram and square in the panels are observational data 
presented in Sec.~\ref{sec:2}.
}
\label{fig:A2}
\end{figure}

Figure~\ref{fig:A2} shows the metallicity distributions, 
gas-phase oxygen abundances and gas mass fractions of galaxies 
with different indices of IMF
($x=1.35$, $1.45$ and $1.55$) and SFEs ($k_{{\rm SF}}=0.010$, $0.050$ and 
$0.090~(\rm{Gyr}^{-1})$).
Similar to the case of the Salpeter IMF, 
when the IMF is identical, galaxies evolve along almost self-similar tracks
on the $\mu$--$Z_{\rm O}$ diagram, regardless of the SFE.
When galaxies of the same SFE but different IMFs are compared, 
galaxies of steeper IMFs have lower 
gas-phase oxygen abundances in the universe at present.
Galaxies of steeper IMFs have slightly smaller present-day gas mass fractions,
because more interstellar gas is locked in low-mass stars.
With regard to the metallicity distributions, 
galaxies of steeper IMFs have lower [Fe/H$]_{\rm peak}$.
For example, a galaxy of $x=1.55$ and $k_{\rm SF}=0.010~({\rm Gyr^{-1}})$
(thinnest blue curve in Fig.~\ref{fig:A2}a) has a metallicity distribution
of [Fe/H$]_{\rm peak}\sim-1.4$, while the metallicity distribution of 
a galaxy of the same SFE but $x=1.35$ (thickest blue curve) has 
[Fe/H$]_{\rm peak}\sim-1.1$.

Fig.~\ref{fig:A2}b suggests that the observed values of 
the gas mass fraction and gas-phase oxygen abundance may be explained 
if the IMF is assumed to be $x\sim1.50-1.55$.
When the SFE is assumed to be $k_{\rm SF}\sim0.040-0.050~(\rm{Gyr}^{-1})$,
the observed values of gas-phase abundance and gas mass fraction can be consistent 
with the model predictions within the range of error.
On the other hand, the metallicity distributions predicted by the models show
higher [Fe/H$]_{\rm peak}$ compared to the observed distribution and
slightly underestimate stars of [Fe/H$]\sim0$.
These results suggest that if this galaxy is a closed system
where steeper IMFs are allowed, 
it may be characterized by physical processes that are not assumed in
model A, such as discontinuous star formation.

\subsection{Accretion-dominated system}
\label{sec:4B}
NGC 6822 is assumed to be a system where the accretion of primordial gas 
is dominant.
The data are compared to model B.

\begin{figure*}
\centering
\begin{tabular}{c}
\begin{minipage}{0.25\hsize}
\centering
\includegraphics[scale=0.46]{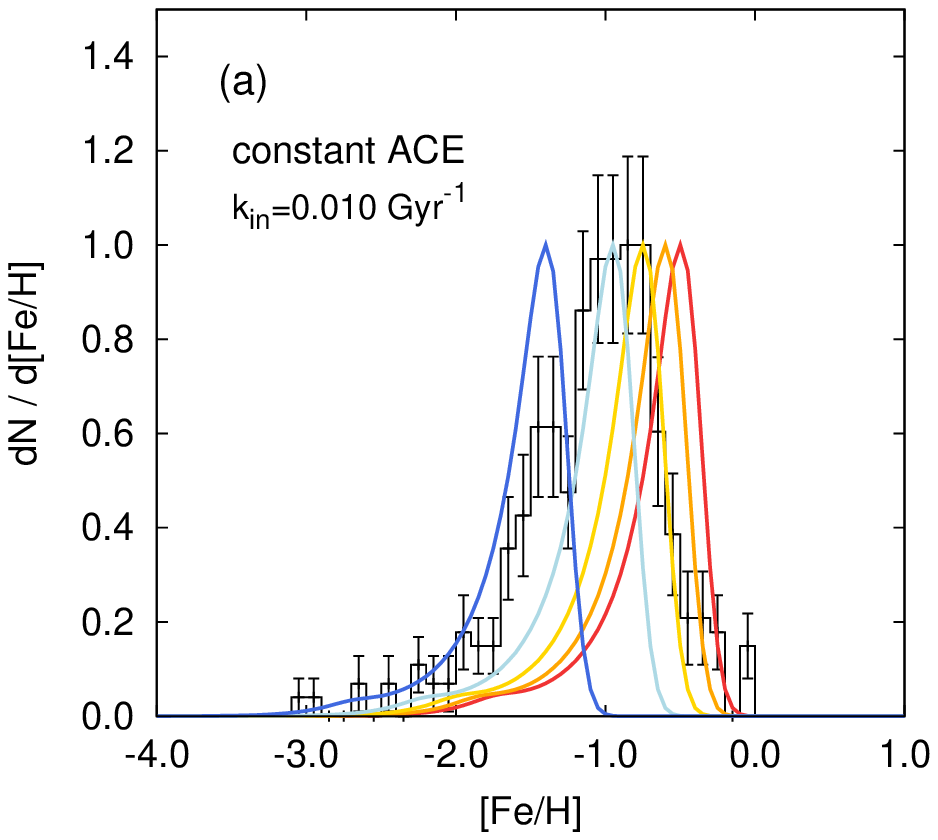}
\end{minipage}
\begin{minipage}{0.25\hsize}
\centering
\includegraphics[scale=0.46]{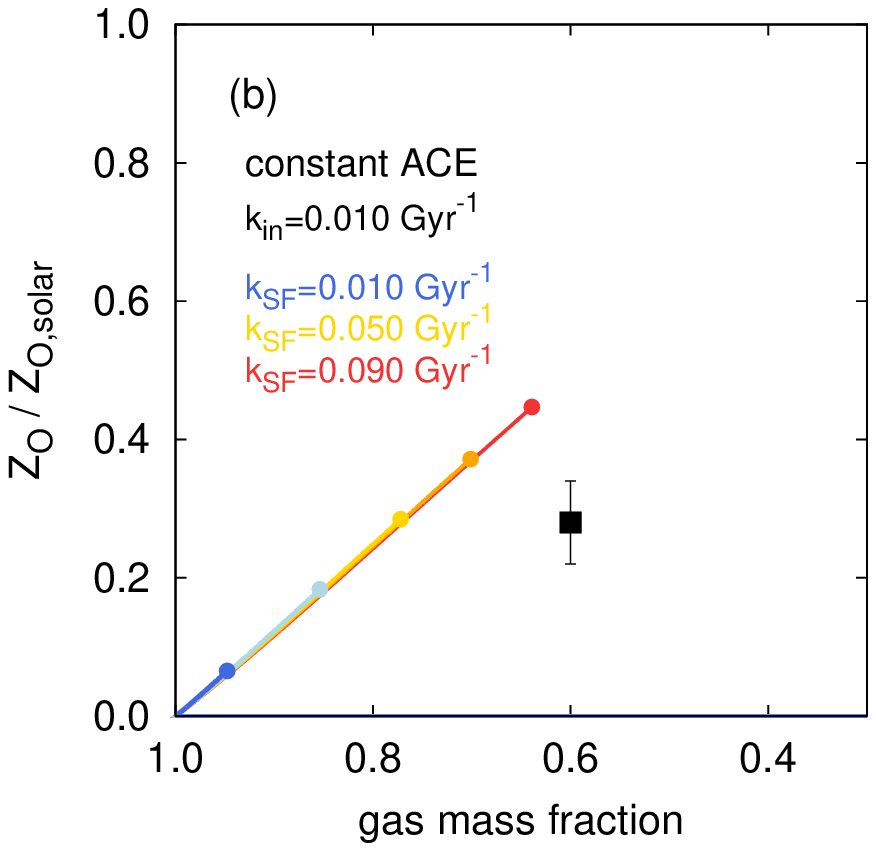}
\end{minipage}
\begin{minipage}{0.25\hsize}
\centering
\includegraphics[scale=0.46]{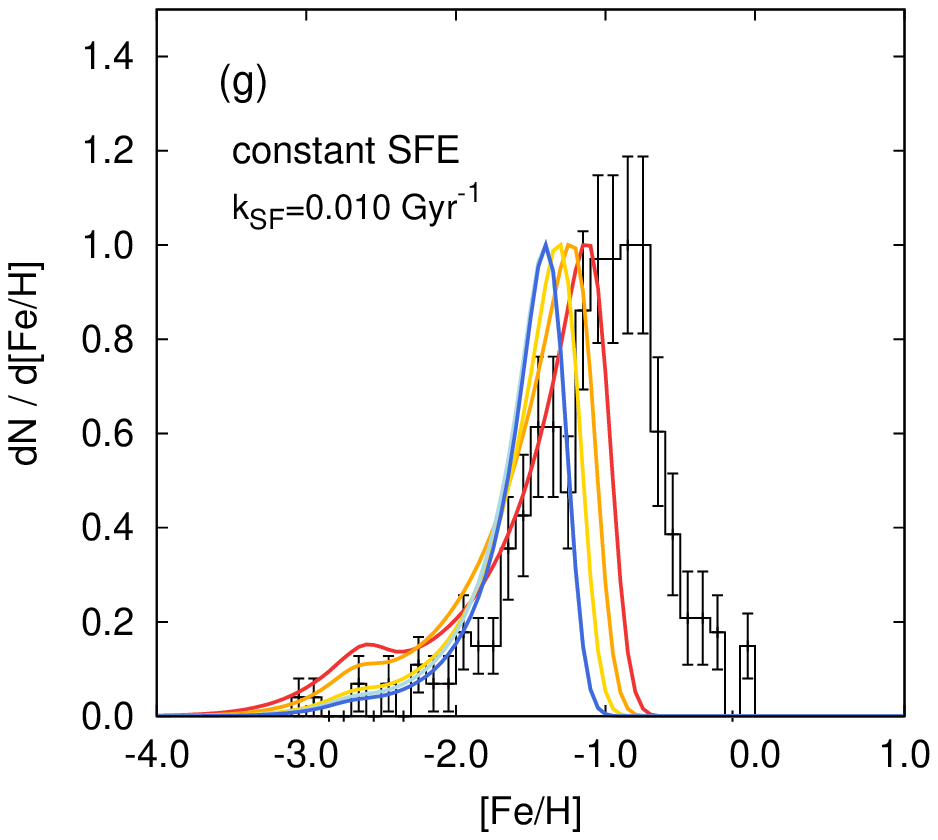} 
\end{minipage}
\begin{minipage}{0.25\hsize}
\centering
\includegraphics[scale=0.46]{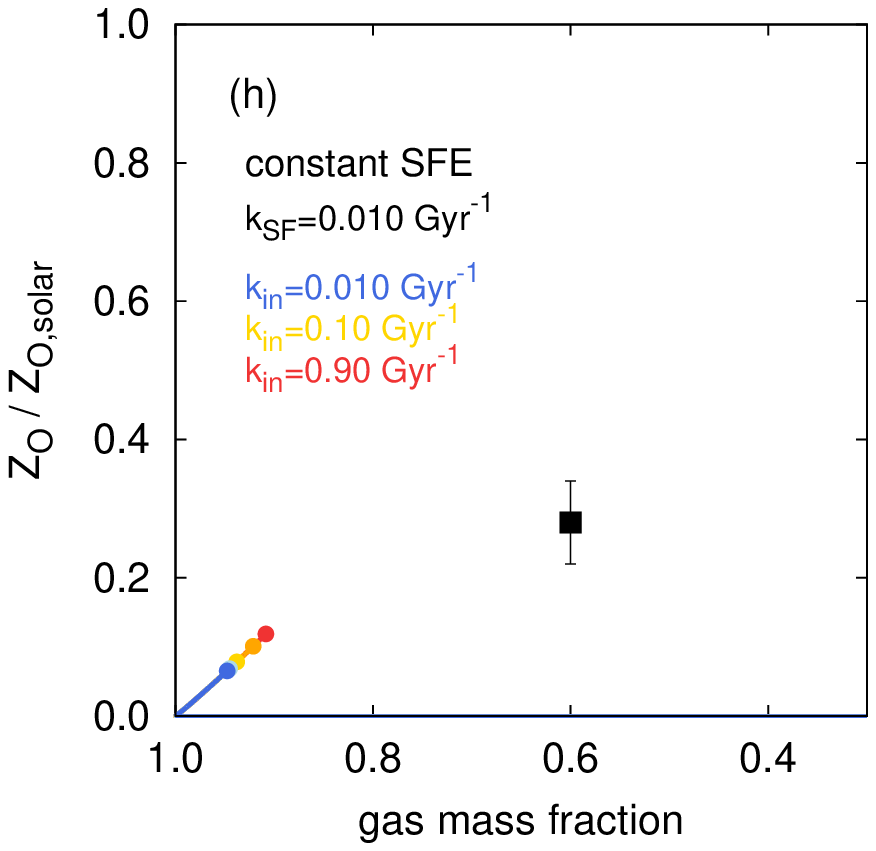} 
\end{minipage} \\
\begin{minipage}{0.25\hsize}
\centering
\includegraphics[scale=0.46]{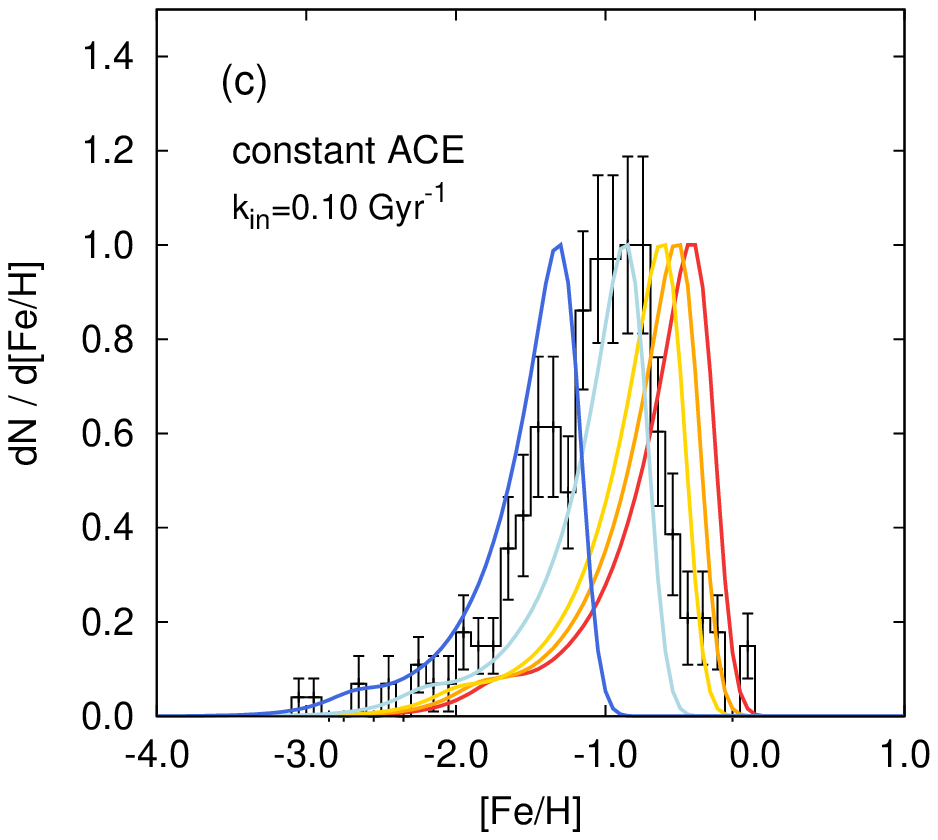} 
\end{minipage}
\begin{minipage}{0.25\hsize}
\centering
\includegraphics[scale=0.46]{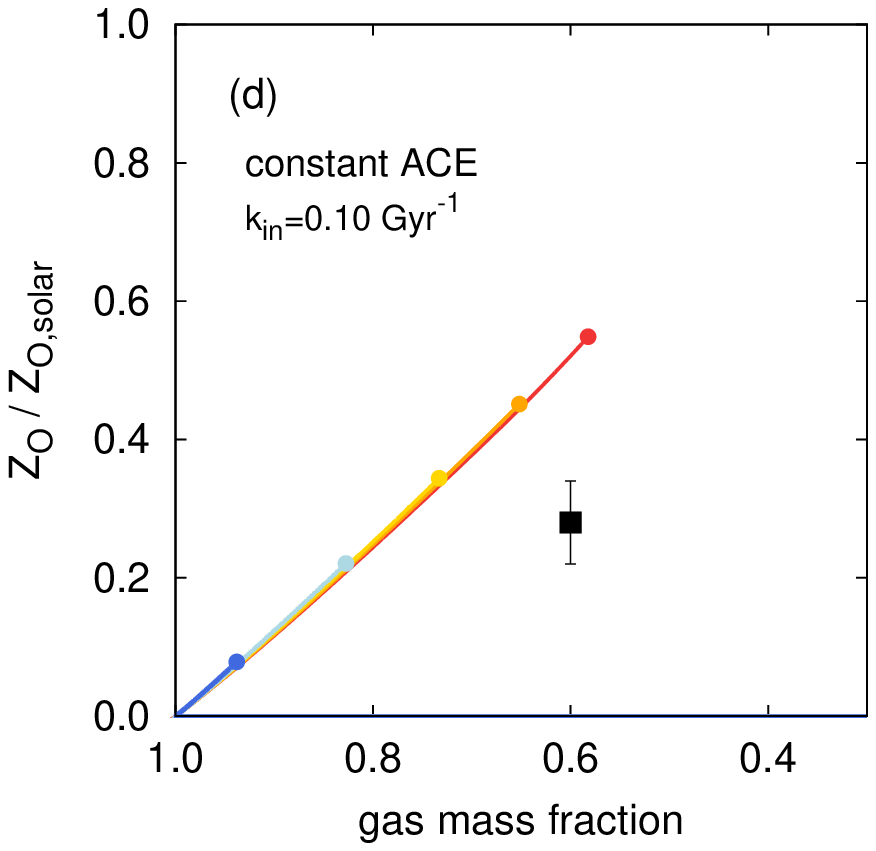} 
\end{minipage}
\begin{minipage}{0.25\hsize}
\centering
\includegraphics[scale=0.46]{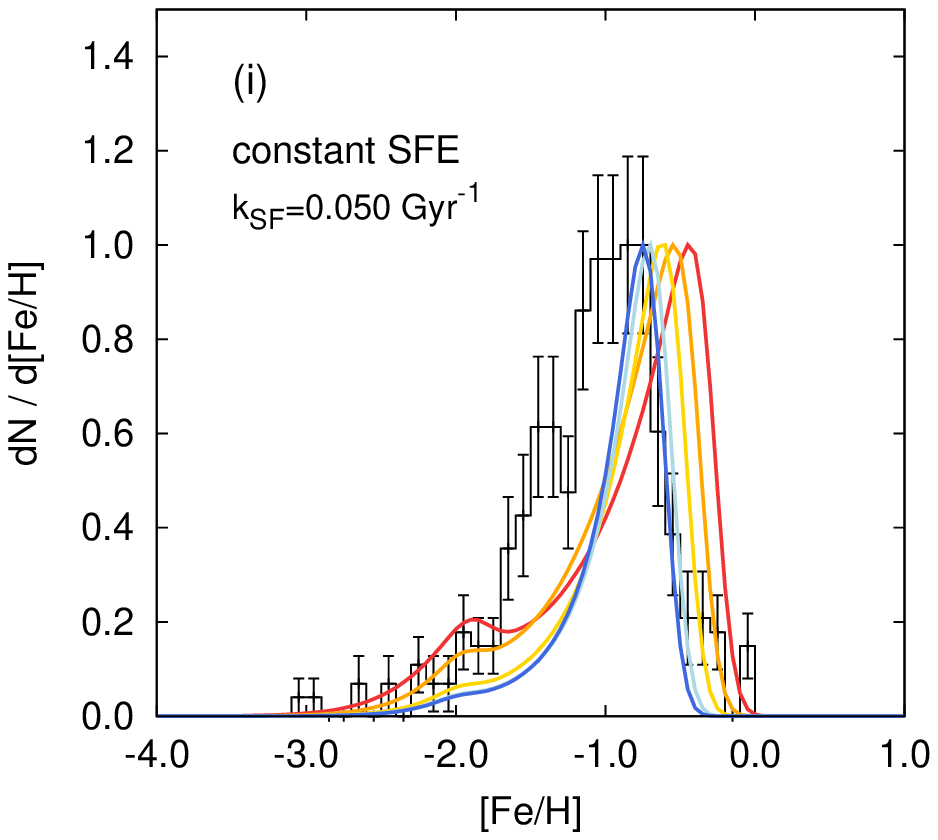}
\end{minipage}
\begin{minipage}{0.25\hsize}
\centering
\includegraphics[scale=0.46]{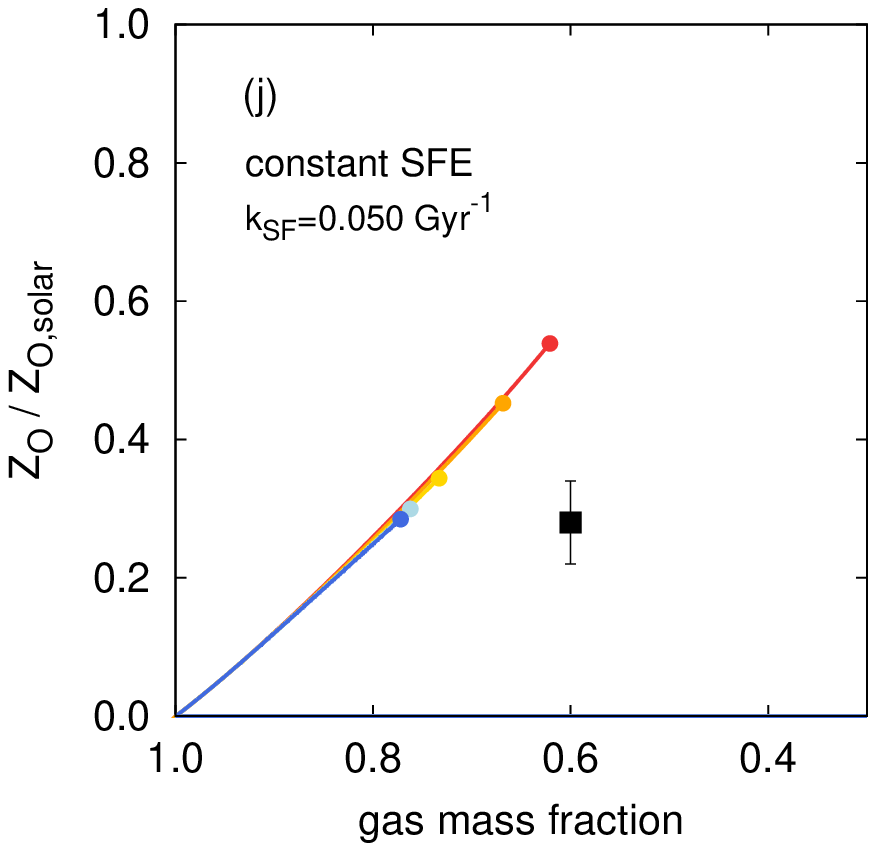}
\end{minipage} \\
\begin{minipage}{0.25\hsize}
\centering
\includegraphics[scale=0.46]{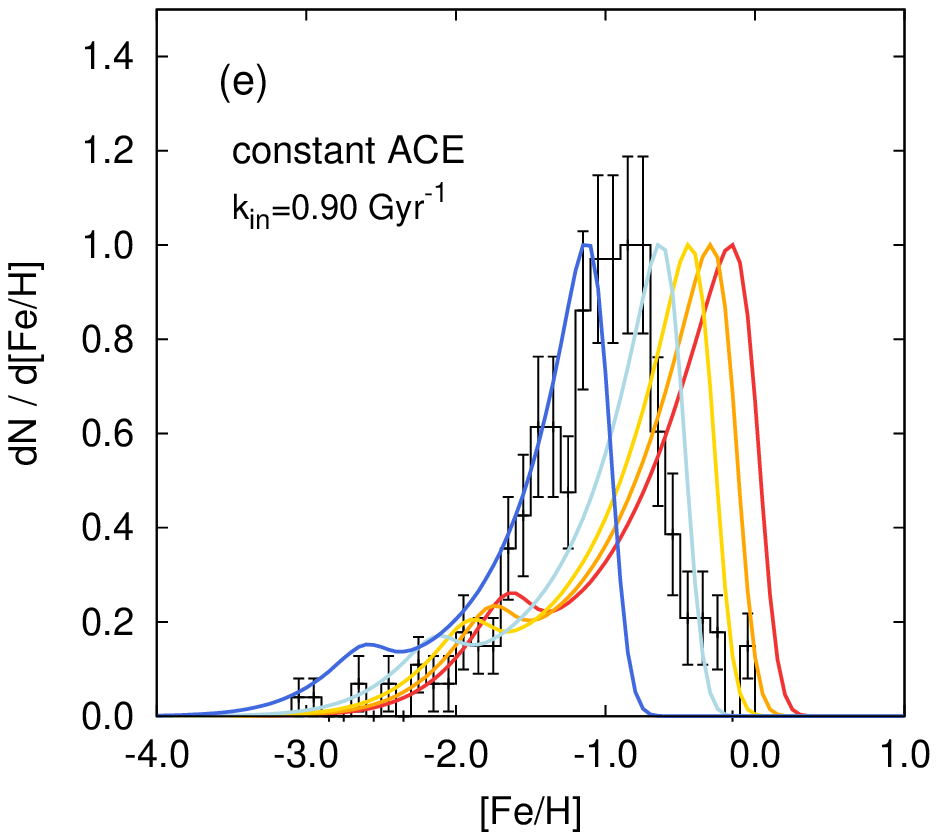}
\end{minipage}
\begin{minipage}{0.25\hsize}
\centering
\includegraphics[scale=0.46]{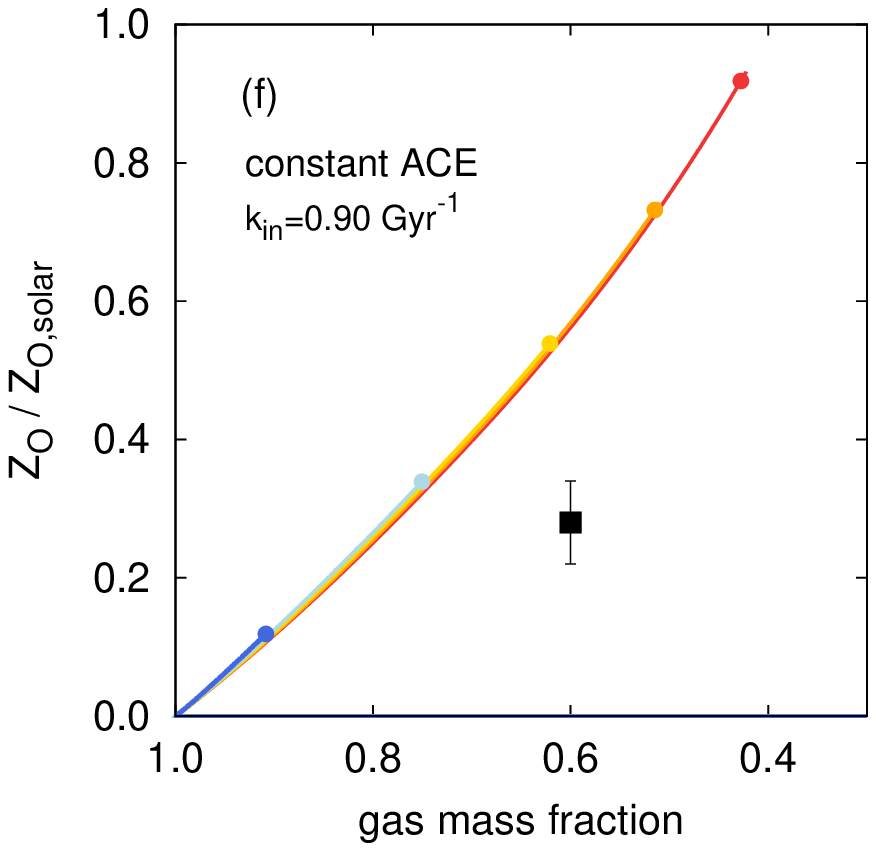}
\end{minipage}
\begin{minipage}{0.25\hsize}
\centering
\includegraphics[scale=0.46]{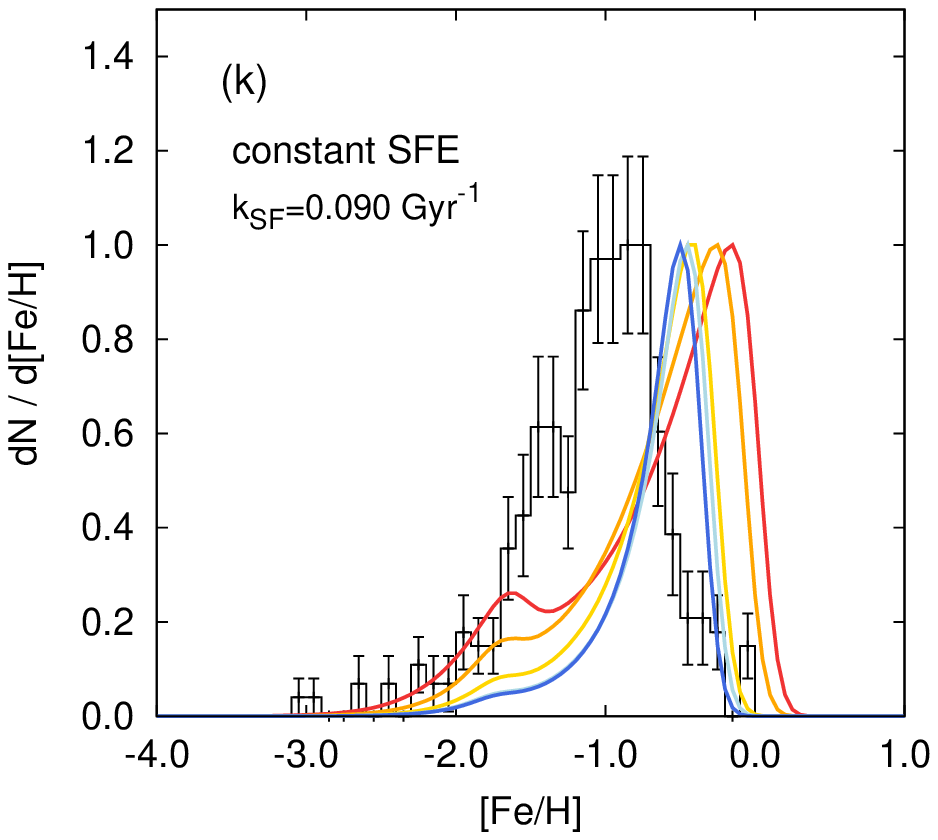} 
\end{minipage}
\begin{minipage}{0.25\hsize}
\centering
\includegraphics[scale=0.46]{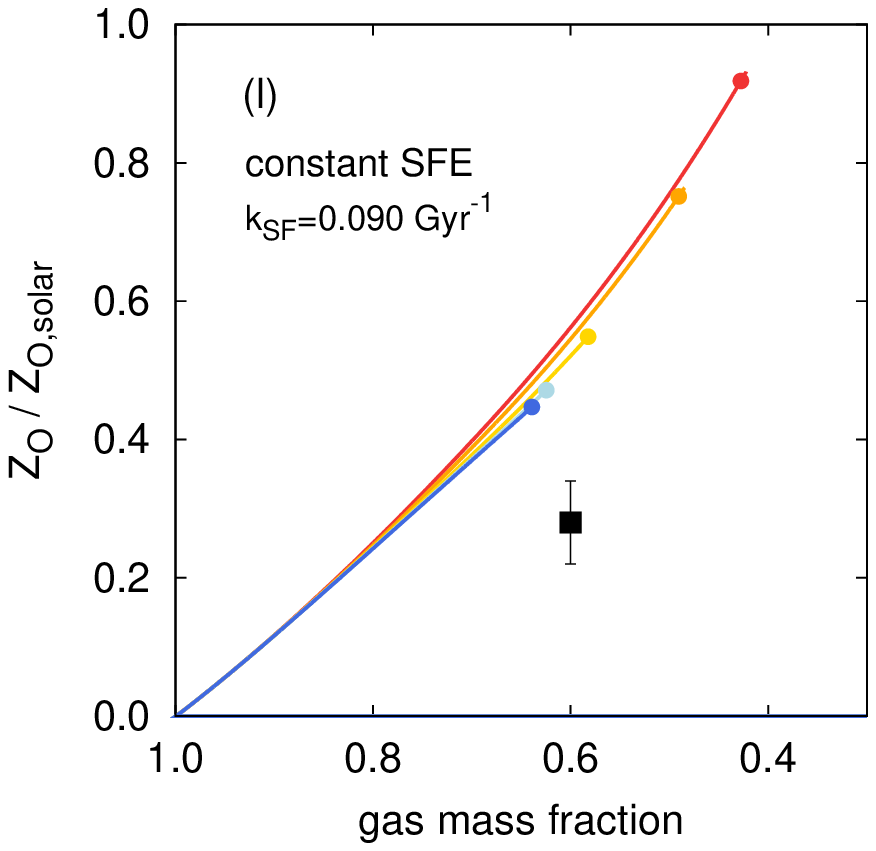} 
\end{minipage}
\end{tabular}
\caption{Comparisons between the observational data 
(black histograms and squares; see Sec.~\ref{sec:2} for the references)
and model B.
The Salpeter IMF ($x=1.35$) is assumed.
(a)--(f) Cases of different SFEs while the ACE is fixed.
The colours of the curves correspond to the SFE:
blue, sky blue, yellow, orange and red curves are cases of 
$k_{\rm SF}=0.010, 0.030, 0.050, 0.070$ and $0.090~({\rm Gyr}^{-1})$,
respectively.
The values of the ACE are $k_{\rm in}=0.010$ (panels a and b), 
$0.10$ (c and d) and $0.90~({\rm Gyr}^{-1})$ (e and f).
(g)--(l) Cases of different ACEs while the SFE is fixed.
The colours of the curves correspond to the ACE:
blue, sky blue, yellow, orange and red curves are cases of 
$k_{\rm in}=0.010, 0.032, 0.10, 0.32$ and $0.90~({\rm Gyr}^{-1})$,
respectively.
The values of the SFE are $k_{\rm SF}=0.010$ (panels g and h), 
$0.050$ (i and j) and $0.090~({\rm Gyr}^{-1})$ (k and l).
Points on the curves show the gas-phase oxygen abundances 
and gas mass fractions in the universe at present.
}
\label{fig:B}
\end{figure*}

Fig.~\ref{fig:B} shows how the quantities predicted by model B vary
with the SFE or the ACE when the Salpeter IMF is assumed.
If different SFEs are applied while keeping the ACE fixed,
similar trends to those of model A are seen (Figs. \ref{fig:B}a--f).
On the $\mu$--$Z_{\rm O}$ plane,
galaxies evolve along almost self-similar tracks.

When different ACEs are applied 
while keeping the SFE fixed (Figs. \ref{fig:B}g--l),
higher ACE results in the accretion of a larger fraction of gas 
in the reservoir in the early stage of the evolution, 
and stars are formed from the gas.
Thus, if a galaxy has a high ACE, 
the metallicity distribution can have a slightly high ${\rm [Fe/H]_{peak}}$ and 
high gas-phase oxygen abundance in the present universe.
Galaxies of different ACEs evolve along almost 
self-similar tracks on the $\mu$--$Z_{\rm O}$ diagram.

As it is known that the accretion of gas is one of solutions of the 
G-dwarf problem in the solar neighbourhood \citep[e.g.][]{PP75},
the overproduction of stars in the low-metallicity tail of 
metallicity distributions seems to be alleviated compared to model A.
When the SFE is fixed (Figs.~\ref{fig:B}g, i and k),
galaxies of smaller values of ACE/SFE have metallicity distributions of 
sharper main peaks, 
in the ranges of the SFE and ACE investigated in this study.
Galaxies of higher ACEs also tend to have stars of 
wider range of the metallicity.

As seen in the $\mu$--$Z_{\rm O}$ diagrams in Fig.~\ref{fig:B},
model B predicts higher gas-phase oxygen abundances at the observed value of
gas mass fraction of NGC 6822.
For instance, a gas mass fraction predicted by a model of 
$k_{\rm SF}=0.050~({\rm Gyr}^{-1})$ and $k_{\rm in}=0.90~({\rm Gyr}^{-1})$
(yellow curves in Figs.~\ref{fig:B}e and f) is almost consistent with
the observed value, 
but the model predicts higher gas-phase oxygen abundance and 
[Fe/H$]_{\rm peak}$ compared to the observed values.
Fig.~\ref{fig:B} also suggests that there are cases where the observed 
gas-phase oxygen abundance and [Fe/H$]_{\rm peak}$ can be explained
(e.g. blue curves in Figs.~\ref{fig:B}i and j), 
but such models predict larger gas mass fractions compared to the 
observed value.
Therefore, it seems that the observed values of gas-phase oxygen abundance
and gas mass fraction are unlikely to be explained by 
different ACEs and SFEs alone.

\subsubsection{Cases of steeper IMF}
As described in Sec.~\ref{subsec:A}, if star formation obeys
a steep IMF, the galaxy may have lower gas-phase oxygen abundance
at a fixed gas fraction.
In this section, the observational data are compared to model B
under the assumption that star formation obeys steeper IMFs
compared to the Salpeter IMF.

\begin{figure*}
\centering
\begin{tabular}{c}
\begin{minipage}{0.25\hsize}
\centering
\includegraphics[scale=0.46]{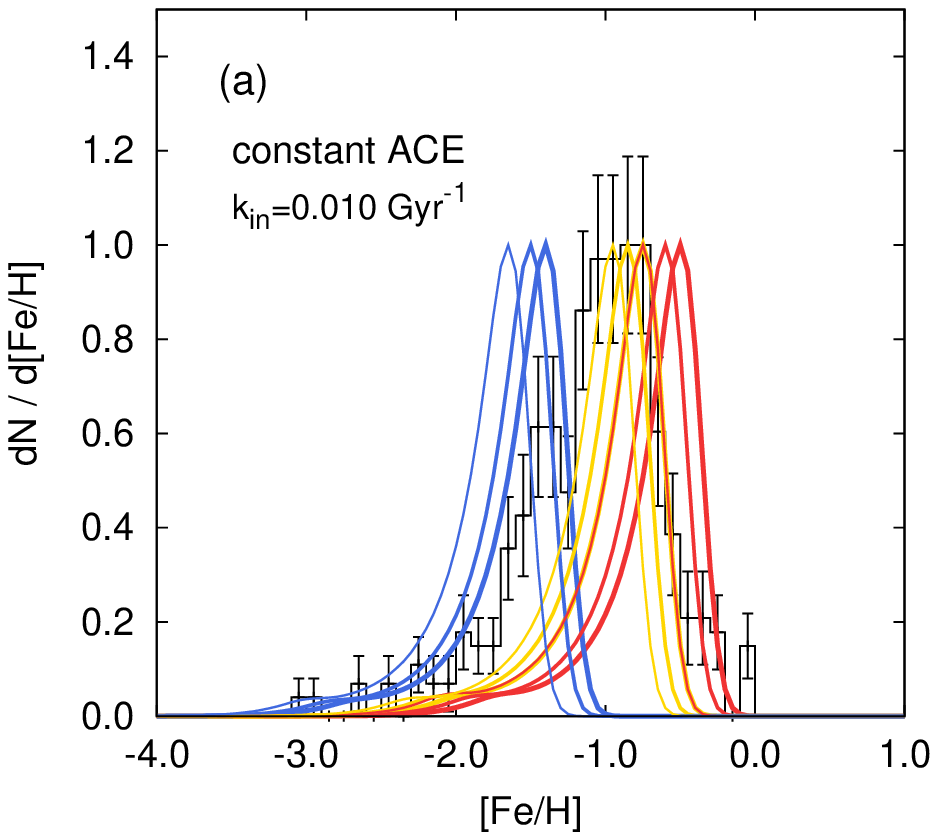}
\end{minipage}
\begin{minipage}{0.25\hsize}
\centering
\includegraphics[scale=0.46]{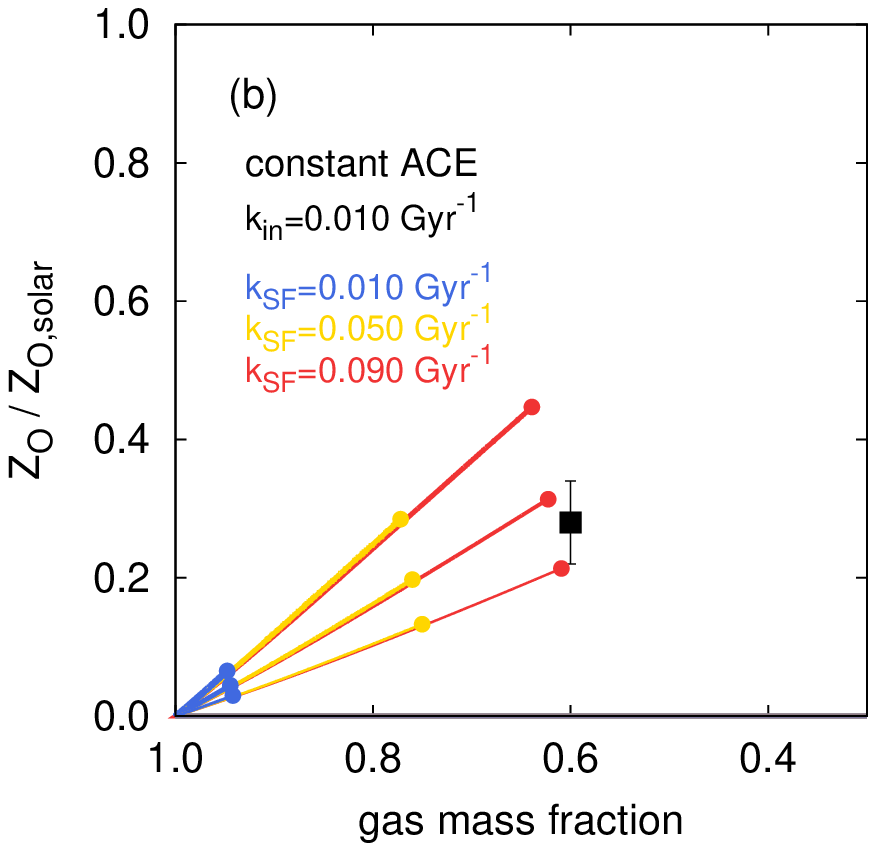}
\end{minipage}
\begin{minipage}{0.25\hsize}
\centering
\includegraphics[scale=0.46]{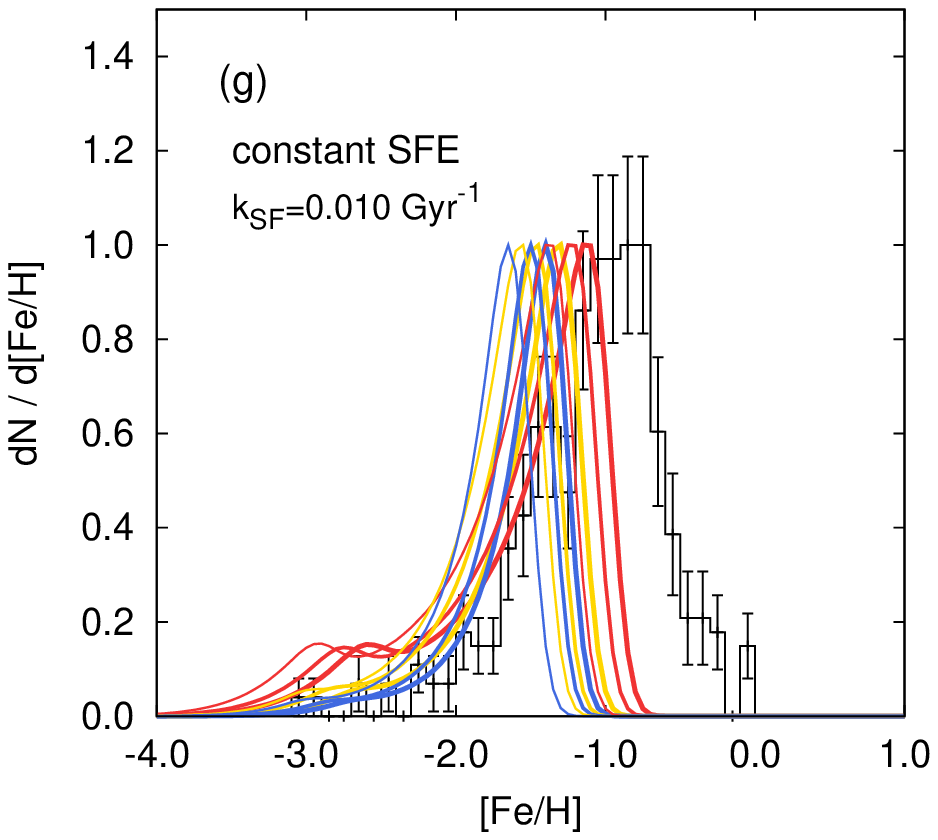} 
\end{minipage}
\begin{minipage}{0.25\hsize}
\centering
\includegraphics[scale=0.46]{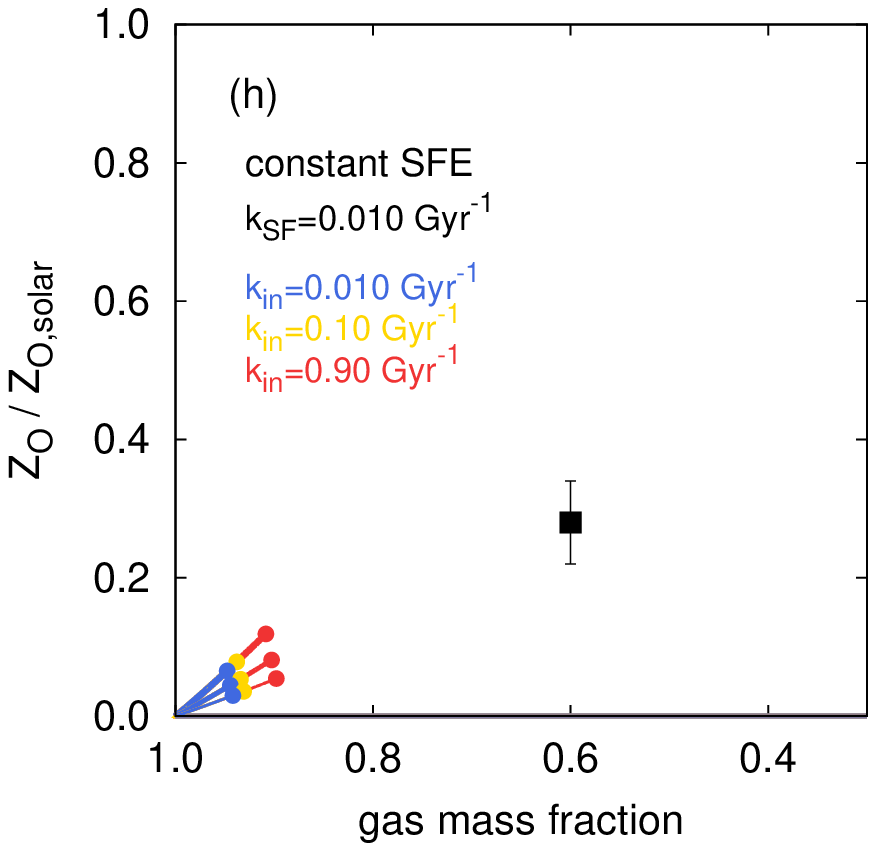} 
\end{minipage} \\
\begin{minipage}{0.25\hsize}
\centering
\includegraphics[scale=0.46]{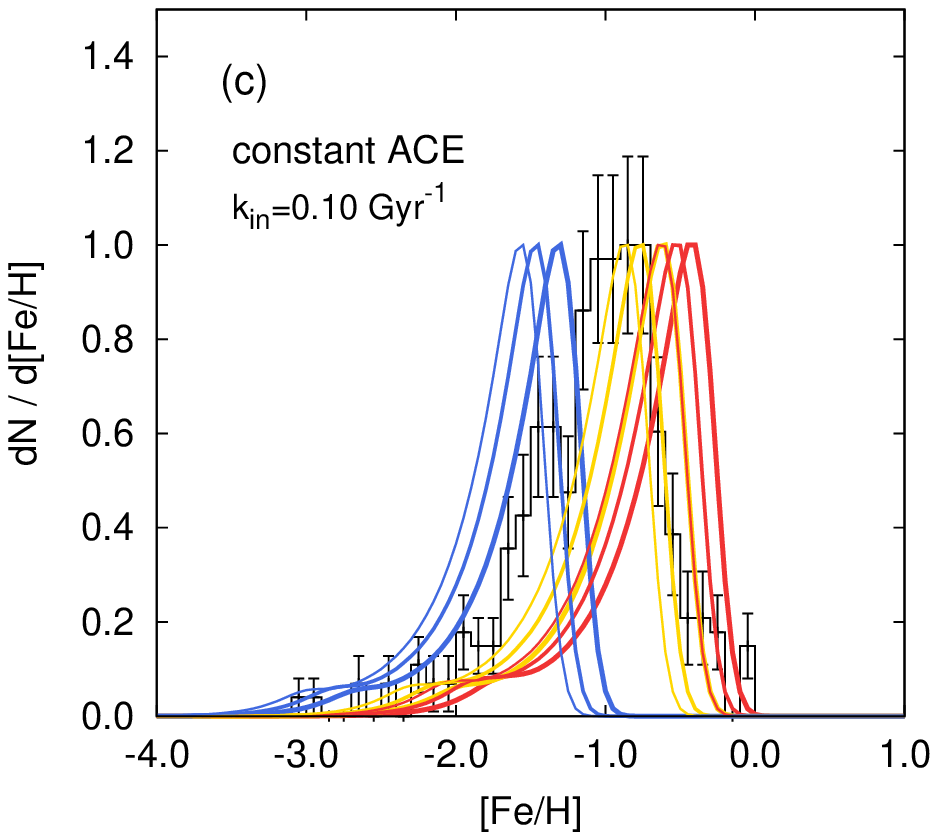} 
\end{minipage}
\begin{minipage}{0.25\hsize}
\centering
\includegraphics[scale=0.46]{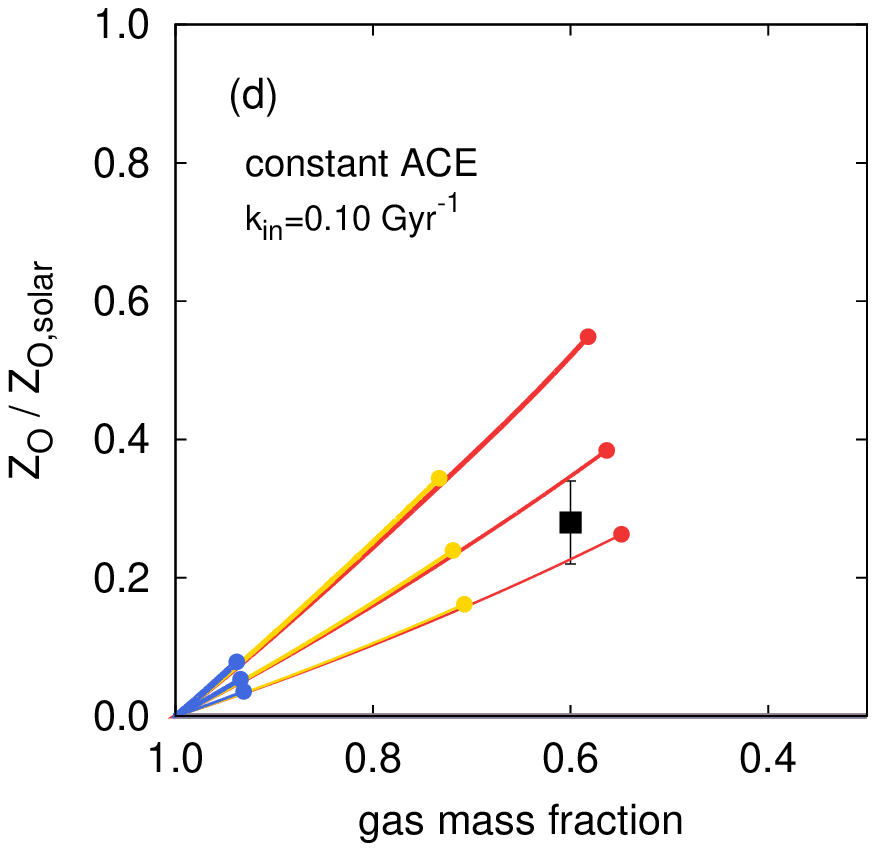} 
\end{minipage}
\begin{minipage}{0.25\hsize}
\centering
\includegraphics[scale=0.46]{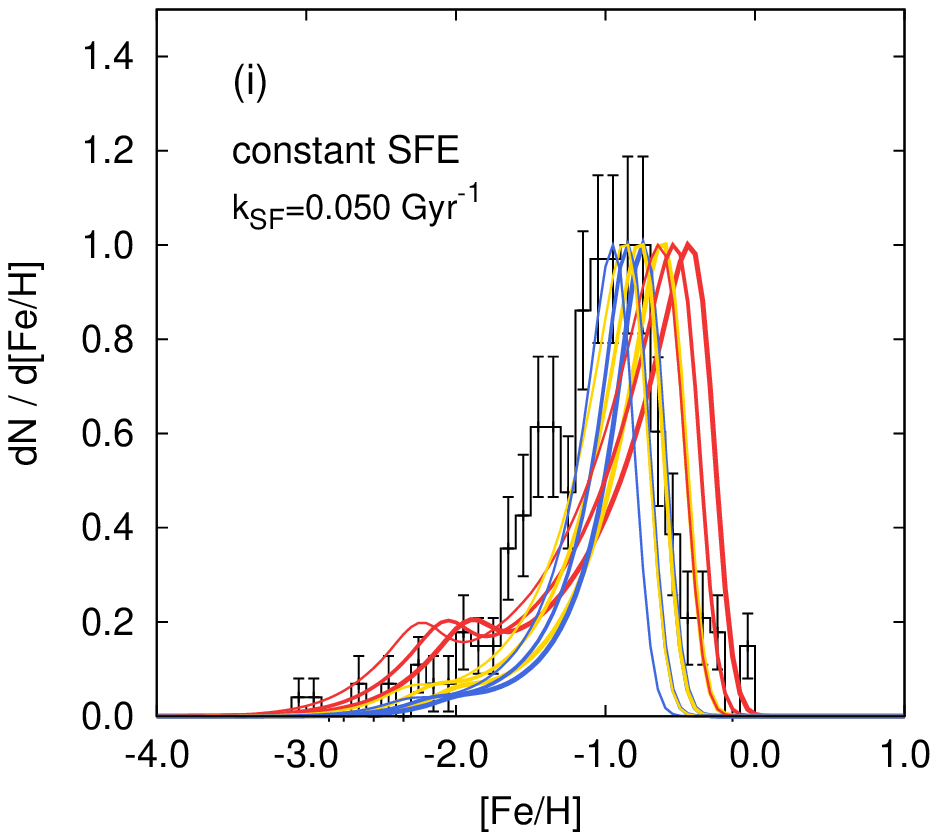}
\end{minipage}
\begin{minipage}{0.25\hsize}
\centering
\includegraphics[scale=0.46]{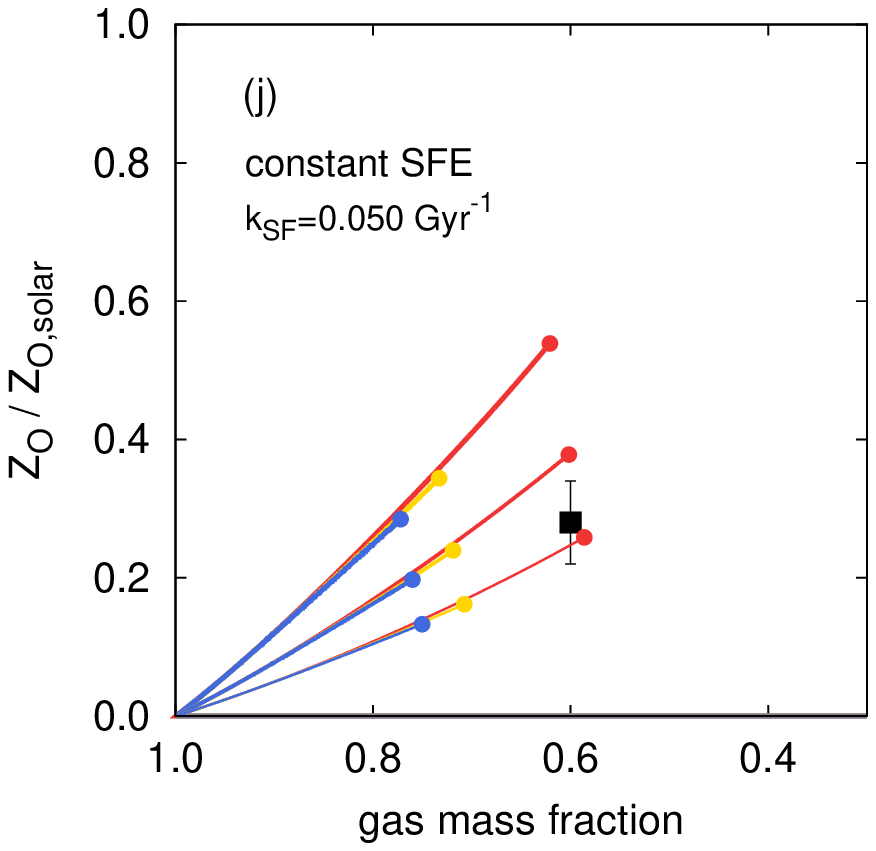}
\end{minipage} \\
\begin{minipage}{0.25\hsize}
\centering
\includegraphics[scale=0.46]{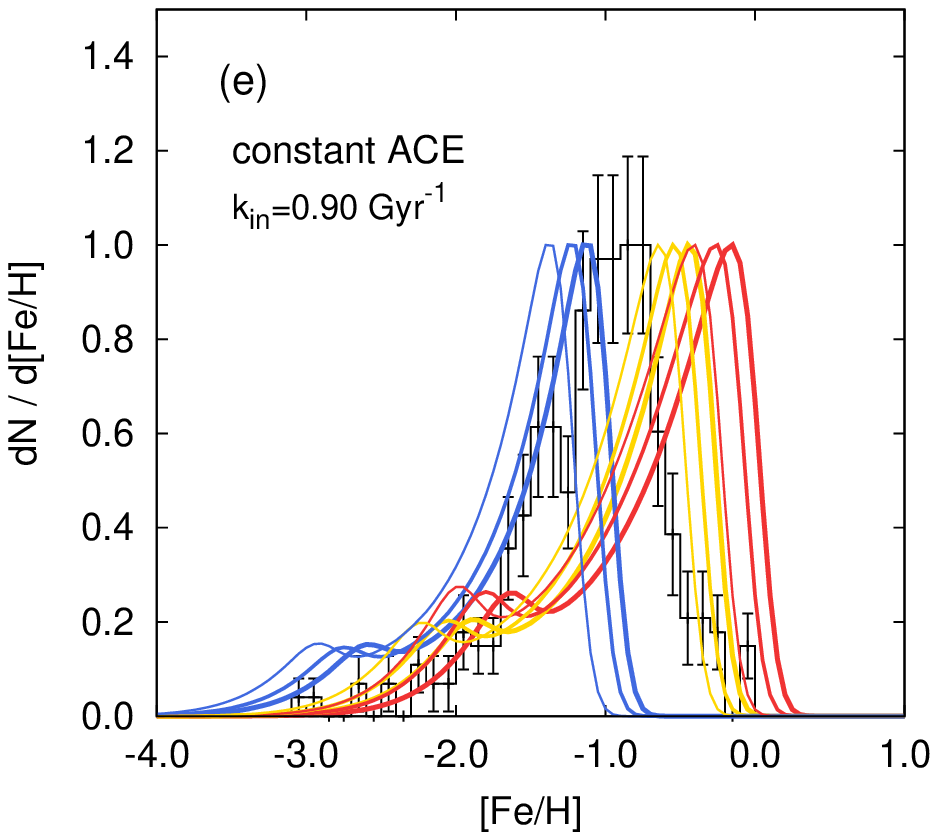}
\end{minipage}
\begin{minipage}{0.25\hsize}
\centering
\includegraphics[scale=0.46]{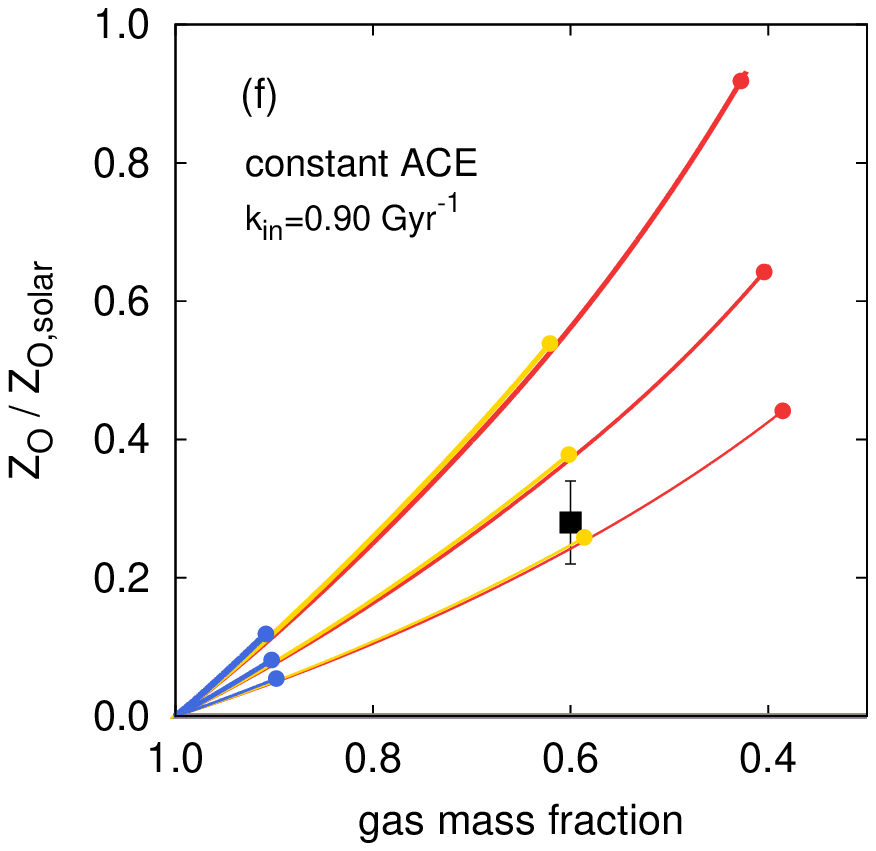}
\end{minipage}
\begin{minipage}{0.25\hsize}
\centering
\includegraphics[scale=0.46]{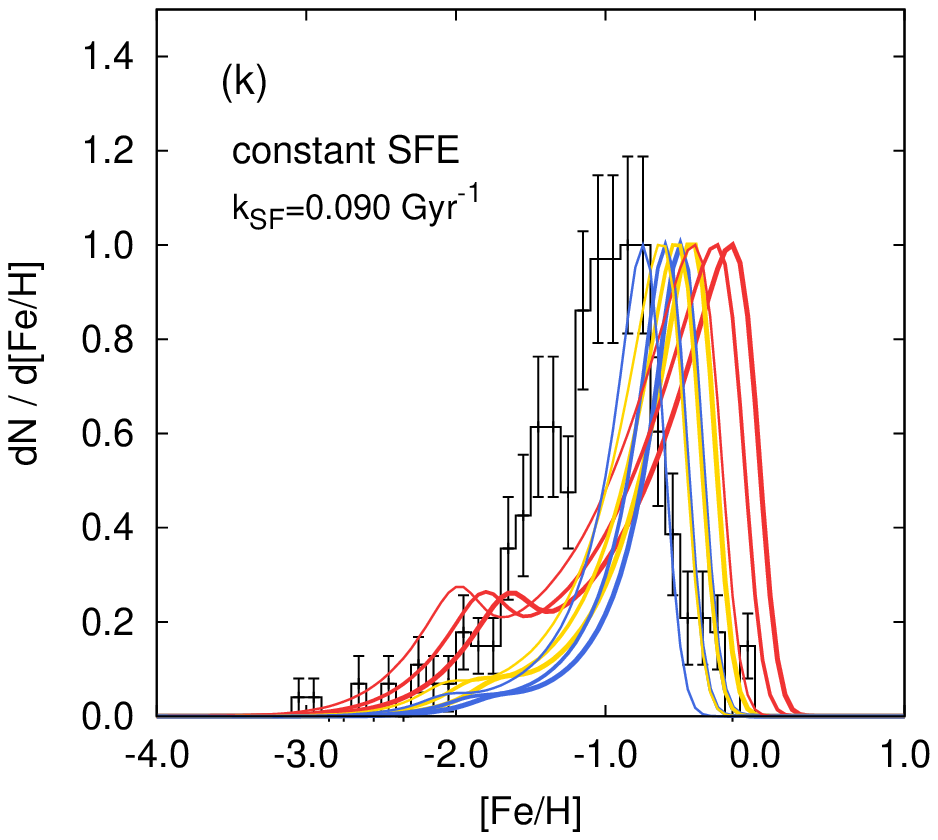} 
\end{minipage}
\begin{minipage}{0.25\hsize}
\centering
\includegraphics[scale=0.46]{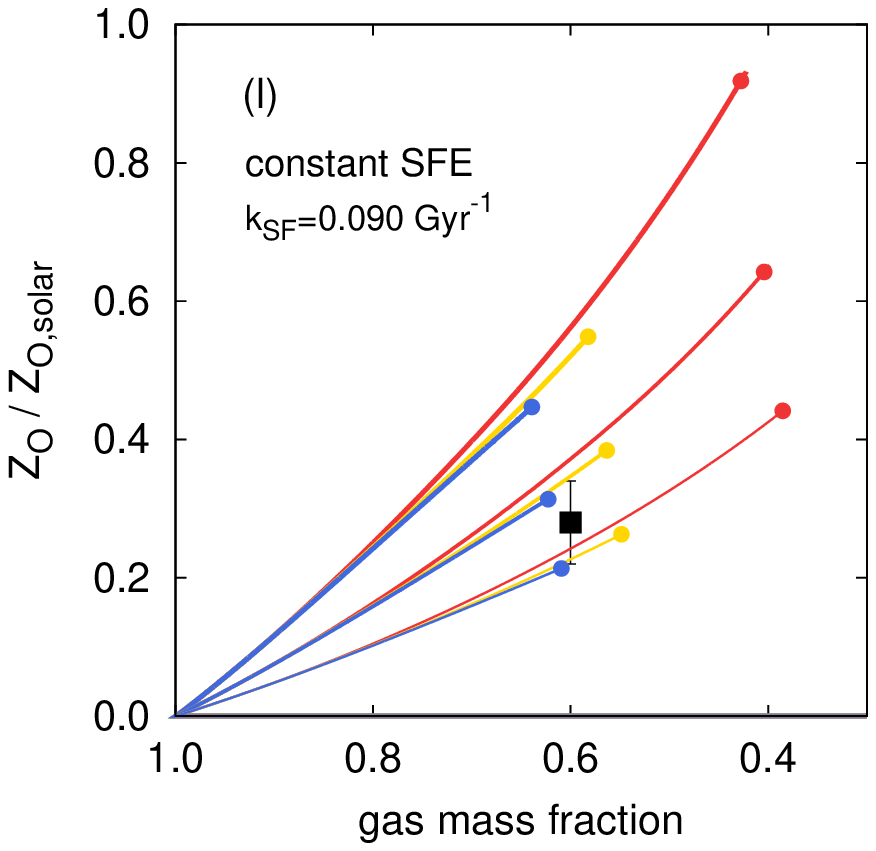} 
\end{minipage}
\end{tabular}
\caption{Metallicity distributions, gas-phase oxygen abundances and 
gas mass fractions of galaxies of different IMFs, SFEs and ACEs predicted 
by model B.
Thinner curves show galaxies of steeper IMF.
The index of IMF is varied by 0.10: $x=1.35$ (thickest curves), $1.45$ 
and $1.55$ (thinnest curves).
(a)--(f) Cases of different SFEs while the ACE is fixed.
The colours of the curves correspond to the SFE:
blue, yellow and red curves show galaxies of $k_{\rm SF}=0.010, 0.050$ 
and $0.090~{\rm Gyr}^{-1}$, respectively.
The ACE is assumed to be $k_{\rm in}=0.010$ (panels a and b), 
$0.10$ (c and d) and $0.90~({\rm Gyr}^{-1})$ (e and f).
(g)--(l) Cases of different ACEs while the SFE is fixed.
The colours of the curves correspond to the value of the ACE:
blue, yellow and red curves show galaxies of $k_{\rm in}=0.010, 0.10$ 
and $0.90~{\rm Gyr}^{-1}$, respectively.
The SFE is assumed to be $k_{\rm SF}=0.010$ (panels g and h), 
$0.050$ (i and j) and $0.090~({\rm Gyr}^{-1})$ (k and l).
Points on the curves in the $\mu$--$Z_{\rm O}$ diagrams show 
the present-day values.
Black histograms and squares in the panels are the observational data 
presented in Sec.~\ref{sec:2}.
}
\label{fig:B2}
\end{figure*}

Figure~\ref{fig:B2} shows the properties of galaxies of different SFEs, ACEs and IMFs.
Generally, at a fixed IMF, the trends are similar to those in the case of the Salpeter IMF.
In the $\mu$--$Z_{\rm O}$ diagrams, 
galaxies evolve along almost self-similar tracks, regardless of the SFE or ACE.
When galaxies have the same SFE and ACE, galaxies of steeper IMFs have 
lower gas-phase oxygen abundances at a given gas mass fraction
and metallicity distributions of lower [Fe/H$]_{\rm peak}$.

In the $\mu$--$Z_{\rm O}$ diagrams shown in Fig.~\ref{fig:B2},
the observed values of NGC 6822 seem to be explained
if the slope of IMF is assumed to be $x\sim1.50-1.55$
(see also Appendix~\ref{app:B} for details).
This result may not disagree with those of a study by
\citet[][]{G13}, who suggest that the gas-phase oxygen abundances and 
gas fractions of dwarf irregular galaxies can be explained by 
the infall model in which no outflow is included.
On the other hand, the models that can explain the gas-phase oxygen
abundance and gas mass fraction predicts a metallicity distribution of 
higher [Fe/H$]_{\rm peak}$ compared to the observed distribution.

In summary, if steeper IMFs are allowed, the observed values of 
gas mass fraction and gas-phase oxygen abundance can be explained 
by model B.
On the other hand, metallicity distributions predicted by the models 
tend to have higher [Fe/H$]_{\rm peak}$ and the shape of the observed
distribution cannot be explained.
If NGC 6822 is an accretion-dominated system,
there may be dominant physical processes 
that are not taken into account in model B.
Although the literature suggests that more massive irregular galaxies
are likely to have more continuous star formation histories
\citep[][]{T98,W12},
assuming fluctuations in gas accretion and multiple starbursts may explain
the metallicity distribution of NGC 6822 \citep[][]{H15}.

\subsection{Outflow-dominated system}
NGC 6822 is assumed to be a system in which the outflow is dominant.
Model C is compared to the data and 
the Salpeter IMF is assumed.

\begin{figure*}
\centering
\begin{tabular}{c}
\begin{minipage}{0.25\hsize}
\centering
\includegraphics[scale=0.46]{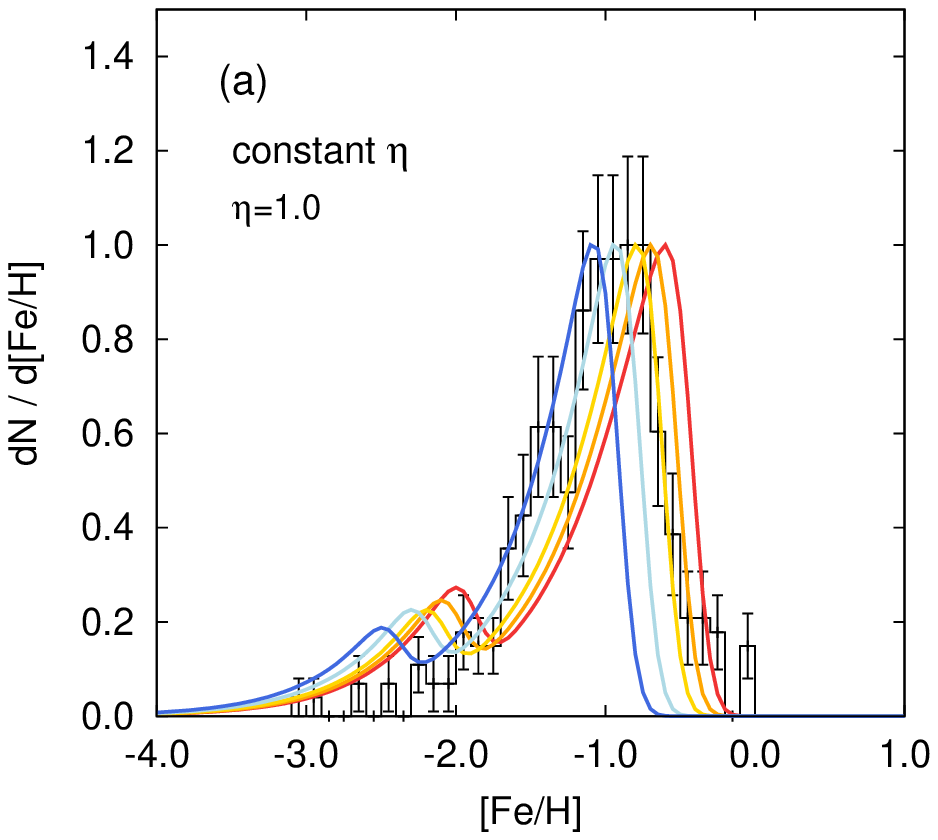}
\end{minipage}
\begin{minipage}{0.25\hsize}
\centering
\includegraphics[scale=0.46]{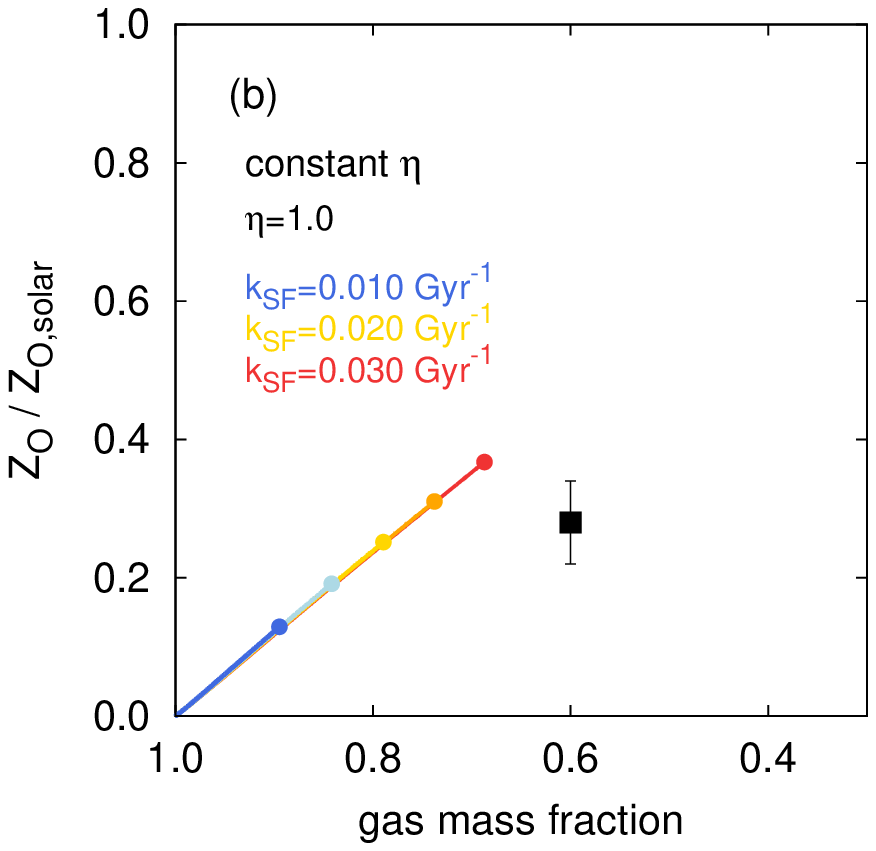}
\end{minipage}
\begin{minipage}{0.25\hsize}
\centering
\includegraphics[scale=0.46]{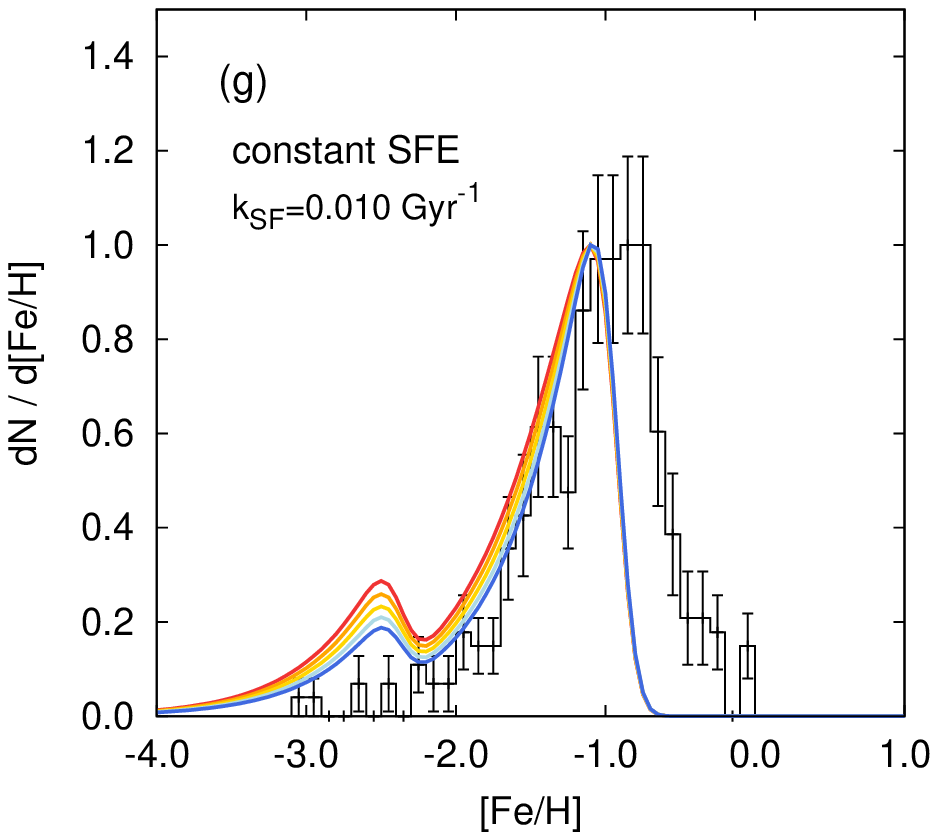} 
\end{minipage}
\begin{minipage}{0.25\hsize}
\centering
\includegraphics[scale=0.46]{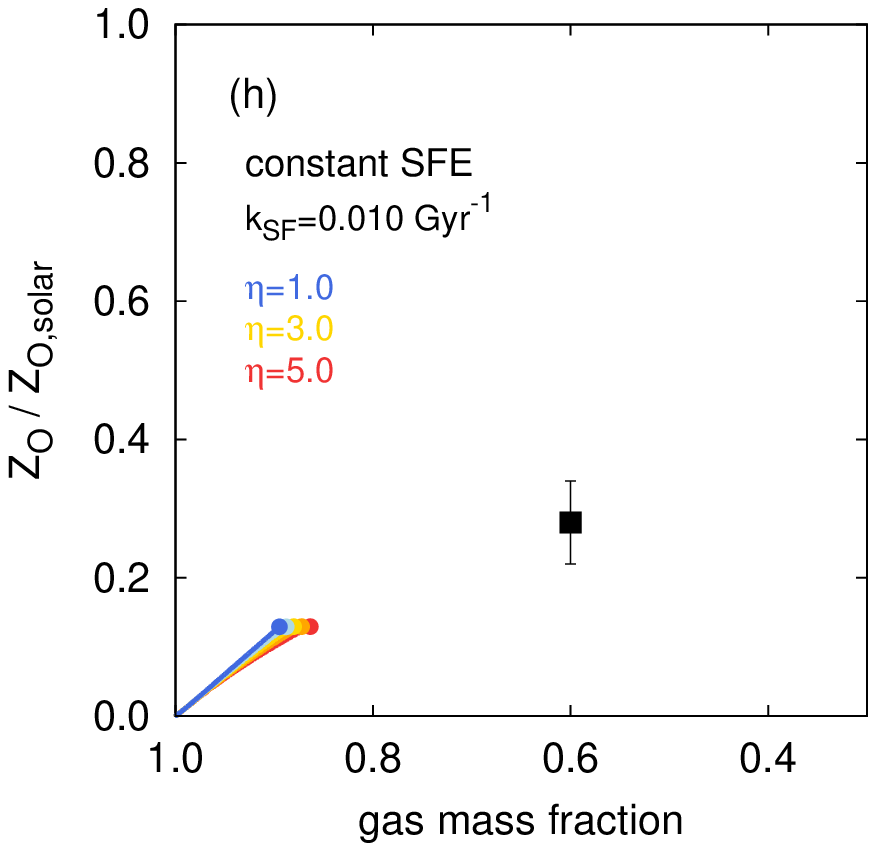} 
\end{minipage} \\
\begin{minipage}{0.25\hsize}
\centering
\includegraphics[scale=0.46]{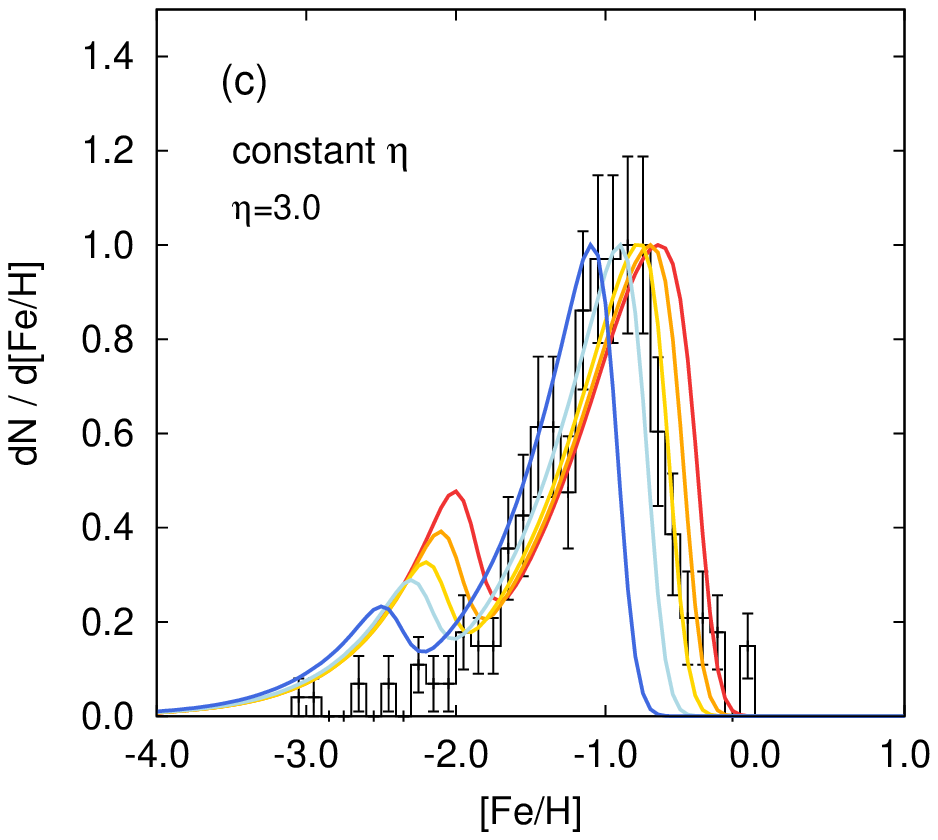} 
\end{minipage}
\begin{minipage}{0.25\hsize}
\centering
\includegraphics[scale=0.46]{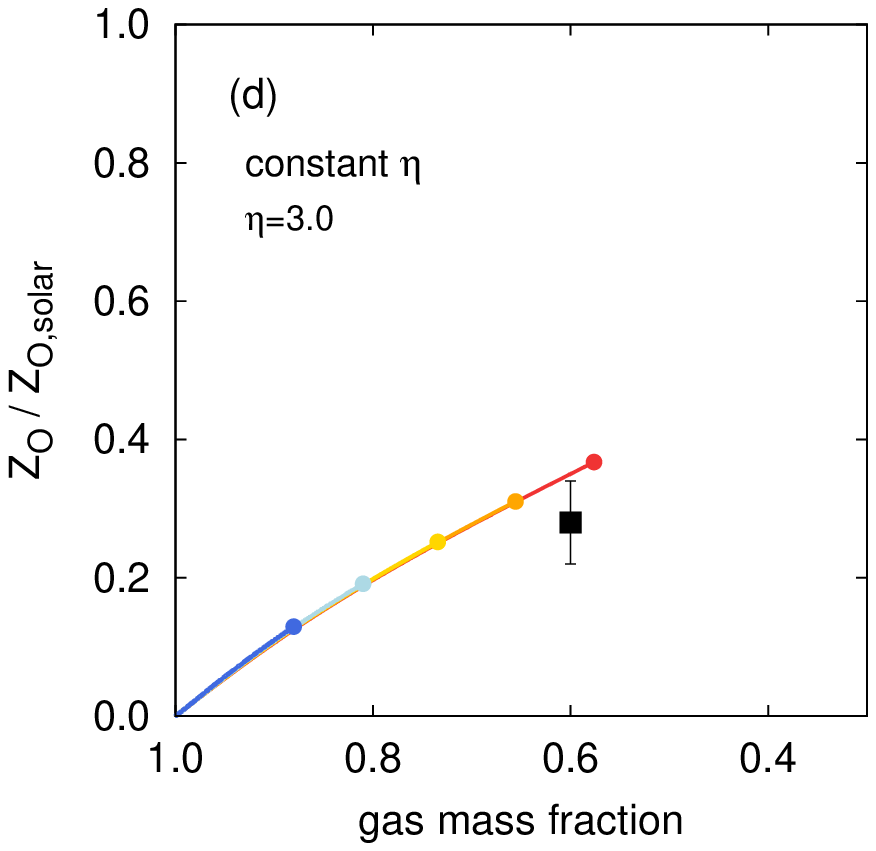} 
\end{minipage}
\begin{minipage}{0.25\hsize}
\centering
\includegraphics[scale=0.46]{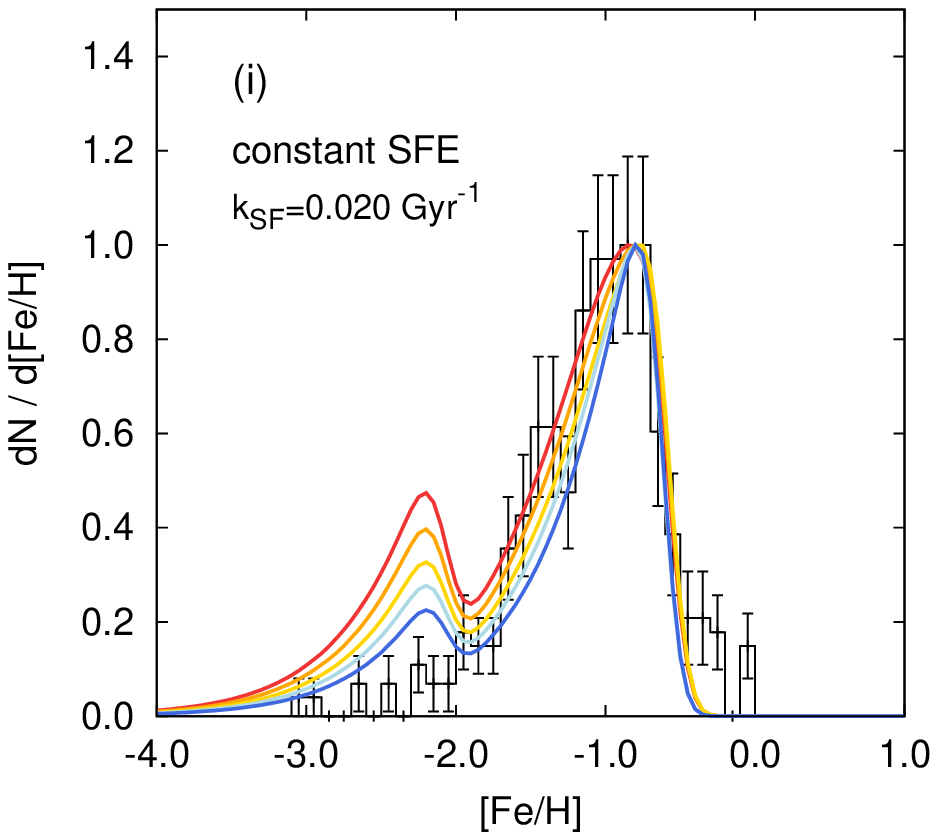}
\end{minipage}
\begin{minipage}{0.25\hsize}
\centering
\includegraphics[scale=0.46]{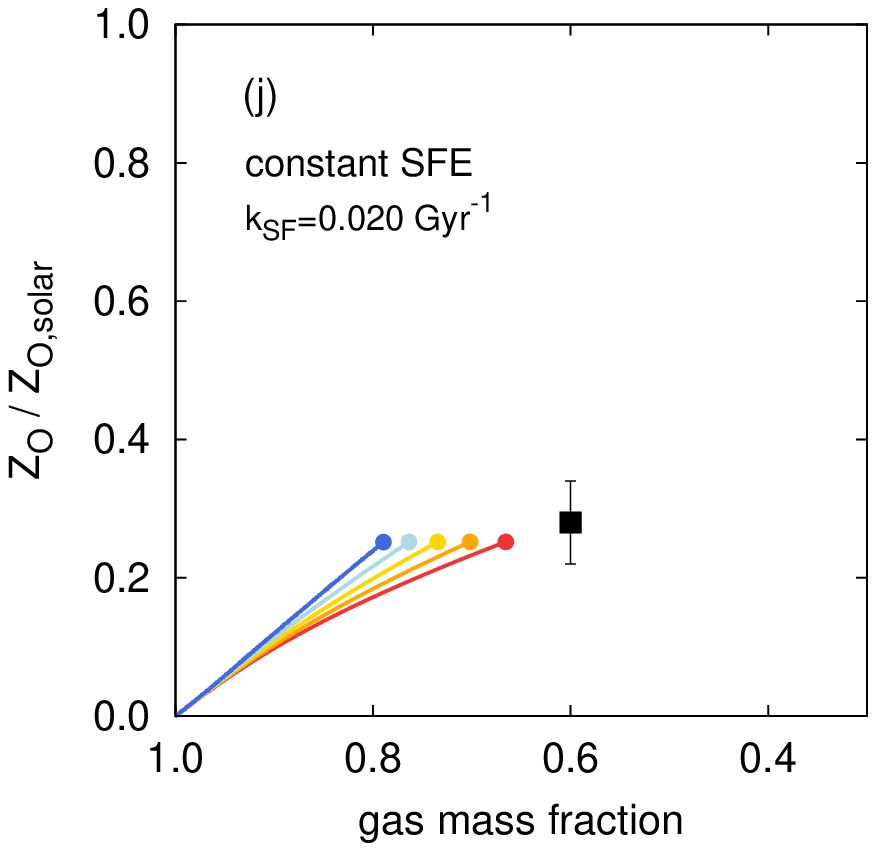}
\end{minipage} \\
\begin{minipage}{0.25\hsize}
\centering
\includegraphics[scale=0.46]{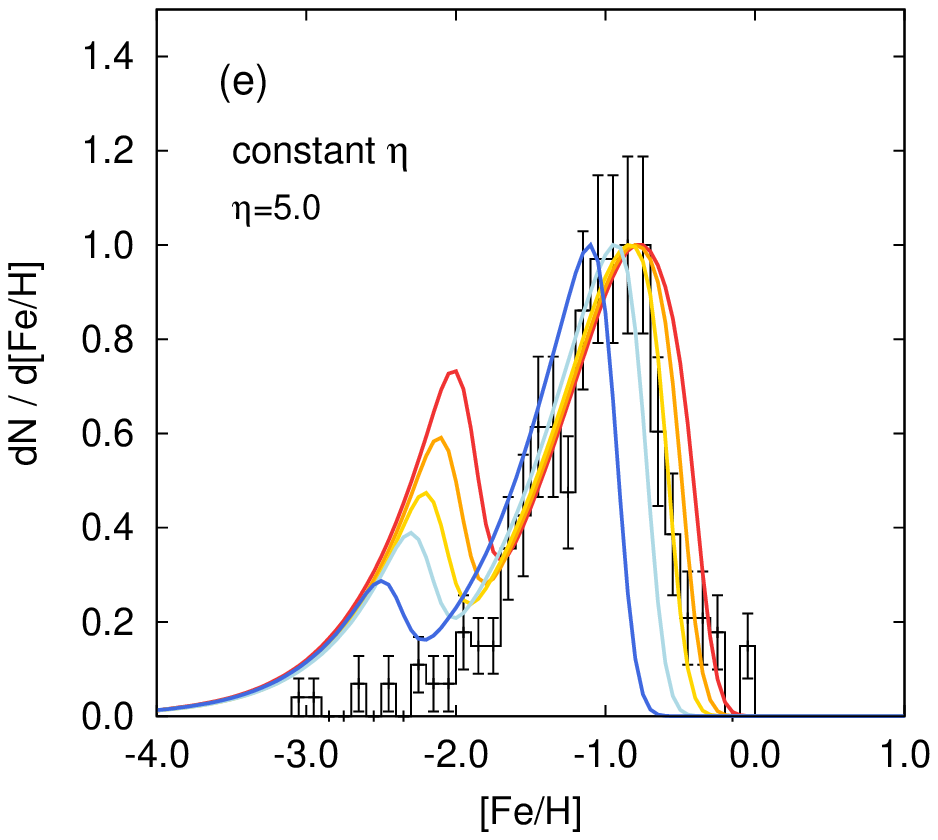}
\end{minipage}
\begin{minipage}{0.25\hsize}
\centering
\includegraphics[scale=0.46]{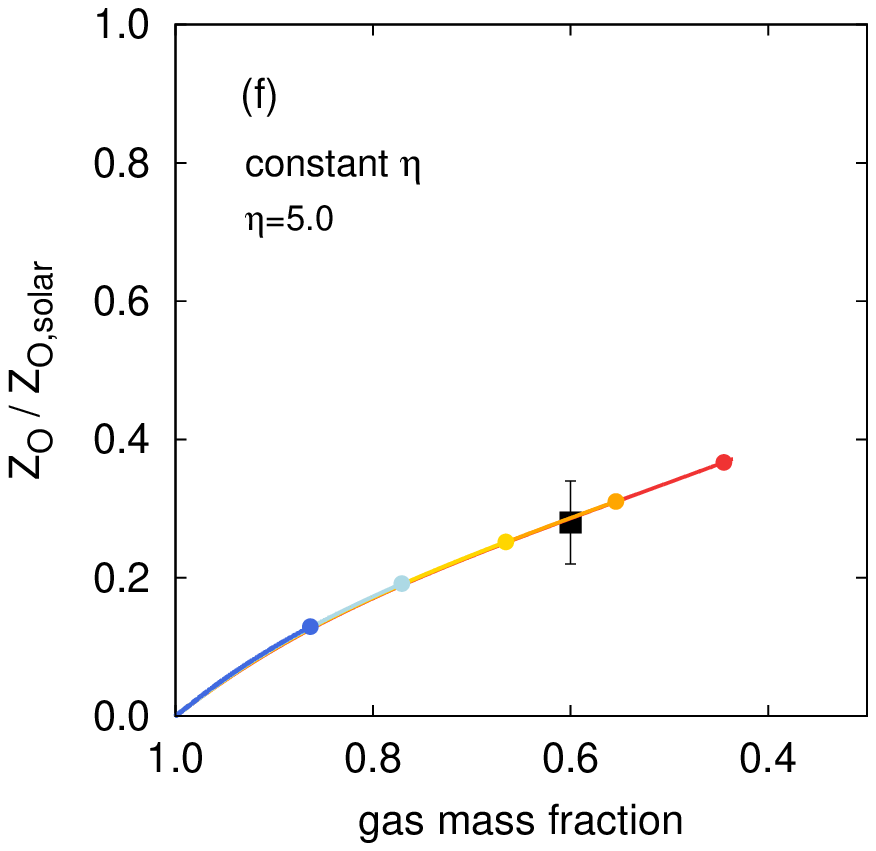}
\end{minipage}
\begin{minipage}{0.25\hsize}
\centering
\includegraphics[scale=0.46]{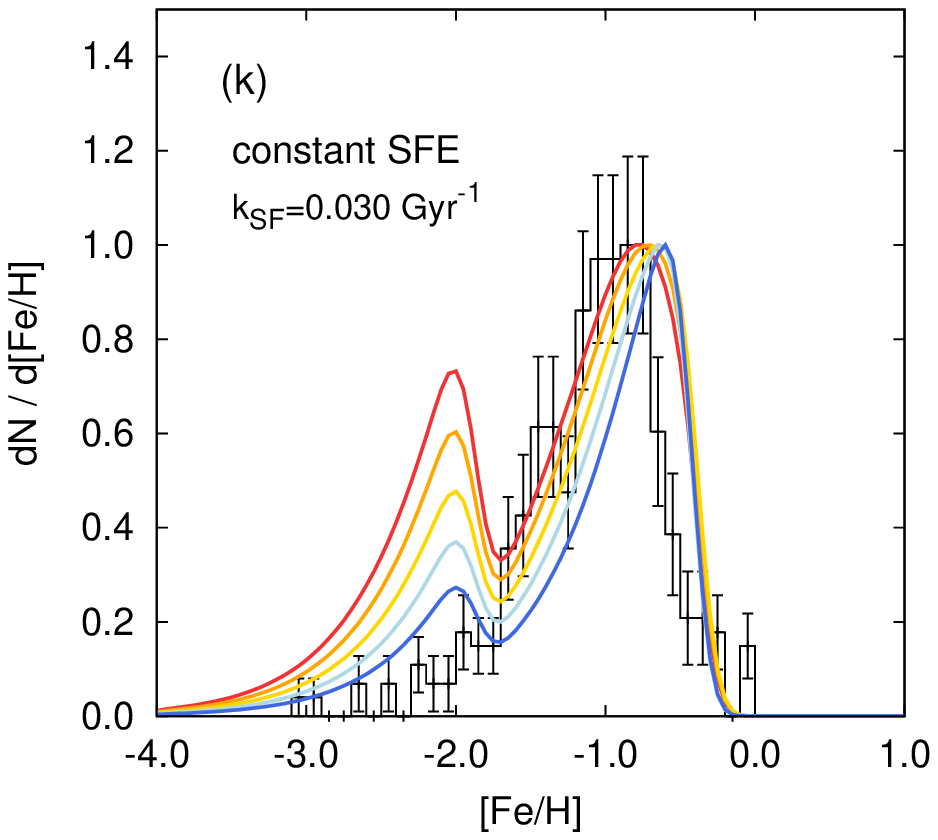} 
\end{minipage}
\begin{minipage}{0.25\hsize}
\centering
\includegraphics[scale=0.46]{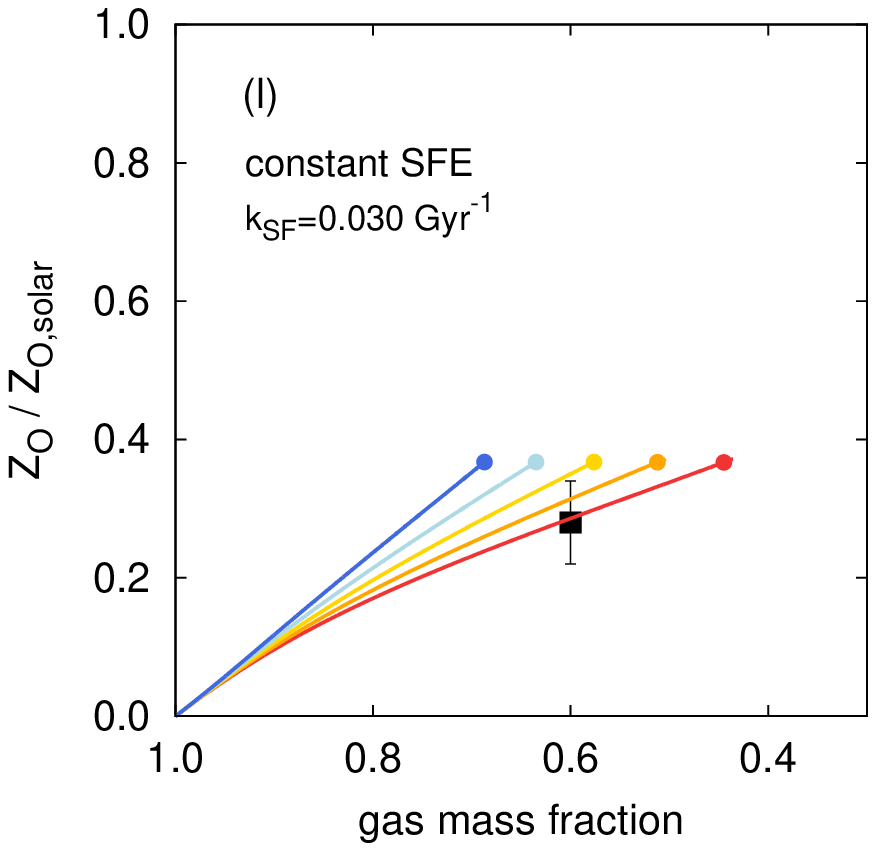} 
\end{minipage}
\end{tabular}
\caption{Comparisons between the observational data 
(black histograms and squares, see Sec.~\ref{sec:2} for references)
and model C.
The Salpeter IMF ($x=1.35$) is assumed.
(a)--(f) Cases of different SFEs while the mass-loading factor is fixed.
The colours of the curves correspond to the SFE:
blue, sky blue, yellow, orange and red curves are cases of 
$k_{\rm SF}=0.010, 0.015, 0.020, 0.025$ and $0.030~({\rm Gyr}^{-1})$,
respectively.
The values of the mass-loading factor are $\eta=1.0$ (panels a and b), 
$3.0$ (c and d) and $5.0$ (e and f).
(g)--(l) Cases of different mass-loading factors while the SFE is fixed.
The colours of the curves correspond to the mass-loading factor:
blue, sky blue, yellow, orange and red curves are cases of 
$\eta=1.0, 2.0, 3.0, 4.0$ and $5.0$, respectively.
The values of the SFE are $k_{\rm SF}=0.010$ (panels g and h), 
$0.020$ (i and j) and $0.030~({\rm Gyr}^{-1})$ (k and l).
Points on the curves show the gas-phase oxygen abundances 
and gas mass fractions in the universe at present.
}
\label{fig:C}
\end{figure*}

Fig.~\ref{fig:C} shows how quantities predicted by model C
vary based on the SFE (Figs. \ref{fig:C}a--f)
or the mass-loading factor (Figs. \ref{fig:C}g--l).
When galaxies of different SFEs are compared,
galaxies of higher SFE have a smaller 
gas mass fraction and higher gas-phase oxygen abundance 
in the present universe (Figs.~\ref{fig:C}b, d and f).
When the mass-loading factor is fixed, galaxies evolve along 
almost self-similar tracks on the $\mu$--$Z_{\rm O}$ plane, 
regardless of the SFE.

In addition, as shown in Figs.~\ref{fig:C}g--l, 
galaxies of larger mass-loading factor have lower gas-phase oxygen
abundances at a given gas mass fraction
\citep[][]{H76,P97}.
If a galaxy has a larger mass-loading factor, 
the gas of that galaxy is expelled more efficiently,
resulting in a smaller gas mass fraction in the universe at present
(Figs.~\ref{fig:C}h, j and l).
With regard to metallicity distributions, galaxies of larger mass-loading factors
tend to have lower ${\rm [Fe/H]_{peak}}$s (Fig.~\ref{fig:C}k).
The metallicities of the most metal-rich stars seem to be
almost consistent among galaxies of different mass-loading factors.

The trends shown in Fig.~\ref{fig:C} suggest
that the observed gas-phase oxygen abundance at the gas mass fraction 
of NGC 6822 can be explained by models of $\eta\gtrsim3$ (Fig.~\ref{fig:C}d).
If the SFE is too low, a galaxy has a large gas mass fraction and 
low gas-phase oxygen abundance in the universe at present compared to
the observed values.
On the other hand, if the SFE is high, a galaxy evolves rapidly,
resulting in a small gas mass fraction and high gas-phase oxygen abundance.
Thus, the range of SFE is also roughly constrained. 

\begin{figure}
\centering
\begin{tabular}{c}
\begin{minipage}{0.5\hsize}
\centering
\includegraphics[scale=0.46]{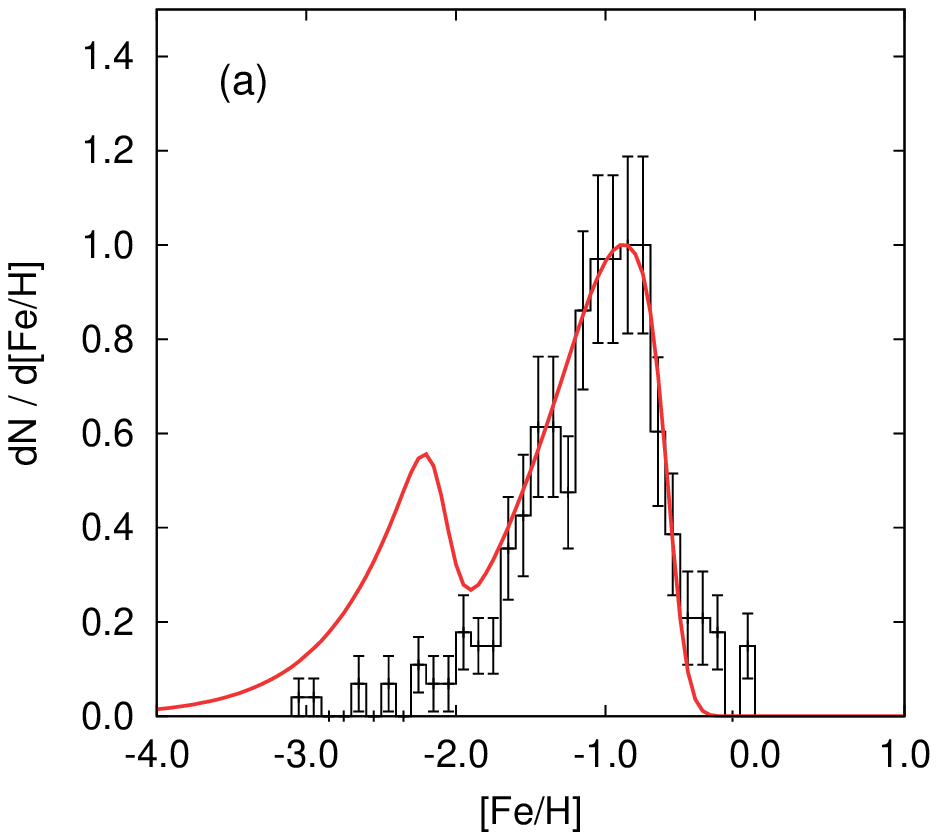}
\end{minipage}
\begin{minipage}{0.5\hsize}
\centering
\includegraphics[scale=0.46]{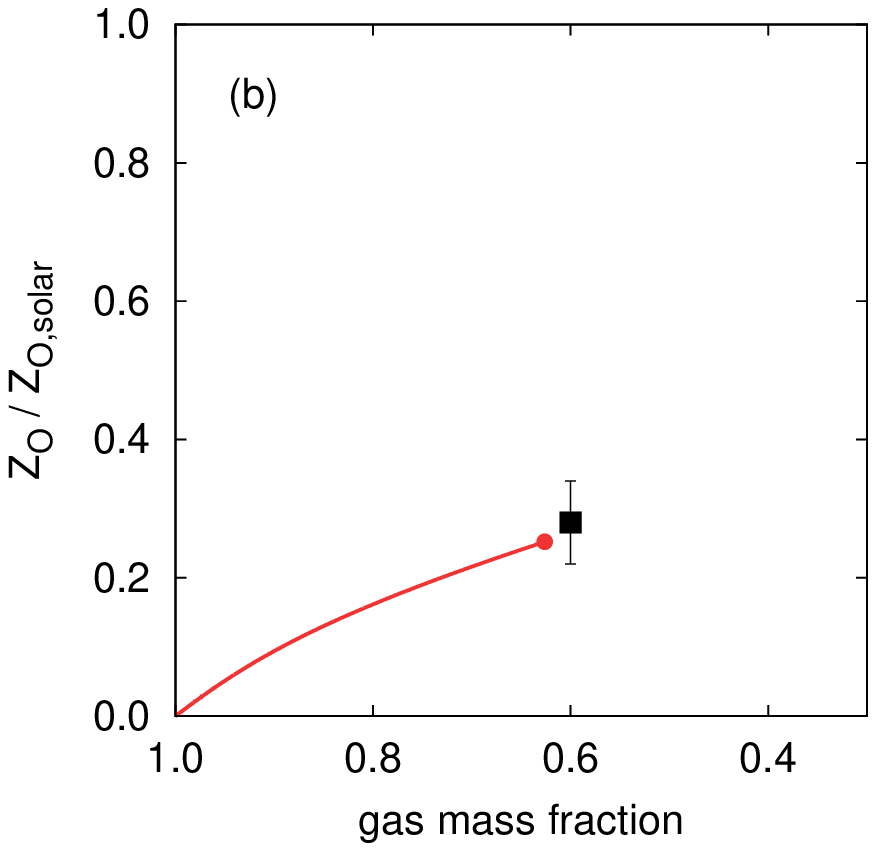}
\end{minipage}
\end{tabular}
\caption{Same as Fig.~\ref{fig:C}, but the case of $\eta=6.0$ 
(red curves) is investigated.
The SFE is assumed to be $k_{\rm SF}=0.020~({\rm Gyr}^{-1})$.
}
\label{fig:C2}
\end{figure}

In Fig.~\ref{fig:C2}, the case of $\eta=6.0$ and 
$k_{{\rm SF}}=0.020~({\rm Gyr}^{-1})$
is examined as an example.
The gas mass fraction and gas-phase oxygen abundance are almost consistent
with the values predicted
by the model
(Fig.~\ref{fig:C2}b).
The predicted metallicity distribution is also roughly 
consistent with the observed distribution within the margin of error
in the metallicity range of [Fe/H]$\sim-2.0$ to $-0.5$.
The values of the mass-loading factor that explain the observational data
are not markedly different from those suggested by other theoretical studies
using simulations
\citep[e.g. $3\lesssim\eta\lesssim10$ for galaxies of ${\rm M_*}\sim10^8~{\rm M_{\odot}}$,][]{M17}.

In summary, the gas-phase oxygen abundance, gas mass fraction and 
metallicity distribution ($-2.0\lesssim$[Fe/H]$\lesssim-0.5$) can be
roughly explained by model C, in which continuous star formation and 
outflow are assumed.
Thus, if NGC 6822 is an outflow-dominated system, 
the history of star formation and outflow of this galaxy 
may be continuous.
Meanwhile, it should be noted that the present study is based on
the simplified models.
In model C, it is assumed that a galaxy is formed from a gas cloud,
while observations suggest the accretion of other systems, such as
dwarf galaxies \citep[e.g.][]{H14}.
More importantly, model C generally overproduces stars of [Fe/H]$\lesssim -2$
(Fig.~\ref{fig:C2}).
As discussed in Sec.~\ref{sec:4B} and previous studies 
\citep[e.g.][]{PP75,P03},
the gas accretion may alleviate the overproduction.
Thus, outflow may be a dominant process, 
but other physical processes, including the accretion of gas or 
gas-rich systems, are not completely excluded.
For further investigation of the history of star formation and gas flows, 
detailed modellings and observational information, 
such as stellar abundance ratios, are needed.

\subsection{Caveats}
\label{subsec:3.4}

It should be noted that observational data are compared to models 
with a few assumptions.
In the models, SNe\,II and SNe\,Ia are assumed to be
the main contributors to chemical enrichment,
but it is possible that other events 
e.g. hypernovae or different types of supernovae,
affect the enrichment of oxygen and iron.
The assumptions about SNe\,Ia are also oversimplified.
The lifetime and fraction of the progenitors are assumed to be 
fixed values.
Relaxation of these assumptions may result in different values of 
predicted gas-phase and stellar abundances,
and thus lead to different conclusions.
In addition, 
the significance of dust depletion in a low-metallicity environment is 
still unclear.
Additional observational and theoretical data are required.

\section{Summary and discussion}
\label{sec:5}

This paper discusses the chemical evolution of a dwarf irregular galaxy 
in the Local Group NGC 6822 from the viewpoints of gas-phase 
and stellar abundances.
Observed gas-phase oxygen abundance, gas mass fraction and stellar 
metallicity distribution are compared to chemical evolution models in which
continuous star formation, gas accretion and/or outflow are taken into account.

In the cases of the closed-box and infall models, 
the gaseous properties can be explained
if steeper IMFs are allowed.
On the other hand, the shapes of the metallicity distributions predicted 
by the models are not consistent with the observed distribution,
suggesting that the galaxy has a more complex history
of star formation, gas flows and/or chemical enrichment 
than assumed in the models.

When the observational data are compared to the outflow model,
the observed values of gas-phase oxygen abundance and gas mass fraction 
can be explained.
The shape of the observed metallicity distribution is also roughly 
consistent with those predicted by some of the models 
in the metallicity range of $-2.0\lesssim$[Fe/H]$\lesssim-0.5$.
Thus, if NGC 6822 is an outflow-dominated system,
the history of star formation and outflow may be continuous.
This result does not mean that the accretion of gas or gas-rich systems 
is completely excluded.
For further discussion about the history of star formation and gas flows, 
additional observables and detailed modelling are needed.
Because stars of different properties (e.g. mass) contribute to the
enrichment of different elements, 
stellar abundance ratios may give insight into the star formation 
history.
With regard to the observed stellar metallicity,
the shape of metallicity distributions might be different
if the sample size and the method of metallicity measurement 
are different \citep[][]{S16}.
More spectroscopic data will help understand dominant physical processes
that affect the chemical evolution.

Finally, each dwarf galaxy has a unique star formation history,
so studies of other dwarf irregular galaxies from the viewpoints of gas-phase 
and stellar metallicities may provide further insight into the variation
in the star formation and gas flow histories.



\section*{Acknowledgements}

Sincere thanks are due to Dr. Yuhri Ishimaru, who passed away in 
November 2017.
This study has been developed from a study
conducted under her guidance.
The reviewer's comments have also greatly improved the manuscript.








\appendix
\section{The fraction and lifetime of progenitors of type-Ia supernovae}
\label{app:A}
Although it is the possible that the mechanisms of SNe\,Ia
differ among galaxies \citep[e.g.][]{K19},
it is still not clear whether or how the mechanism of SNe\,Ia
varies from galaxy to galaxy.
In the metallicity range of $-2.5\lesssim$[Fe/H]$\lesssim0$,
the binary fraction of field stars may not significantly varied with
the metallicity \citep{C05}.
In this study, the fraction $f_{{\rm I_a}}$ and the lifetime $\tau_{{\rm I_a}}$ 
in models A--C are determined based on the assumption that 
the mechanism of SNe\,Ia in dwarf galaxies 
is not significantly different from that in the Milky Way.

The accretion of gas may be one way in which
the metallicity distribution of long-lived stars 
in the solar neighbourhood can be explained \citep[e.g.][]{PP75}.
Assuming that the formation of the thin disc of the Milky Way is dominated 
by the accretion of primordial gas, the metallicity distribution of long-lived
stars in the thin disc and stellar abundance ([Mg/Fe]) of the Milky Way 
are compared to the infall model (model B).
The one-zone model is probably too simple to completely 
reproduce the observational data of the Milky Way \citep[][]{P18}.
Previous studies have indicated that there are two or three families 
of stars on the [Mg/Fe]--[Fe/H] plane at 
[Fe/H]$\gtrsim-0.7$ \citep[e.g.][]{A11},
and their origins are still a matter of debate.
In this study, $\tau_{{\rm I_a}}$ and $f_{{\rm I_a}}$ are determined, and 
trends of the observational data are broadly explained using model B.
With regard to the [Mg/Fe]--[Fe/H] diagram, it is assumed that $\alpha$-elements are produced
mainly by massive stars in the early stages of evolution, and then [Mg/Fe]
begins to decline due to the enrichment of Fe by SNe\,Ia.

\begin{figure}
\centering
\begin{tabular}{c}
\begin{minipage}{0.50\hsize}
\centering
\includegraphics[scale=0.42]{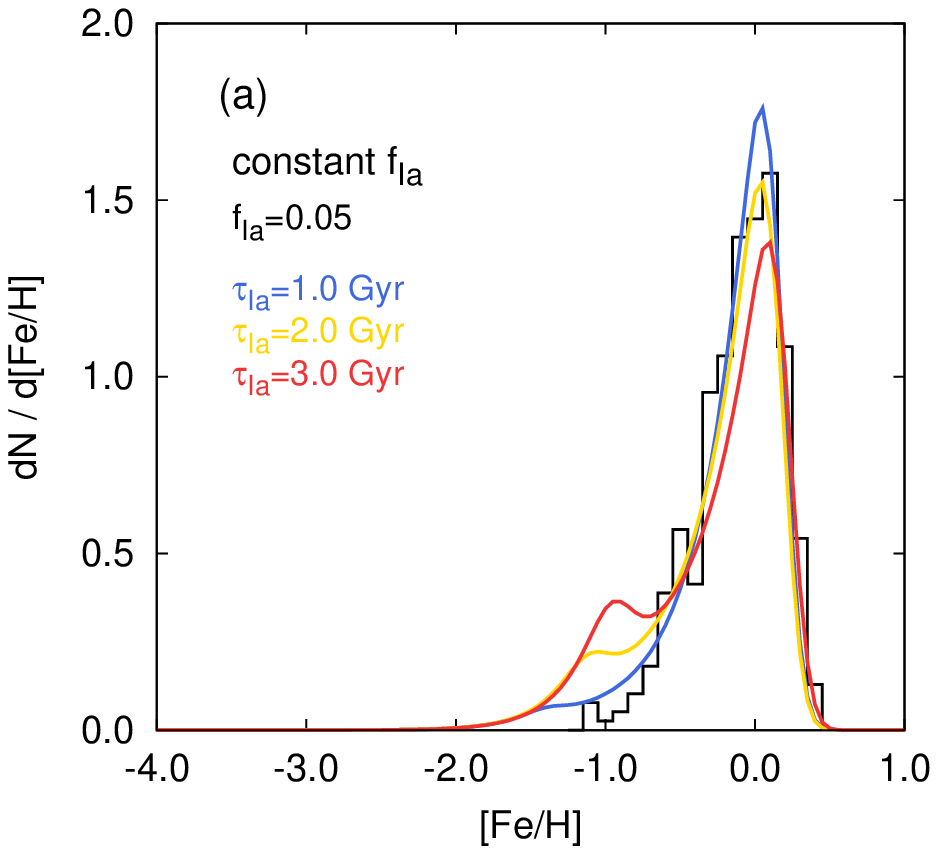}
\end{minipage}
\begin{minipage}{0.50\hsize}
\centering
\includegraphics[scale=0.42]{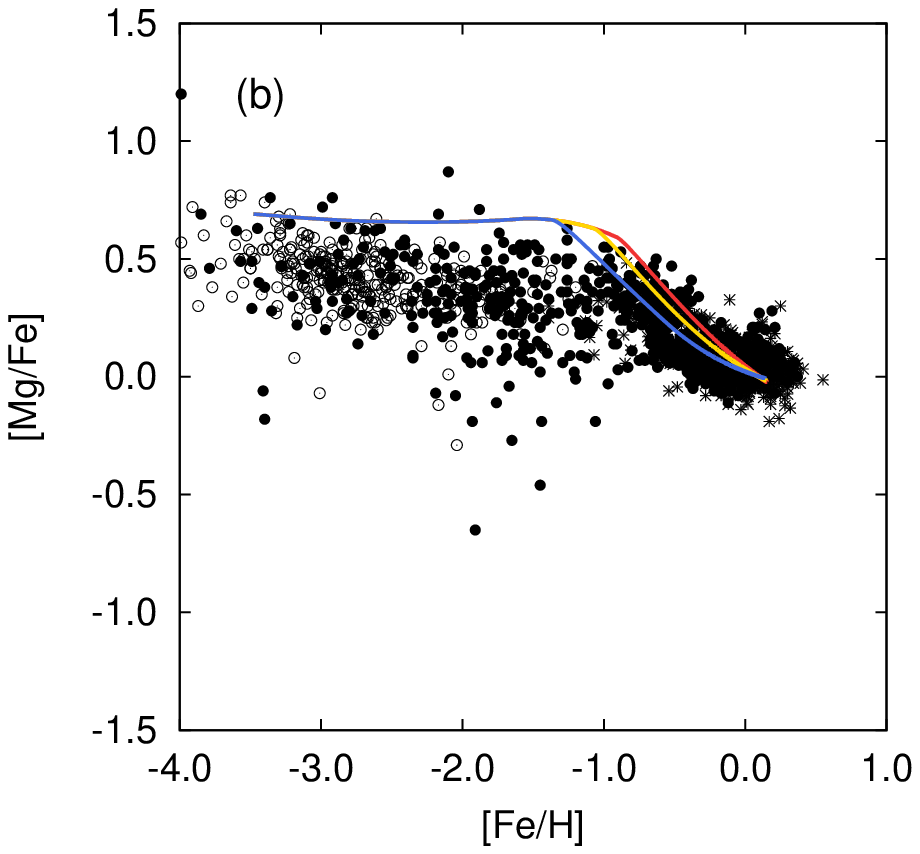}
\end{minipage}\\
\begin{minipage}{0.50\hsize}
\centering
\includegraphics[scale=0.42]{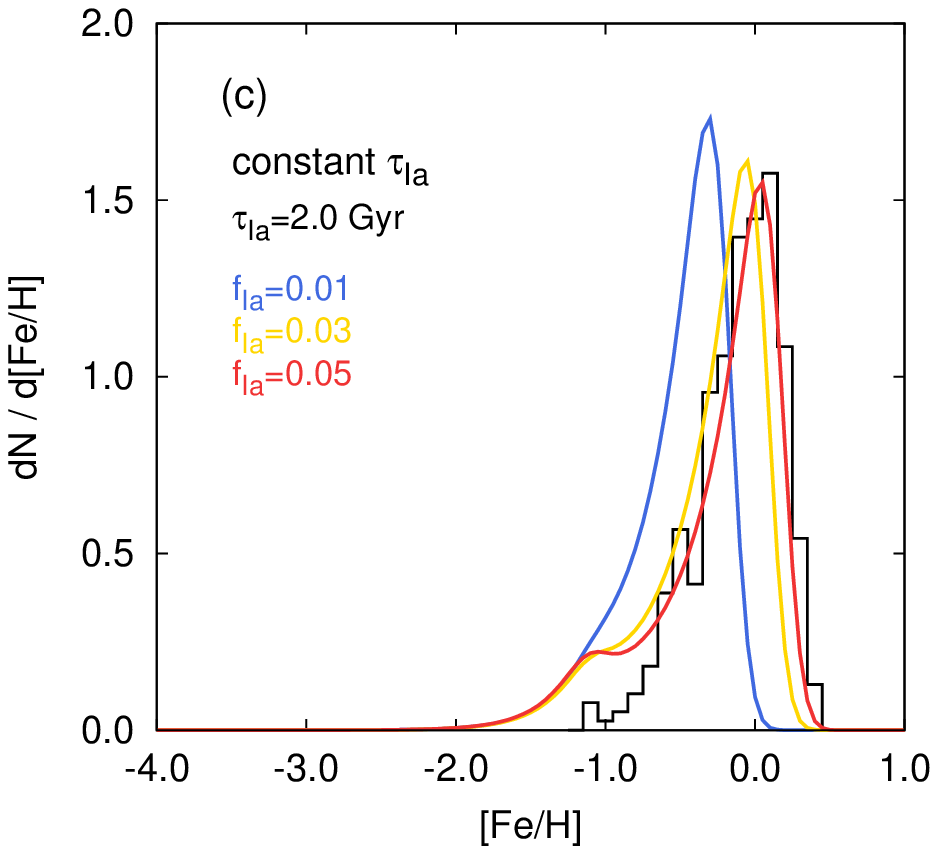}
\end{minipage}
\begin{minipage}{0.50\hsize}
\centering
\includegraphics[scale=0.42]{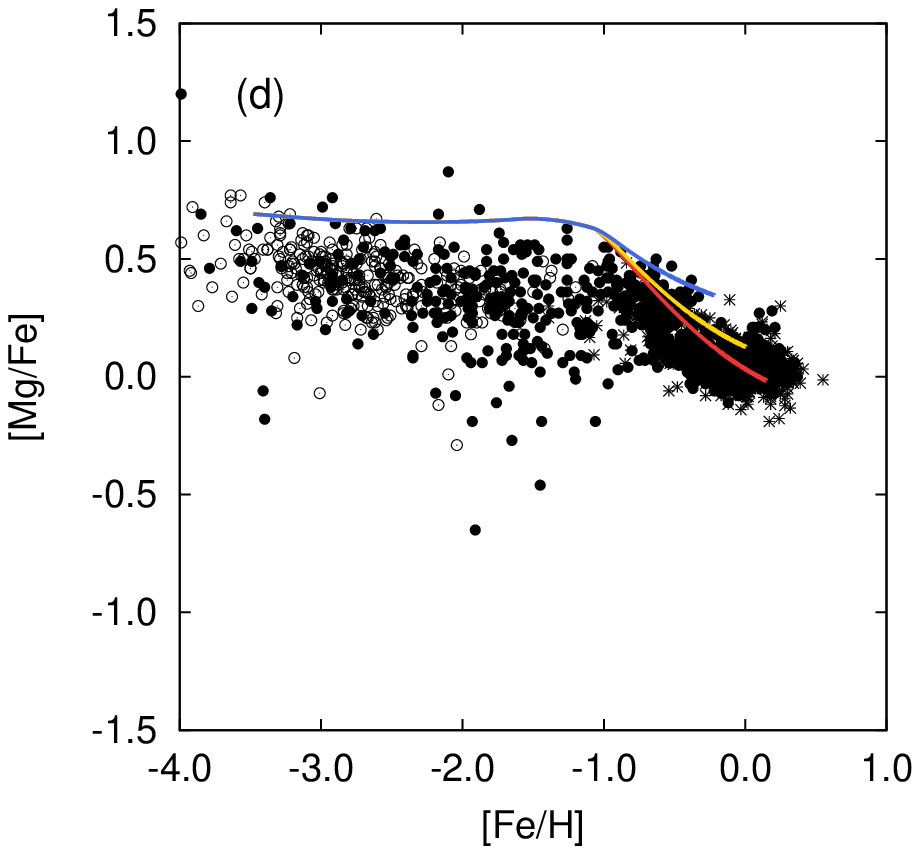}
\end{minipage}
\end{tabular}
\caption{Metallicity distributions (panels a and c) and [Mg/Fe] 
(b and d) predicted by model B.
(Panels a and b) The case where the lifetime of progenitors of SNe\,Ia 
$\tau_{{\rm I_a}}$ is varied while the fraction $f_{\rm I_a}$
is fixed at $f_{{\rm I_a}}=0.05$.
Blue, yellow and red curves show cases of $\tau_{{\rm I_a}}=1.0, 2.0$
and $3.0$~(Gyr), respectively.
(Panels c and d) The case where $f_{{\rm I_a}}$ is varied while $\tau_{{\rm I_a}}$
is fixed at $\tau_{{\rm I_a}}=2.0$~(Gyr).
Blue, yellow and red curves show cases of $f_{{\rm I_a}}=0.01, 0.03$
and $0.05$.
For all panels, the SFE and ACE
are assumed to be $k_{{\rm SF}}=0.40~({\rm Gyr}^{-1})$ and 
$k_{{\rm in}}=0.10~({\rm Gyr}^{-1})$, respectively.
The Salpeter IMF is assumed.
The black histograms in panels a and c show the observed metallicity
distribution, as reported by \citet{B14}.
Data of [Mg/Fe] of individual stars in the Milky Way (panels b and d)
are taken from
\citet[][dots]{V04}, \citet[][open circles]{R14} 
and \citet[][asterisks]{A12}.
}
\label{fig:MW1}
\end{figure}

Fig.~\ref{fig:MW1} shows how the metallicity distribution and 
[Mg/Fe] can be varied by $\tau_{{\rm I_a}}$ and $f_{{\rm I_a}}$.
Figs.~\ref{fig:MW1}a and b show the case where different $\tau_{{\rm I_a}}$ 
are assumed while $f_{{\rm I_a}}$ is fixed.
As shown in the [Mg/Fe]-[Fe/H] diagram,
[Mg/Fe] starts to decline at higher [Fe/H] with longer $\tau_{{\rm I_a}}$.
In model B, when the lifetime of progenitors of SNe\,Ia is 
set to $\tau_{{\rm I_a}}$~(Gyr), the first SN\,Ia explosion occurs
$\tau_{{\rm I_a}}$~Gyr after star formation starts in the system.
Until the first SN\,Ia explodes, the chemical enrichment proceeds by
SNe\,II.
Therefore, if $\tau_{{\rm I_a}}$ is assumed to be longer,
the metallicity at the time when SNe\,Ia starts to contribute 
to the chemical enrichment can be high.
This effect also causes a larger number of stars of [Fe/H]$\sim-1.0$
in metallicity distributions for cases of longer $\tau_{{\rm I_a}}$ 
(Fig.~\ref{fig:MW1}a).

Figs.~\ref{fig:MW1}c and d show cases where $f_{{\rm I_a}}$ is varied
while $\tau_{{\rm I_a}}$ is fixed.
If $f_{{\rm I_a}}$ is large, the number of SNe\,Ia is also large,
resulting in more enrichment of Fe by SNe\,Ia.
Thus, if a larger $f_{{\rm I_a}}$ is assumed, 
the [Fe/H] at the peak of metallicity distributions is high
(Fig.~\ref{fig:MW1}c).
In addition, larger $f_{{\rm I_a}}$ is associated with more significant 
declines in [Mg/Fe] at 
[Fe/H]$\gtrsim-1.0$ in the [Mg/Fe]-[Fe/H] diagram (Fig.~\ref{fig:MW1}d).

\begin{figure}
\centering
\begin{tabular}{c}
\begin{minipage}{0.50\hsize}
\centering
\includegraphics[scale=0.42]{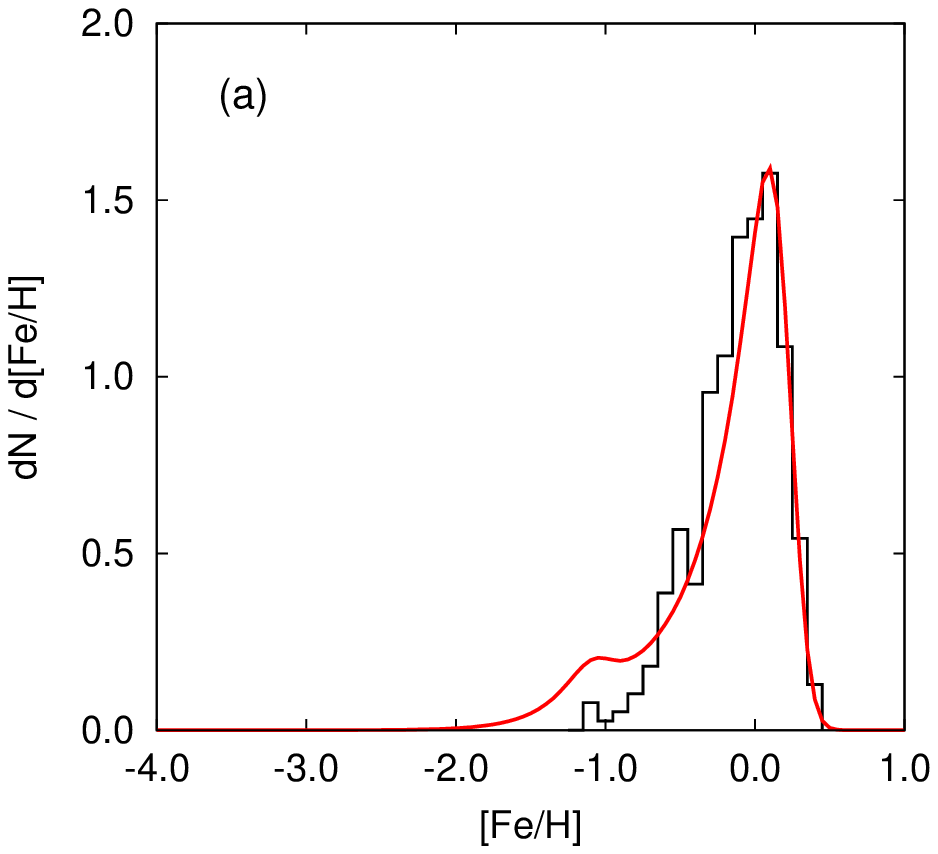}
\end{minipage}
\begin{minipage}{0.50\hsize}
\centering
\includegraphics[scale=0.42]{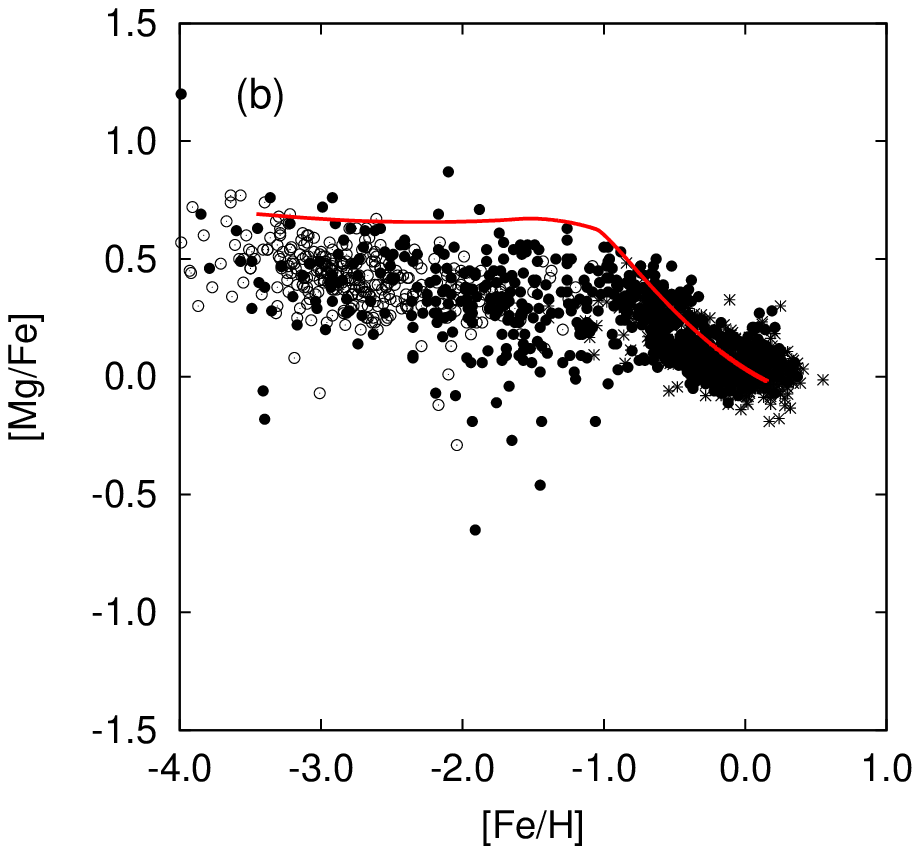}
\end{minipage}
\end{tabular}
\caption{Same as Fig.~\ref{fig:MW1}, but comparing a model of
$\tau_{{\rm I_a}}=2.0$~(Gyr), $f_{{\rm I_a}}=0.05$,
$k_{{\rm SF}}=0.42~({\rm Gyr^{-1}})$ and 
$k_{{\rm in}}=0.10~({\rm Gyr^{-1}})$ (red curves)
to observational data.
The Salpeter IMF is adopted.
The data are the same as in Fig.~\ref{fig:MW1}.
}
\label{fig:MW2}
\end{figure}

The curves shown in Figs. \ref{fig:MW2}a and b are predictions made by model B, 
where $\tau_{{\rm I_a}}=2.0$~(Gyr), $f_{{\rm I_a}}=0.05$, $k_{{\rm SF}}=0.42~({\rm Gyr^{-1}})$ and 
$k_{{\rm in}}=0.10~({\rm Gyr^{-1}})$.
This model broadly explains the shape of the metallicity distribution
(for the overproduction of metal-poor stars, see Sec.~\ref{subsec:modelA}),
the metallicity around the point where [Mg/Fe] starts to decline 
([Fe/H]$\sim-1$) and the solar abundance ([Mg/Fe]$\sim$[Fe/H]$\sim$0).
The gas to stellar fraction predicted by the model is also 
consistent with that of the Milky Way (about 0.1--0.15).
These values of the model are roughly within the ranges suggested in the literature
\citep[e.g.][]{Y96}.
Based on these observations, it is assumed that 5 per cent of stars of 3--8 ${\rm M_{\odot}}$
explode as SNe\,Ia 2~Gyr after their formation.
The observational data can also be explained by different values of $\tau_{{\rm I_a}}$
and $f_{{\rm I_a}}$, such as $\tau_{{\rm I_a}}=1.5$~(Gyr) and $f_{{\rm I_a}}=0.05$, but if the Kroupa IMF
\citep{K01} is adopted, $\tau_{{\rm I_a}}=2.0$~(Gyr) and $f_{{\rm I_a}}=0.05$ are preferred.
The variations in the lifetime and fraction with IMF are not clear,
so this study assumes that these quantities are not correlated with IMF, and 
that these quantities are the same in 
the Milky Way and other galaxies. 

As seen in Fig.~\ref{fig:MW1}a, [Fe/H$]_{\rm peak}$ of metallicity 
distributions seems not to be greatly varied with $\tau_{\rm I_a}$.
While the assumption about $\tau_{\rm Ia}$ is simple, other uncertainty
included in the models, e.g. assumptions about the chemical 
enrichment, may more greatly affect the general results of this study.

\section{Comparison between the observational data and model B}
\label{app:B}
\begin{figure*}
\centering
\begin{tabular}{c}
\begin{minipage}{0.25\hsize}
\centering
\includegraphics[scale=0.46]{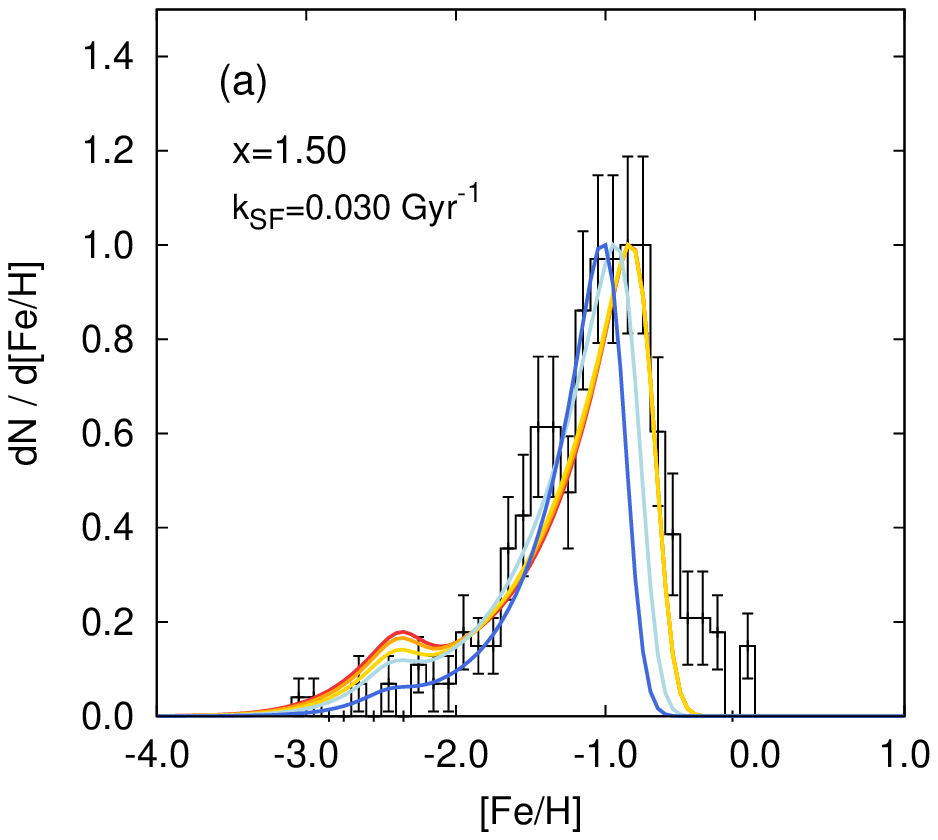}
\end{minipage}
\begin{minipage}{0.25\hsize}
\centering
\includegraphics[scale=0.46]{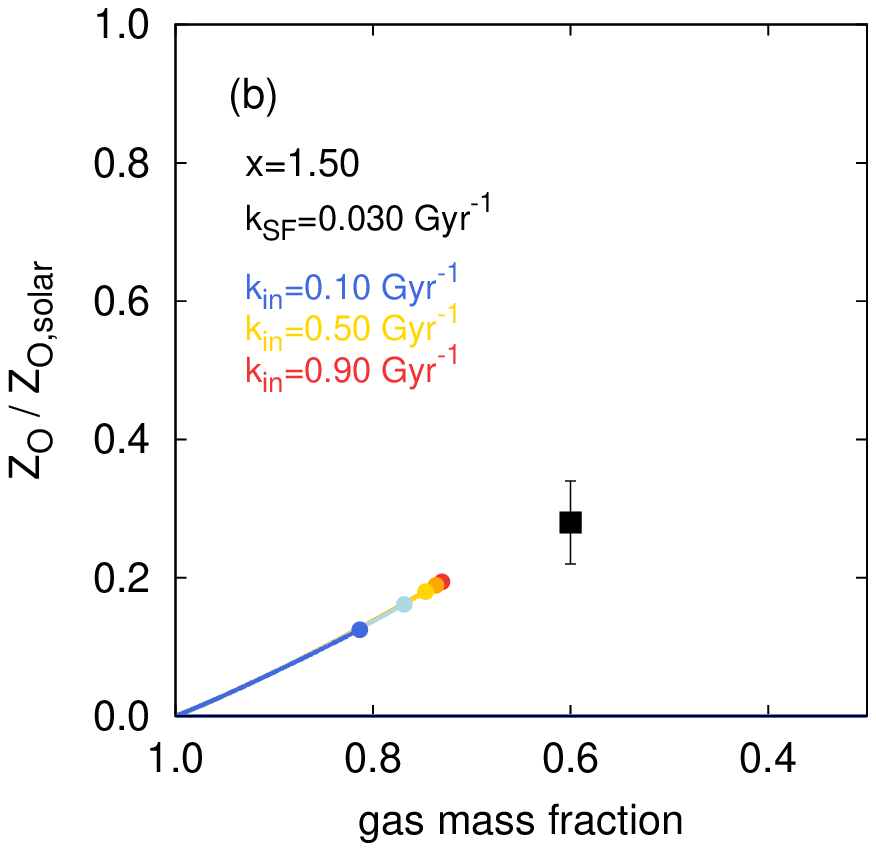}
\end{minipage}
\begin{minipage}{0.25\hsize}
\centering
\includegraphics[scale=0.46]{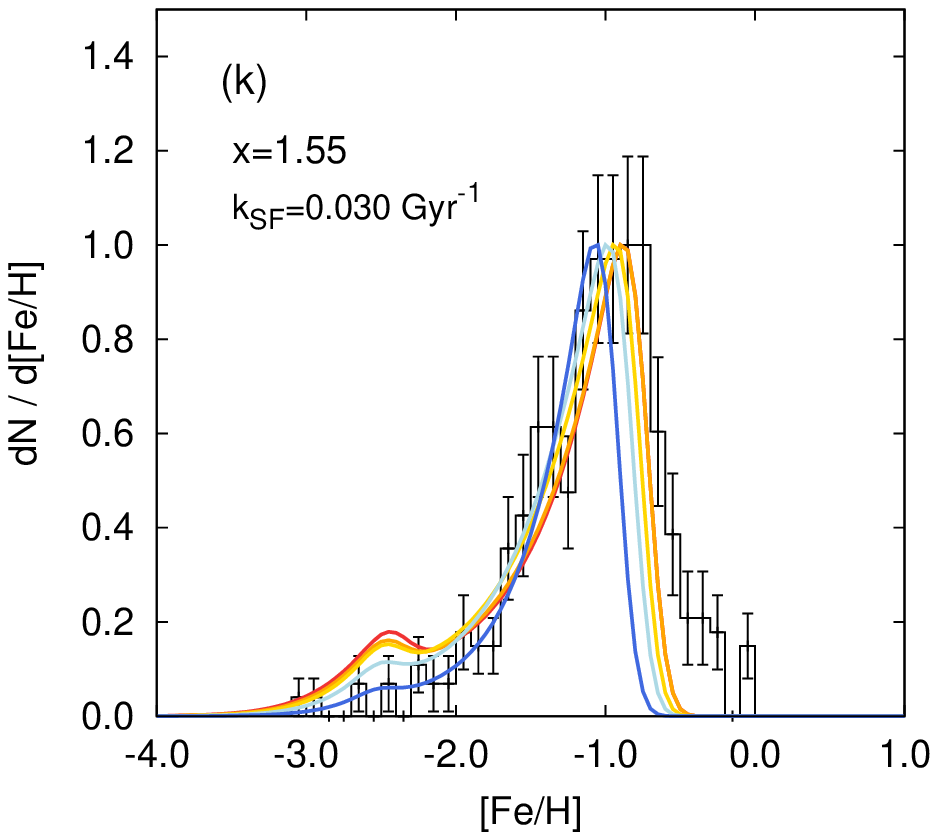} 
\end{minipage}
\begin{minipage}{0.25\hsize}
\centering
\includegraphics[scale=0.46]{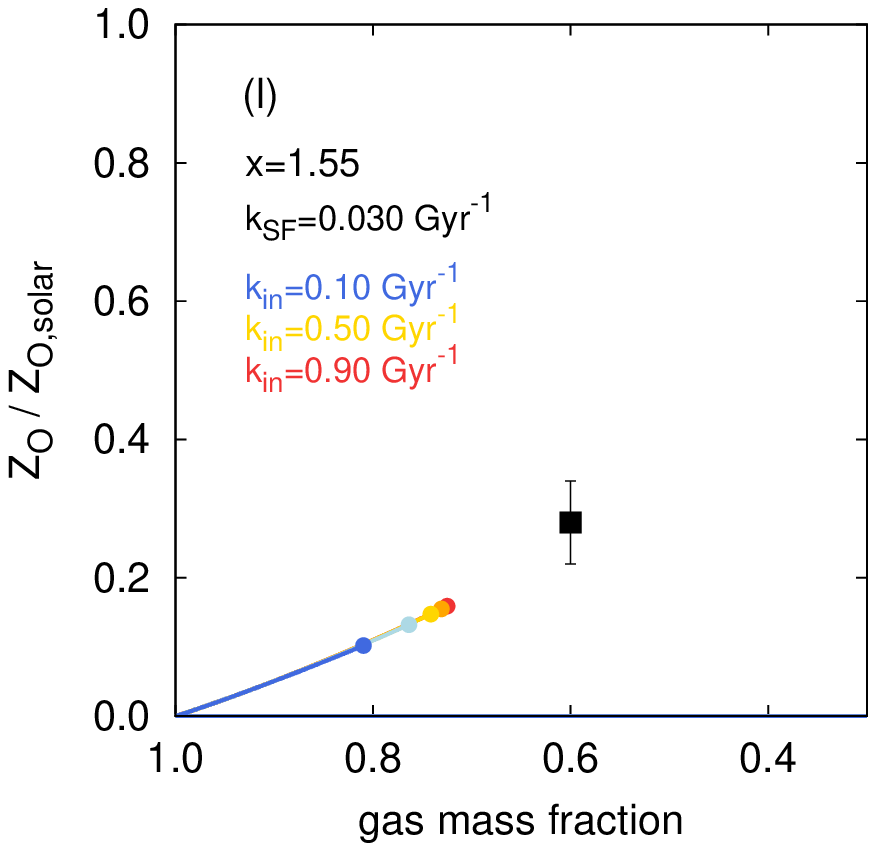} 
\end{minipage} \\
\begin{minipage}{0.25\hsize}
\centering
\includegraphics[scale=0.46]{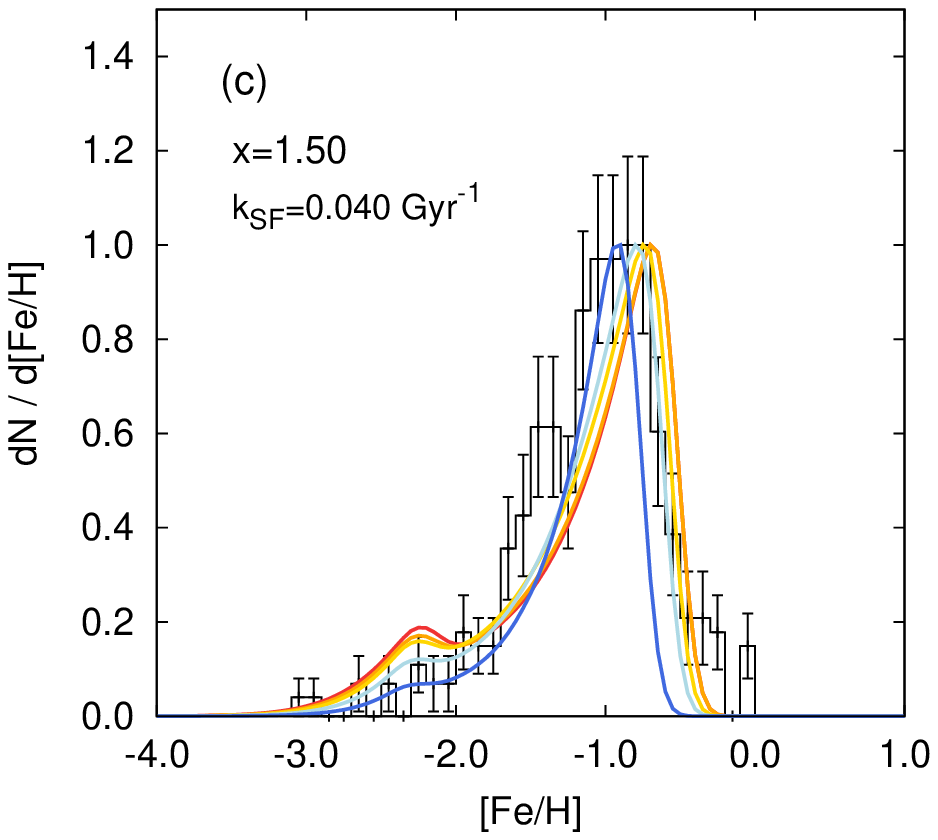} 
\end{minipage}
\begin{minipage}{0.25\hsize}
\centering
\includegraphics[scale=0.46]{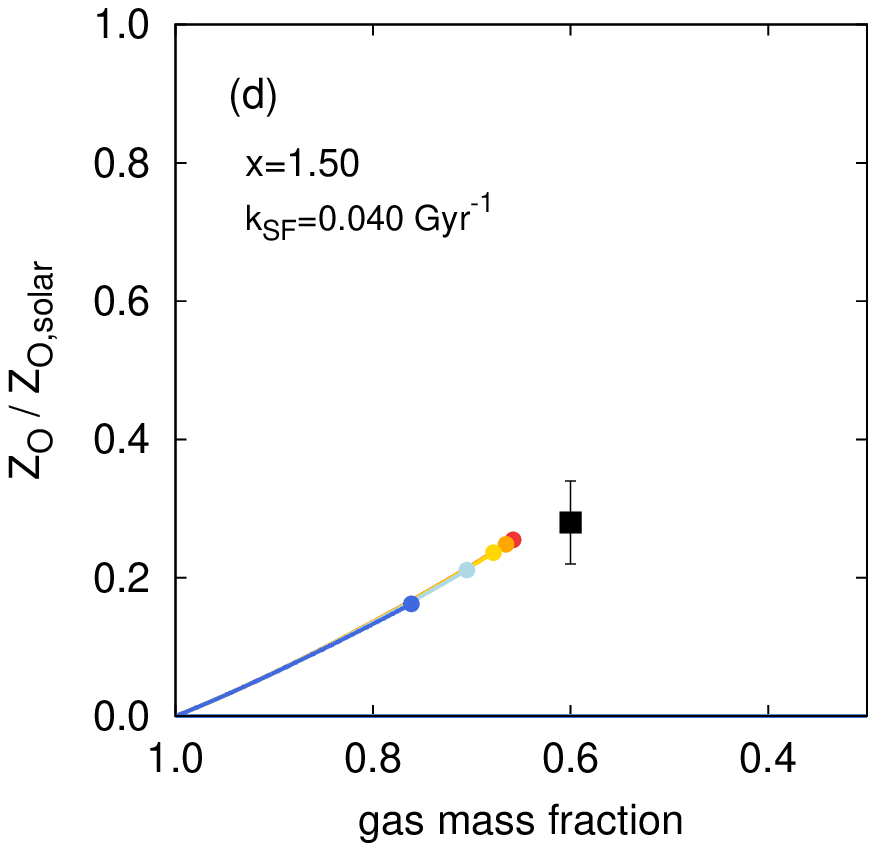} 
\end{minipage}
\begin{minipage}{0.25\hsize}
\centering
\includegraphics[scale=0.46]{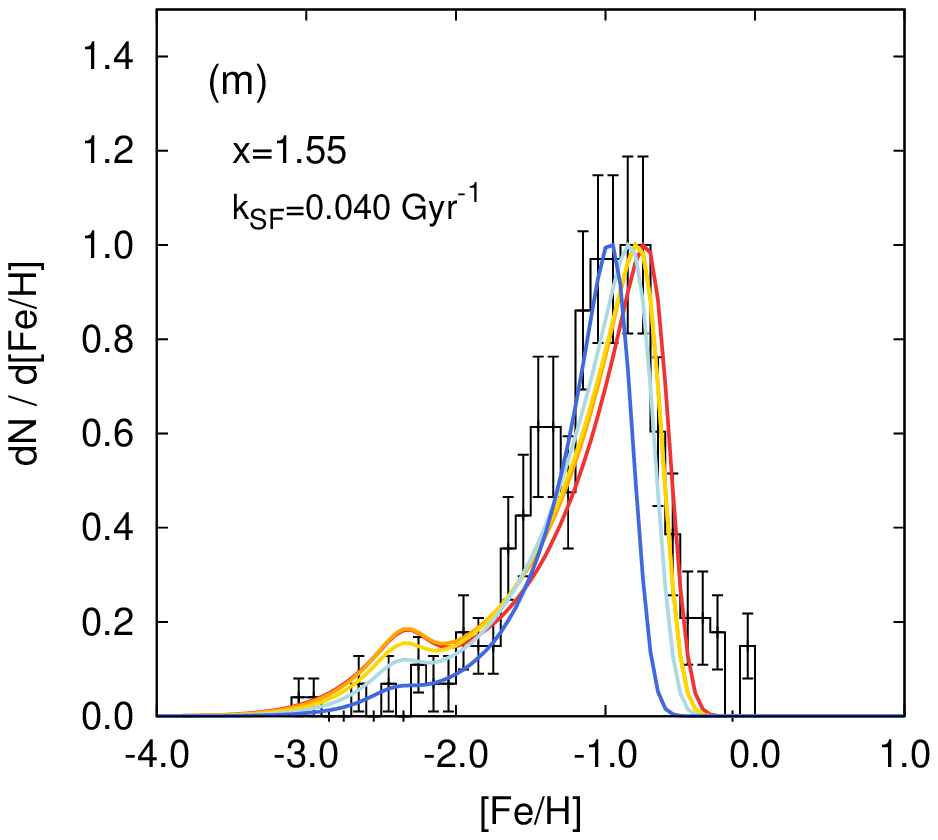} 
\end{minipage}
\begin{minipage}{0.25\hsize}
\centering
\includegraphics[scale=0.46]{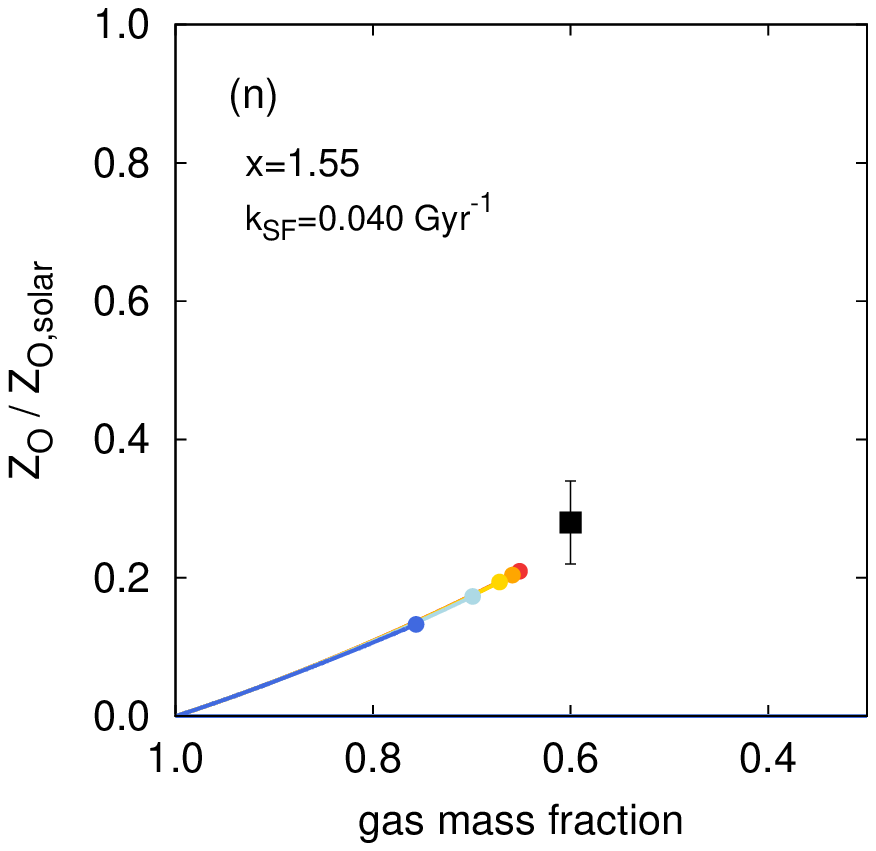} 
\end{minipage} \\
\begin{minipage}{0.25\hsize}
\centering
\includegraphics[scale=0.46]{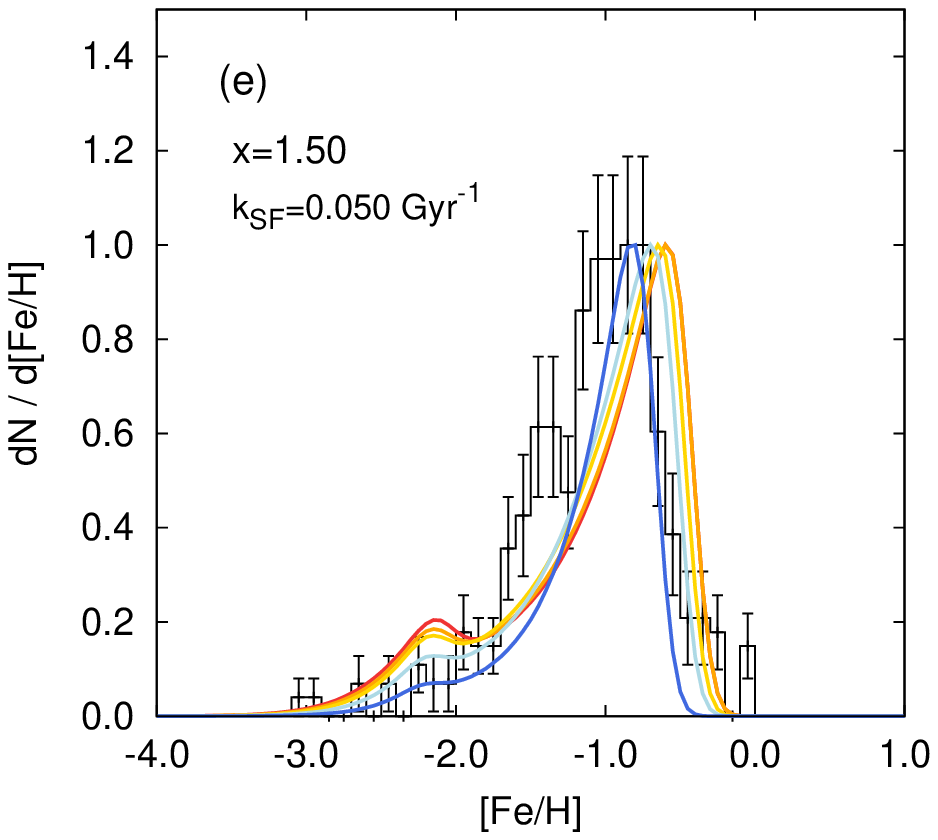}
\end{minipage}
\begin{minipage}{0.25\hsize}
\centering
\includegraphics[scale=0.46]{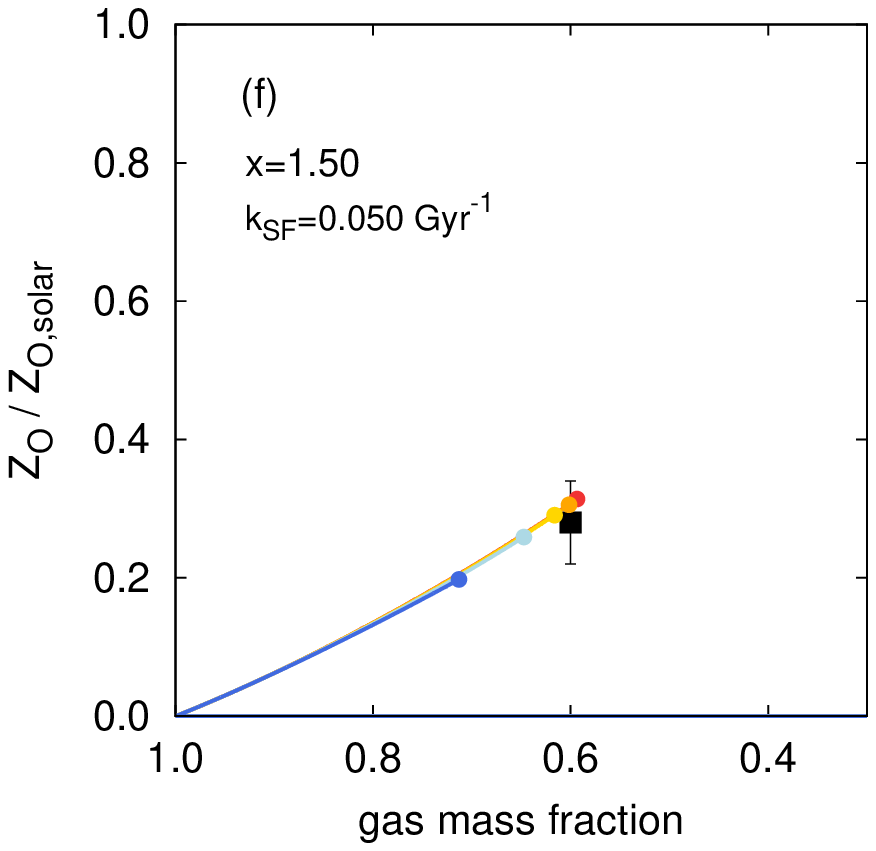}
\end{minipage}
\begin{minipage}{0.25\hsize}
\centering
\includegraphics[scale=0.46]{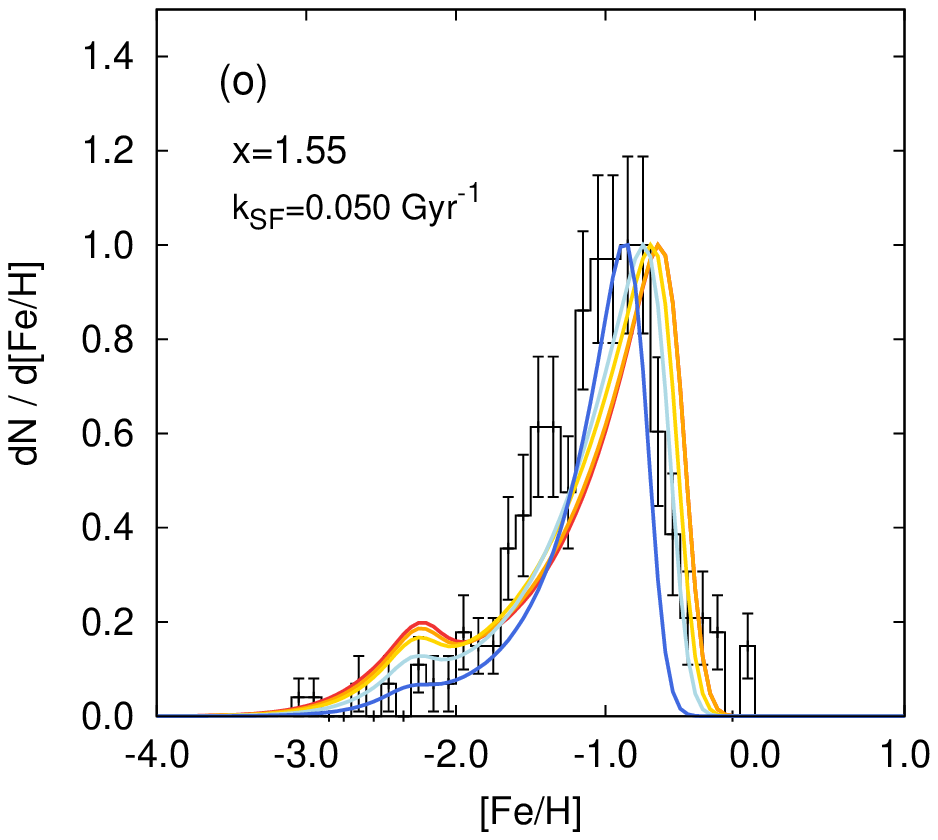} 
\end{minipage}
\begin{minipage}{0.25\hsize}
\centering
\includegraphics[scale=0.46]{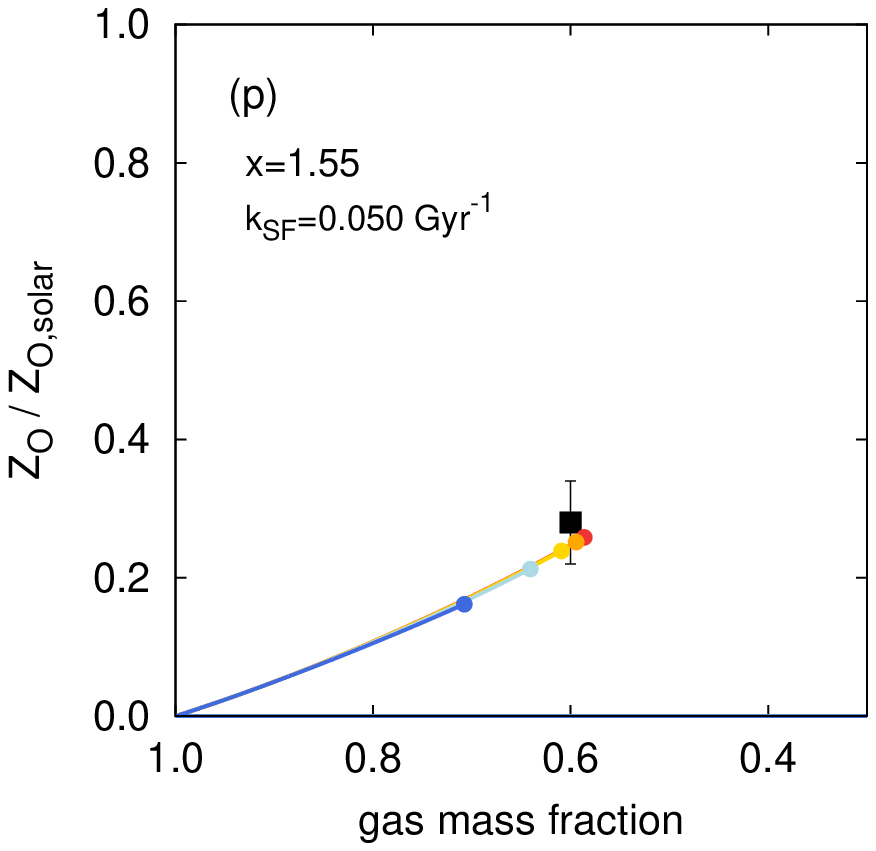} 
\end{minipage} \\
\begin{minipage}{0.25\hsize}
\centering
\includegraphics[scale=0.46]{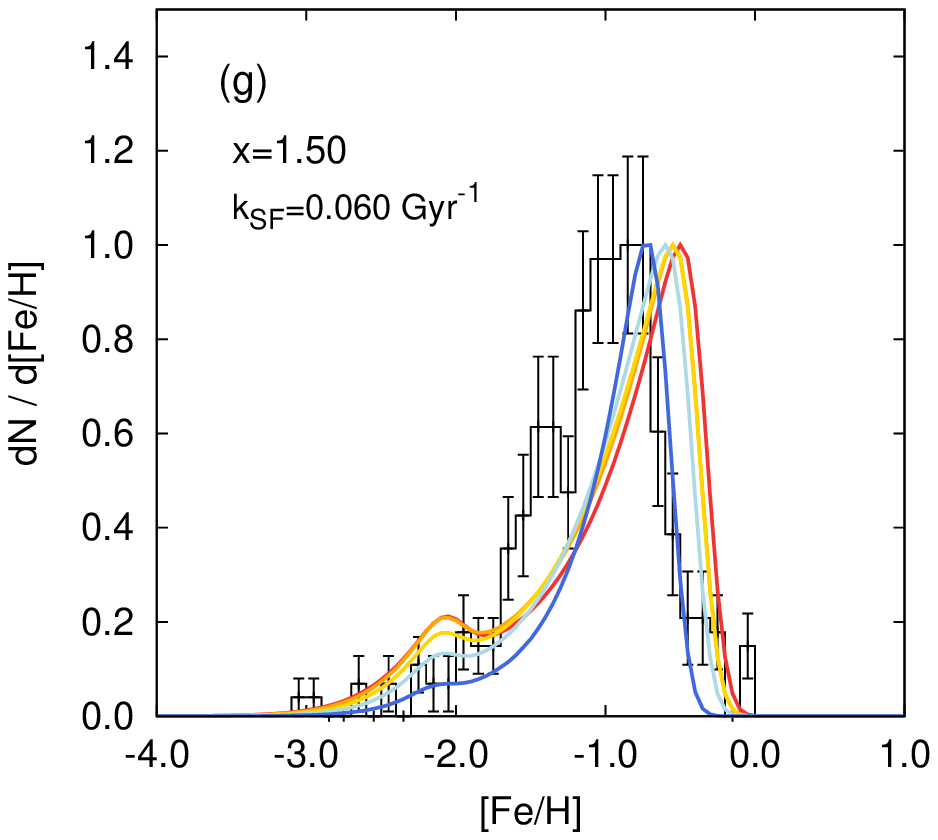} 
\end{minipage}
\begin{minipage}{0.25\hsize}
\centering
\includegraphics[scale=0.46]{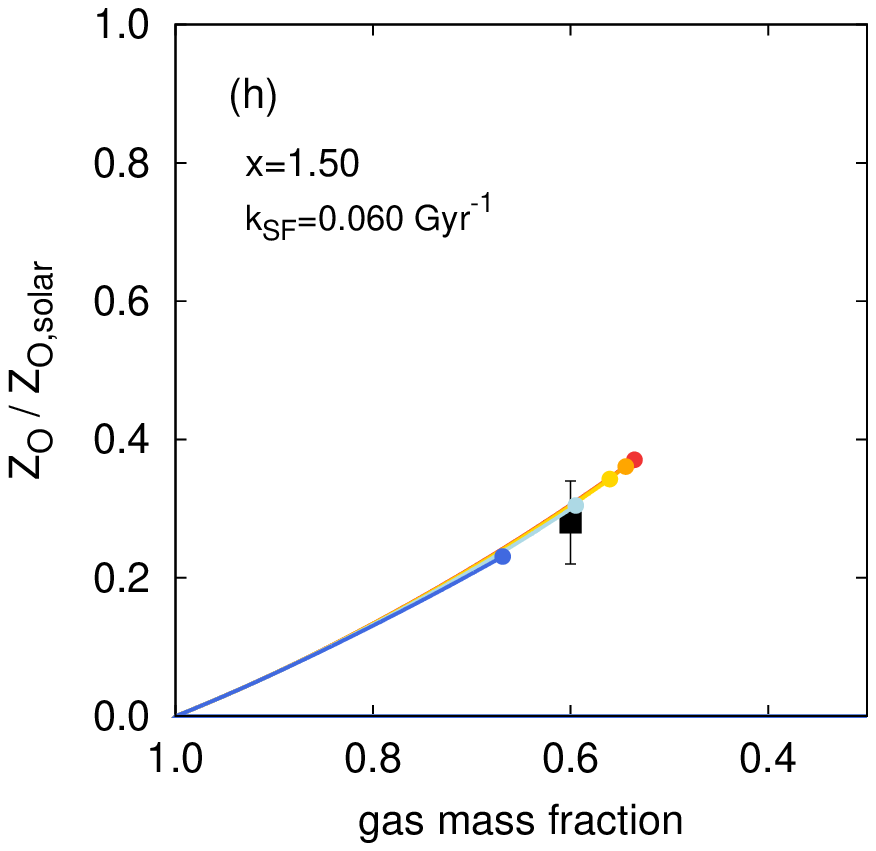} 
\end{minipage}
\begin{minipage}{0.25\hsize}
\centering
\includegraphics[scale=0.46]{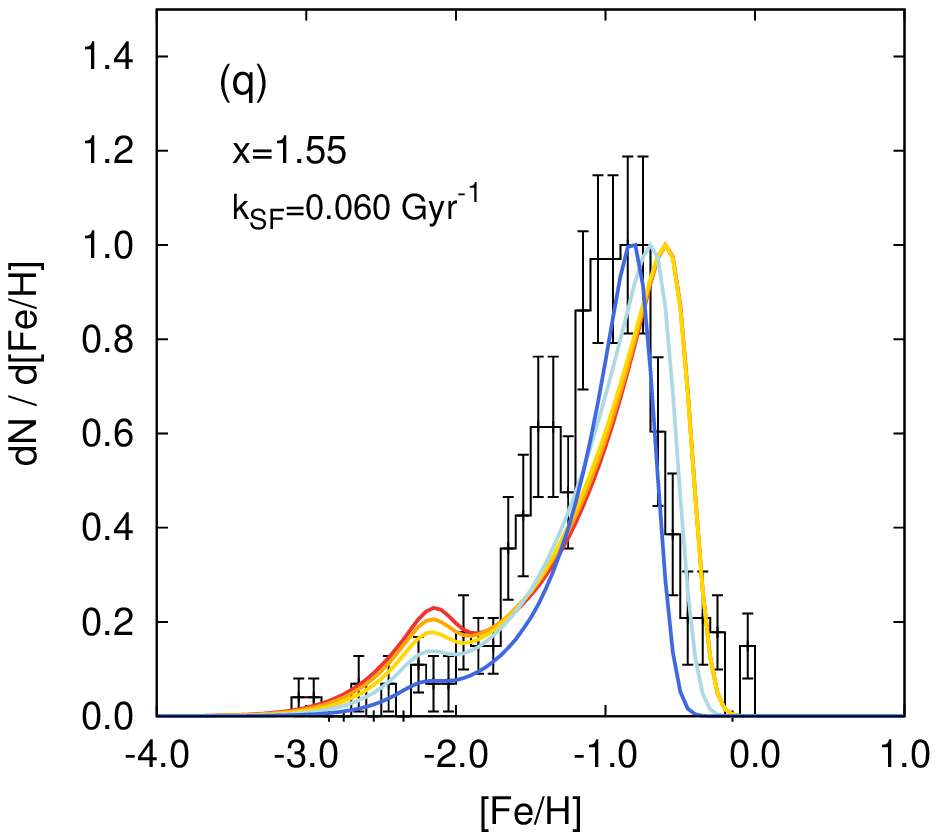} 
\end{minipage}
\begin{minipage}{0.25\hsize}
\centering
\includegraphics[scale=0.46]{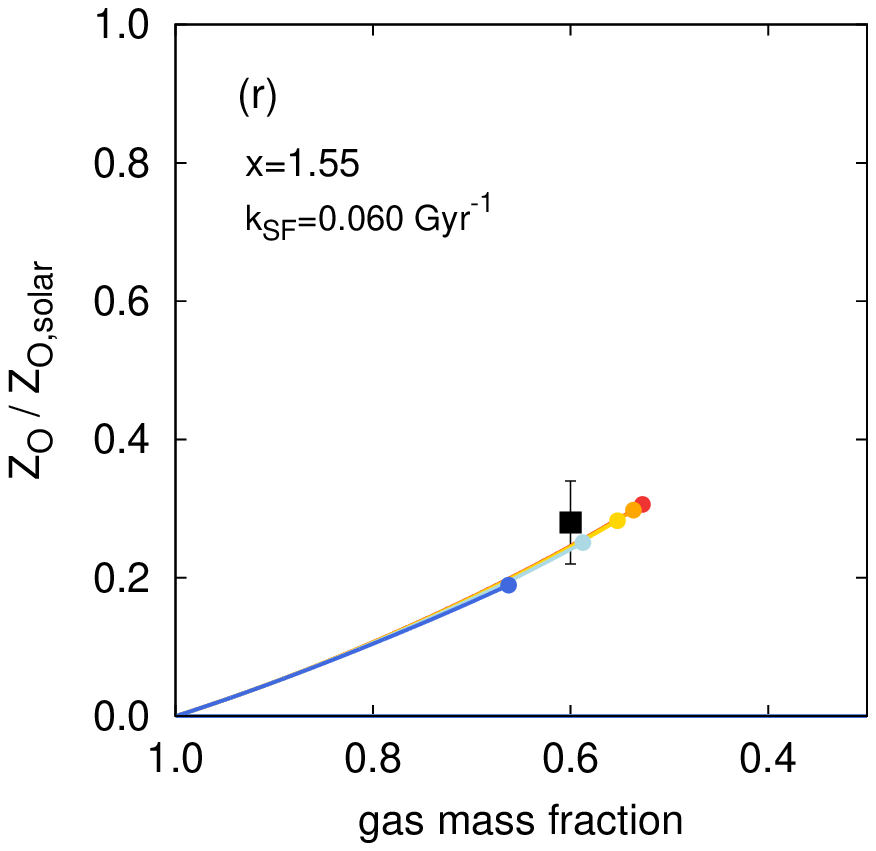} 
\end{minipage} \\
\begin{minipage}{0.25\hsize}
\centering
\includegraphics[scale=0.46]{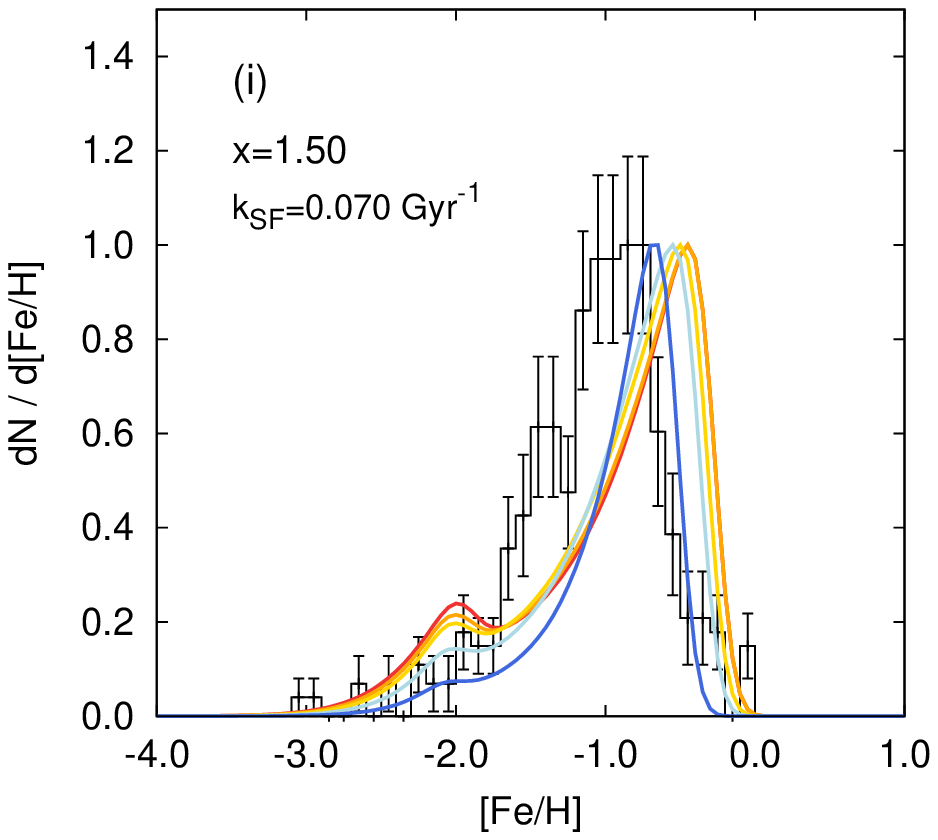}
\end{minipage}
\begin{minipage}{0.25\hsize}
\centering
\includegraphics[scale=0.46]{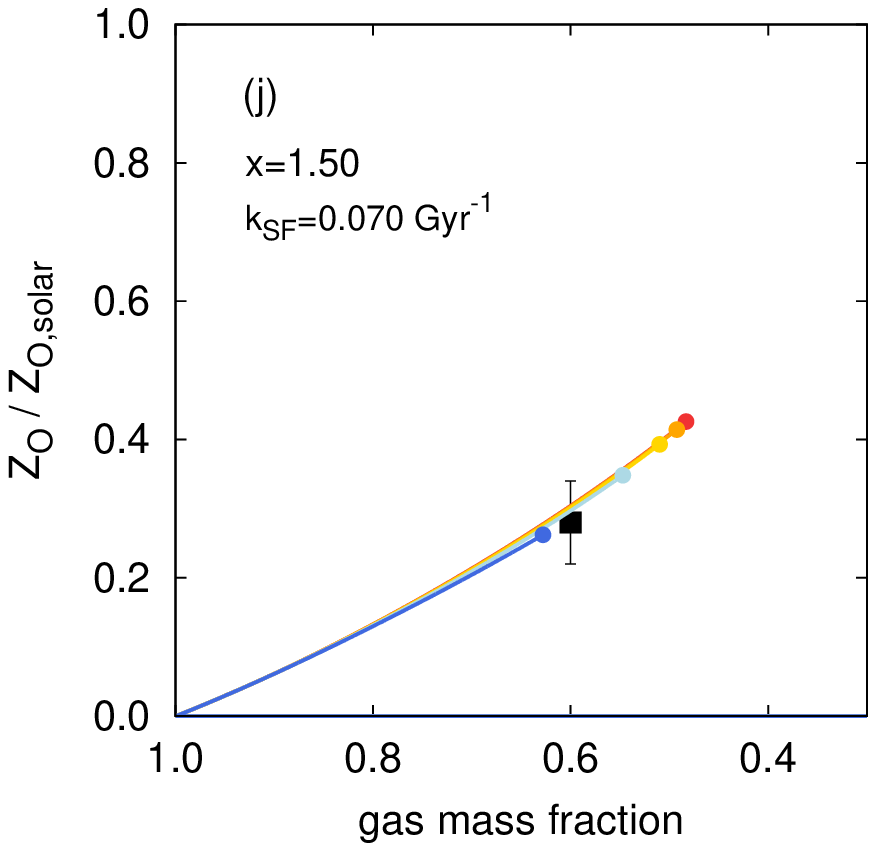}
\end{minipage}
\begin{minipage}{0.25\hsize}
\centering
\includegraphics[scale=0.46]{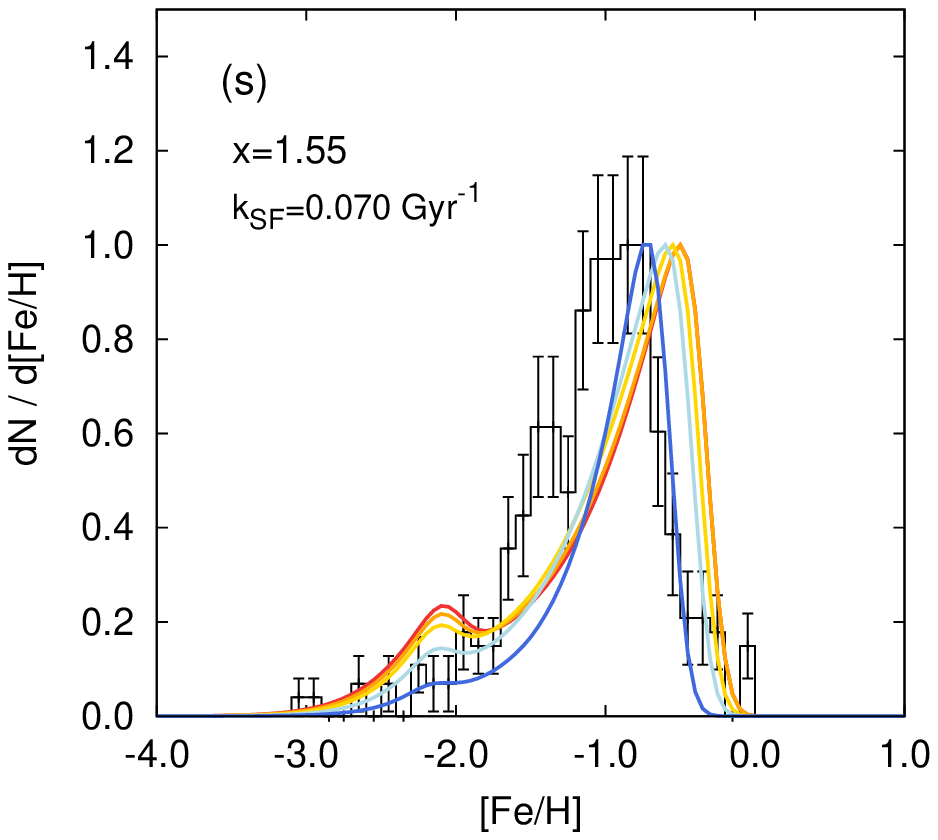} 
\end{minipage}
\begin{minipage}{0.25\hsize}
\centering
\includegraphics[scale=0.46]{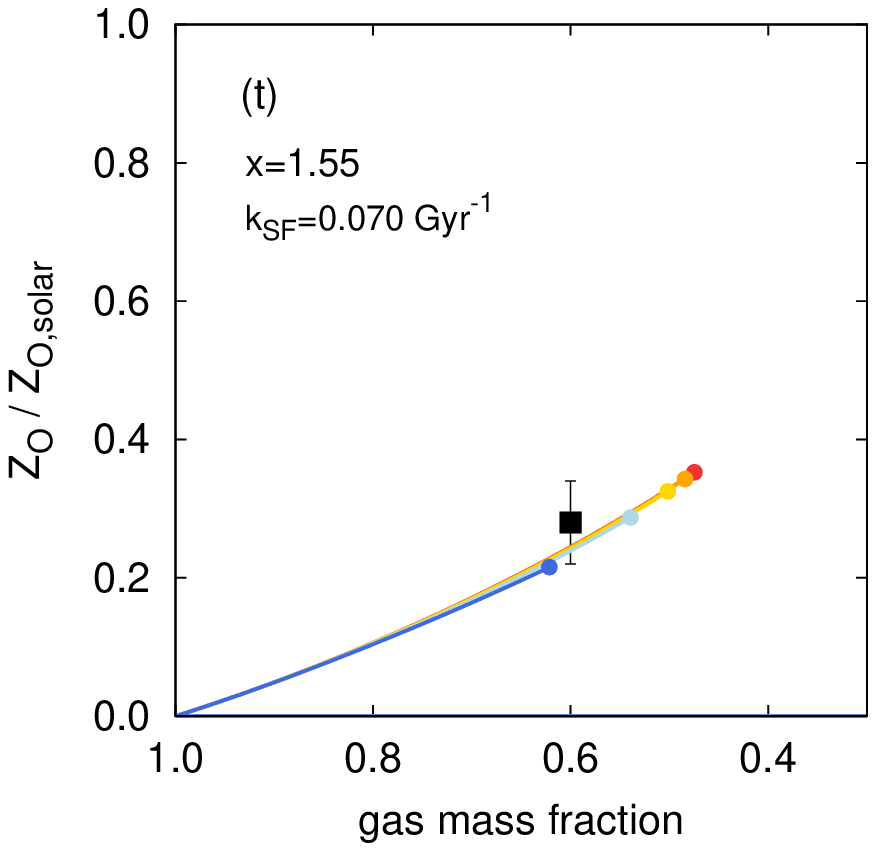} 
\end{minipage}
\end{tabular}
\caption{Comparison between observational data and model B.
Cases of $x=1.50$ and $x=1.55$ are examined in detail.
(a)--(j) Cases of $x=1.50$. SFEs are 
$k_{{\rm SF}}=0.030$ (panels a and b), $0.040$ (c and d), $0.050$ (e and f), 
$0.060$ (g and h) and $0.070~(\rm{Gyr}^{-1})$ (i and j).
The colours of the curves correspond to the value of the ACE:
blue, sky blue, yellow, orange and red curves show cases of
$k_{\rm in}=0.10, 0.30, 0.50, 0.70$ and $0.90~({\rm Gyr^{-1}})$,
respectively.
(k)--(t) Cases of $x=1.55$. 
The SFE is assumed to be 
$k_{{\rm SF}}=0.030$ (panels k and l), $0.040$ (m and n), $0.050$ (o and p), 
$0.060$ (q and r) and $0.070~(\rm{Gyr}^{-1})$ (s and t).
ACEs of the models are the same as those for the case $x=1.50$.
Points on the curves in the $\mu$--$Z_{\rm O}$
diagrams represent the present-day values.
Black histograms and squares in the panels are observational data 
presented in Sec.~\ref{sec:2}.
}
\label{fig:B3}
\end{figure*}

Fig.~\ref{fig:B3} compares the observational data to models of 
$x=1.50$ (Figs.~\ref{fig:B3}a--j) and $1.55$ (Figs.~\ref{fig:B3}k--t)
in detail.
As explained in Sec.~\ref{sec:4B}, 
the observed values of the gas-phase oxygen abundance and 
gas mass fraction of NGC 6822 can be explained by model B, 
if steeper IMFs are allowed.
For example, the observed values are consistent with models of
$x=1.50$, $k_{{\rm SF}}=0.050~\rm{Gyr}^{-1}$ and 
$k_{{\rm in}}\sim0.70~\rm{Gyr}^{-1}$
within the margin of error for the observed gas-phase oxygen abundance
(orange curve and point in Fig.~\ref{fig:B3}f).
However, the shape of the metallicity distributions predicted by the models
seem not to be consistent with the observed distribution.
Similar trends can be seen in Fig.~\ref{fig:B3} for other models.

\bsp	
\label{lastpage}
\end{document}